
\message
{JNL.TEX version 0.92 as of 6/9/87.  Report bugs and problems to Doug Eardley.}

\catcode`@=11
\expandafter\ifx\csname inp@t\endcsname\relax\let\inp@t=\input
\def\input#1 {\expandafter\ifx\csname #1IsLoaded\endcsname\relax
\inp@t#1%
\expandafter\def\csname #1IsLoaded\endcsname{(#1 was previously loaded)}
\else\message{\csname #1IsLoaded\endcsname}\fi}\fi
\catcode`@=12



\font\twelverm=cmr10 scaled 1200    \font\twelvei=cmmi10 scaled 1200
\font\twelvesy=cmsy10 scaled 1200   \font\twelveex=cmex10 scaled 1200
\font\twelvebf=cmbx10 scaled 1200   \font\twelvesl=cmsl10 scaled 1200
\font\twelvett=cmtt10 scaled 1200   \font\twelveit=cmti10 scaled 1200
\font\twelvesc=cmcsc10 scaled 1200  \font\twelvesf=cmcsc10 scaled 1200
\skewchar\twelvei='177   \skewchar\twelvesy='60


\def\twelvepoint{\normalbaselineskip=12.4pt plus 0.1pt minus 0.1pt
  \abovedisplayskip 12.4pt plus 3pt minus 9pt
  \belowdisplayskip 12.4pt plus 3pt minus 9pt
  \abovedisplayshortskip 0pt plus 3pt
  \belowdisplayshortskip 7.2pt plus 3pt minus 4pt
  \smallskipamount=3.6pt plus1.2pt minus1.2pt
  \medskipamount=7.2pt plus2.4pt minus2.4pt
  \bigskipamount=14.4pt plus4.8pt minus4.8pt
  \def\rm{\fam0\twelverm}          \def\it{\fam\itfam\twelveit}%
  \def\sl{\fam\slfam\twelvesl}     \def\bf{\fam\bffam\twelvebf}%
  \def\mit{\fam 1}                 \def\cal{\fam 2}%
  \def\sc{\twelvesc}		   \def\tt{\twelvett}
  \def\sf{\twelvesf}
  \textfont0=\twelverm   \scriptfont0=\tenrm   \scriptscriptfont0=\sevenrm
  \textfont1=\twelvei    \scriptfont1=\teni    \scriptscriptfont1=\seveni
  \textfont2=\twelvesy   \scriptfont2=\tensy   \scriptscriptfont2=\sevensy
  \textfont3=\twelveex   \scriptfont3=\twelveex  \scriptscriptfont3=\twelveex
  \textfont\itfam=\twelveit
  \textfont\slfam=\twelvesl
  \textfont\bffam=\twelvebf \scriptfont\bffam=\tenbf
  \scriptscriptfont\bffam=\sevenbf
  \normalbaselines\rm}



\def\beginlinemode{\endmode
  \begingroup\parskip=0pt \obeylines\def\\{\par}\def\endmode{\par\endgroup}}
\def\beginparmode{\endmode
  \begingroup \def\endmode{\par\endgroup}}
\let\endmode=\par
{\obeylines\gdef\
{}}
\def\singlespace{\baselineskip=\normalbaselineskip}

\def\oneandahalfspace{\baselineskip=\normalbaselineskip
  \multiply\baselineskip by 3 \divide\baselineskip by 2}
\def\doublespace{\baselineskip=\normalbaselineskip \multiply\baselineskip by 2}

\newcount\firstpageno
\firstpageno=2
\footline={\ifnum\pageno<\firstpageno{\hfil}\else{\hfil\twelverm\folio\hfil}\fi}
\def\toppageno{\global\footline={\hfil}\global\headline
  ={\ifnum\pageno<\firstpageno{\hfil}\else{\hfil\twelverm\folio\hfil}\fi}}
\let\rawfootnote=\footnote		
\def\footnote#1#2{{\rm\singlespace\parindent=0pt\parskip=0pt
  \rawfootnote{#1}{#2\hfill\vrule height 0pt depth 6pt width 0pt}}}
\def\raggedcenter{\leftskip=4em plus 12em \rightskip=\leftskip
  \parindent=0pt \parfillskip=0pt \spaceskip=.3333em \xspaceskip=.5em
  \pretolerance=9999 \tolerance=9999
  \hyphenpenalty=9999 \exhyphenpenalty=9999 }
\def\dateline{\rightline{\ifcase\month\or
  January\or February\or March\or April\or May\or June\or
  July\or August\or September\or October\or November\or December\fi
  \space\number\year}}
\def\received{\vskip 3pt plus 0.2fill
 \centerline{\sl (Received\space\ifcase\month\or
  January\or February\or March\or April\or May\or June\or
  July\or August\or September\or October\or November\or December\fi
  \qquad, \number\year)}}


\hsize=6.5truein
\hoffset=0truein
\vsize=8.9truein
\voffset=0truein
\parskip=\medskipamount
\def\\{\cr}
\twelvepoint		
\doublespace		
\overfullrule=0pt	




\def\title			
  {\null\vskip 3pt plus 0.2fill
   \beginlinemode \doublespace \raggedcenter \bf}

\def\author			
  {\vskip 3pt plus 0.2fill \beginlinemode
   \singlespace \raggedcenter\sc}

\def\affil			
  {\vskip 3pt plus 0.1fill \beginlinemode
   \oneandahalfspace \raggedcenter \sl}

\def\abstract			
  {\vskip 3pt plus 0.3fill \beginparmode
   \oneandahalfspace ABSTRACT: }

\def\endtitlepage		
  {\endpage			
   \body}

\def\body			
  {\beginparmode}		

\def\head#1{			
  \goodbreak\vskip 0.5truein	
  {\immediate\write16{#1}
   \raggedcenter \uppercase{#1}\par}
   \nobreak\vskip 0.25truein\nobreak}

\def\subhead#1{			
  \vskip 0.25truein		
  {\raggedcenter {#1} \par}
   \nobreak\vskip 0.25truein\nobreak}

\def\beginitems{
\par\medskip\bgroup\def\i##1 {\item{##1}}\def\ii##1 {\itemitem{##1}}
\leftskip=36pt\parskip=0pt}
\def\enditems{\par\egroup}

\def\beneathrel#1\under#2{\mathrel{\mathop{#2}\limits_{#1}}}

\def\refto#1{$^{#1}$}		

\def\references			
  {\head{References}		
   \beginparmode
   \frenchspacing \parindent=0pt \leftskip=1truecm
   \parskip=8pt plus 3pt \everypar{\hangindent=\parindent}}

\def\referencesnohead   	
  {                     	
   \beginparmode
   \frenchspacing \parindent=0pt \leftskip=1truecm
   \parskip=8pt plus 3pt \everypar{\hangindent=\parindent}}

\gdef\refis#1{\item{#1.\ }}			

\gdef\journal#1, #2, #3, 1#4#5#6{		
    {\sl #1~}{\bf #2}, #3 (1#4#5#6)}		

\def\pr{\journal Phys. Rev., }

\def\prb{\journal Phys. Rev. B, }

\def\prl{\journal Phys. Rev. Lett., }

\def\np{\journal Nucl. Phys., }

\def\pl{\journal Phys. Lett., }

\def\endreferences{\body}

\def\figurecaptions		
  {\endpage
   \beginparmode
   \head{Figure Captions}
}

\def\endpage			
  {\vfill\eject}

\def\endpaper			
  {\endmode\vfill\supereject}


\def\heading				
  {\vskip 0.5truein plus 0.1truein	
   \beginparmode \def\\{\par} \parskip=0pt \singlespace \raggedcenter}

\def\subheading				
  {\vskip 0.25truein plus 0.1truein	
   \beginlinemode \singlespace \parskip=0pt \def\\{\par}\raggedcenter}

\def\tag#1$${\eqno(#1)$$}

\def\align#1$${\eqalign{#1}$$}

\def\aligntag#1$${\gdef\tag##1\\{&(##1)\cr}\eqalignno{#1\\}$$
  \gdef\tag##1$${\eqno(##1)$$}}

\def\endaligntag{}

\def\overset #1\to#2{{\mathop{#2}\limits^{#1}}}
\def\underset#1\to#2{{\let\next=#1\mathpalette\undersetpalette#2}}
\def\undersetpalette#1#2{\vtop{\baselineskip0pt
\ialign{$\mathsurround=0pt #1\hfil##\hfil$\crcr#2\crcr\next\crcr}}}


\def\ref#1{Ref.~#1}			
\def\Ref#1{Ref.~#1}			
\def\[#1]{[\cite{#1}]}
\def\cite#1{{#1}}
\def\(#1){(\call{#1})}
\def\call#1{{#1}}
\def\taghead#1{}
\def\frac#1#2{{#1 \over #2}}

\def\12{{1\over2}}
\def\eg{{\it e.g.,\ }}

\def\ie{{\it i.e.,\ }}

\def\etal{{\it et al.\ }}
\def\etc{{\it etc.\ }}

\def\sla{\raise.15ex\hbox{$/$}\kern-.57em}
\def\leaderfill{\leaders\hbox to 1em{\hss.\hss}\hfill}
\def\twiddle{\lower.9ex\rlap{$\kern-.1em\scriptstyle\sim$}}
\def\bigtwiddle{\lower1.ex\rlap{$\sim$}}
\def\gtwid{\mathrel{\raise.3ex\hbox{$>$\kern-.75em\lower1ex\hbox{$\sim$}}}}
\def\ltwid{\mathrel{\raise.3ex\hbox{$<$\kern-.75em\lower1ex\hbox{$\sim$}}}}
\def\square{\kern1pt\vbox{\hrule height 1.2pt\hbox{\vrule width 1.2pt\hskip 3pt
   \vbox{\vskip 6pt}\hskip 3pt\vrule width 0.6pt}\hrule height 0.6pt}\kern1pt}
\def\tdot#1{\mathord{\mathop{#1}\limits^{\kern2pt\ldots}}}

\def\pmb#1{\setbox0=\hbox{#1}%
  \kern-.025em\copy0\kern-\wd0
  \kern  .05em\copy0\kern-\wd0
  \kern-.025em\raise.0433em\box0 }

\catcode`@=11
\newcount\r@fcount \r@fcount=0
\newcount\r@fcurr
\immediate\newwrite\reffile
\newif\ifr@ffile\r@ffilefalse
\def\w@rnwrite#1{\ifr@ffile\immediate\write\reffile{#1}\fi\message{#1}}

\def\writer@f#1>>{}
\def\referencefile{
  \r@ffiletrue\immediate\openout\reffile=\jobname.ref%
  \def\writer@f##1>>{\ifr@ffile\immediate\write\reffile%
    {\noexpand\refis{##1} = \csname r@fnum##1\endcsname = %
     \expandafter\expandafter\expandafter\strip@t\expandafter%
     \meaning\csname r@ftext\csname r@fnum##1\endcsname\endcsname}\fi}%
  \def\strip@t##1>>{}}

\def\citeall#1{\xdef#1##1{#1{\noexpand\cite{##1}}}}
\def\cite#1{\each@rg\citer@nge{#1}}	

\def\each@rg#1#2{{\let\thecsname=#1\expandafter\first@rg#2,\end,}}
\def\first@rg#1,{\thecsname{#1}\apply@rg}	
\def\apply@rg#1,{\ifx\end#1\let\next=\relax
\else,\thecsname{#1}\let\next=\apply@rg\fi\next}

\def\citer@nge#1{\citedor@nge#1-\end-}	
\def\citer@ngeat#1\end-{#1}
\def\citedor@nge#1-#2-{\ifx\end#2\r@featspace#1 
  \else\citel@@p{#1}{#2}\citer@ngeat\fi}	
\def\citel@@p#1#2{\ifnum#1>#2{\errmessage{Reference range #1-#2\space is bad.}%
    \errhelp{If you cite a series of references by the notation M-N, then M and
    N must be integers, and N must be greater than or equal to M.}}\else%
 {\count0=#1\count1=#2\advance\count1 by1\relax\expandafter\r@fcite\the\count0,%
  \loop\advance\count0 by1\relax
    \ifnum\count0<\count1,\expandafter\r@fcite\the\count0,%
  \repeat}\fi}

\def\r@featspace#1#2 {\r@fcite#1#2,}	
\def\r@fcite#1,{\ifuncit@d{#1}
    \newr@f{#1}%
    \expandafter\gdef\csname r@ftext\number\r@fcount\endcsname%
                     {\message{Reference #1 to be supplied.}%
                      \writer@f#1>>#1 to be supplied.\par}%
 \fi%
 \csname r@fnum#1\endcsname}
\def\ifuncit@d#1{\expandafter\ifx\csname r@fnum#1\endcsname\relax}%
\def\newr@f#1{\global\advance\r@fcount by1%
    \expandafter\xdef\csname r@fnum#1\endcsname{\number\r@fcount}}

\let\r@fis=\refis			
\def\refis#1#2#3\par{\ifuncit@d{#1}
   \newr@f{#1}%
   \w@rnwrite{Reference #1=\number\r@fcount\space is not cited up to now.}\fi%
  \expandafter\gdef\csname r@ftext\csname r@fnum#1\endcsname\endcsname%
  {\writer@f#1>>#2#3\par}}

\def\ignoreuncited{
   \def\refis##1##2##3\par{\ifuncit@d{##1}%
     \else\expandafter\gdef\csname r@ftext\csname r@fnum##1\endcsname\endcsname%
     {\writer@f##1>>##2##3\par}\fi}}

\def\r@ferr{\endreferences\errmessage{I was expecting to see
\noexpand\endreferences before now;  I have inserted it here.}}
\let\r@ferences=\references
\def\references{\r@ferences\def\endmode{\r@ferr\par\endgroup}}

\let\endr@ferences=\endreferences
\def\endreferences{\r@fcurr=0
  {\loop\ifnum\r@fcurr<\r@fcount
    \advance\r@fcurr by 1\relax\expandafter\r@fis\expandafter{\number\r@fcurr}%
    \csname r@ftext\number\r@fcurr\endcsname%
  \repeat}\gdef\r@ferr{}\endr@ferences}


\let\r@fend=\endpaper\gdef\endpaper{\ifr@ffile
\immediate\write16{Cross References written on []\jobname.REF.}\fi\r@fend}

\catcode`@=12

\citeall\refto		
\citeall\ref		%
\citeall\Ref		%

\catcode`@=11
\newcount\tagnumber\tagnumber=0

\immediate\newwrite\eqnfile
\newif\if@qnfile\@qnfilefalse
\def\write@qn#1{}
\def\writenew@qn#1{}
\def\w@rnwrite#1{\write@qn{#1}\message{#1}}
\def\@rrwrite#1{\write@qn{#1}\errmessage{#1}}

\def\taghead#1{\gdef\t@ghead{#1}\global\tagnumber=0}
\def\t@ghead{}

\expandafter\def\csname @qnnum-3\endcsname
  {{\t@ghead\advance\tagnumber by -3\relax\number\tagnumber}}
\expandafter\def\csname @qnnum-2\endcsname
  {{\t@ghead\advance\tagnumber by -2\relax\number\tagnumber}}
\expandafter\def\csname @qnnum-1\endcsname
  {{\t@ghead\advance\tagnumber by -1\relax\number\tagnumber}}
\expandafter\def\csname @qnnum0\endcsname
  {\t@ghead\number\tagnumber}
\expandafter\def\csname @qnnum+1\endcsname
  {{\t@ghead\advance\tagnumber by 1\relax\number\tagnumber}}
\expandafter\def\csname @qnnum+2\endcsname
  {{\t@ghead\advance\tagnumber by 2\relax\number\tagnumber}}
\expandafter\def\csname @qnnum+3\endcsname
  {{\t@ghead\advance\tagnumber by 3\relax\number\tagnumber}}

\def\equationfile{%
  \@qnfiletrue\immediate\openout\eqnfile=\jobname.eqn%
  \def\write@qn##1{\if@qnfile\immediate\write\eqnfile{##1}\fi}
  \def\writenew@qn##1{\if@qnfile\immediate\write\eqnfile
    {\noexpand\tag{##1} = (\t@ghead\number\tagnumber)}\fi}
}

\def\callall#1{\xdef#1##1{#1{\noexpand\call{##1}}}}
\def\call#1{\each@rg\callr@nge{#1}}

\def\each@rg#1#2{{\let\thecsname=#1\expandafter\first@rg#2,\end,}}
\def\first@rg#1,{\thecsname{#1}\apply@rg}
\def\apply@rg#1,{\ifx\end#1\let\next=\relax%
\else,\thecsname{#1}\let\next=\apply@rg\fi\next}

\def\callr@nge#1{\calldor@nge#1-\end-}
\def\callr@ngeat#1\end-{#1}
\def\calldor@nge#1-#2-{\ifx\end#2\@qneatspace#1 %
  \else\calll@@p{#1}{#2}\callr@ngeat\fi}
\def\calll@@p#1#2{\ifnum#1>#2{\@rrwrite{Equation range #1-#2\space is bad.}
\errhelp{If you call a series of equations by the notation M-N, then M and
N must be integers, and N must be greater than or equal to M.}}\else%
 {\count0=#1\count1=#2\advance\count1 by1\relax\expandafter\@qncall\the\count0,%
  \loop\advance\count0 by1\relax%
    \ifnum\count0<\count1,\expandafter\@qncall\the\count0,%
  \repeat}\fi}

\def\@qneatspace#1#2 {\@qncall#1#2,}
\def\@qncall#1,{\ifunc@lled{#1}{\def\next{#1}\ifx\next\empty\else
  \w@rnwrite{Equation number \noexpand\(>>#1<<) has not been defined yet.}
  >>#1<<\fi}\else\csname @qnnum#1\endcsname\fi}

\let\eqnono=\eqno
\def\eqno(#1){\tag#1}
\def\tag#1$${\eqnono(\displayt@g#1 )$$}

\def\aligntag#1\endaligntag
  $${\gdef\tag##1\\{&(##1 )\cr}\eqalignno{#1\\}$$
  \gdef\tag##1$${\eqnono(\displayt@g##1 )$$}}

\def\eqalignno#1{\displ@y \tabskip\centering
  \halign to\displaywidth{\hfil$\displaystyle{##}$\tabskip\z@skip
    &$\displaystyle{{}##}$\hfil\tabskip\centering
    &\llap{$\displayt@gpar##$}\tabskip\z@skip\crcr
    #1\crcr}}

\def\displayt@gpar(#1){(\displayt@g#1 )}

\def\displayt@g#1 {\rm\ifunc@lled{#1}\global\advance\tagnumber by1
        {\def\next{#1}\ifx\next\empty\else\expandafter
        \xdef\csname @qnnum#1\endcsname{\t@ghead\number\tagnumber}\fi}%
  \writenew@qn{#1}\t@ghead\number\tagnumber\else
        {\edef\next{\t@ghead\number\tagnumber}%
        \expandafter\ifx\csname @qnnum#1\endcsname\next\else
        \w@rnwrite{Equation \noexpand\tag{#1} is a duplicate number.}\fi}%
  \csname @qnnum#1\endcsname\fi}

\def\ifunc@lled#1{\expandafter\ifx\csname @qnnum#1\endcsname\relax}

\let\@qnend=\end\gdef\end{\if@qnfile
\immediate\write16{Equation numbers written on []\jobname.EQN.}\fi\@qnend}

\catcode`@=12

\def\eg{{\it e.g.,\ }}
\def\ie{{\it i.e.,\ }}
\def\etal{{\it et al.}}
\def\etc{{\it etc.\ }}

\def\>{\rangle}
\def\<{\langle}
\def\o{\over}
\def\sh{{\rm sh}}
\def\ch{{\rm ch}}
\def\t{\tilde}
\def\prop{\propto}

\def\slD{\raise.15ex\hbox{$/$}\kern-.57em\hbox{$D$}}
\def\dsl{\raise.15ex\hbox{$/$}\kern-.57em\hbox{$\Delta$}}
\def\slp{{\raise.15ex\hbox{$/$}\kern-.57em\hbox{$\partial$}}}
\def\nsl{\raise.15ex\hbox{$/$}\kern-.57em\hbox{$\nabla$}}
\def\sla{\raise.15ex\hbox{$/$}\kern-.57em\hbox{$\rightarrow$}}
\def\slla{\raise.15ex\hbox{$/$}\kern-.57em\hbox{$\lambda$}}
\def\slb{\raise.15ex\hbox{$/$}\kern-.57em\hbox{$b$}}
\def\lnp{\raise.15ex\hbox{$/$}\kern-.57em\hbox{$p$}}
\def\lnk{\raise.15ex\hbox{$/$}\kern-.57em\hbox{$k$}}
\def\lnK{\raise.15ex\hbox{$/$}\kern-.57em\hbox{$K$}}
\def\lnq{\raise.15ex\hbox{$/$}\kern-.57em\hbox{$q$}}

\def\a{\alpha}
\def\be{{\beta}}
\def\ga{{\gamma}}
\def\de{{\delta}}
\def\eps{{\epsilon}}
\def\veps{{\varepsilon}}

\def\th{{\theta}}
\def\ka{{\kappa}}
\def\la{\lambda}
\def\si{{\sigma}}

\def\om{{\omega}}

\def\De{{\Delta}}                         

\def\La{{\Lambda}}

\def\Om{{\Omega}}

\def\cA{{\cal A}}

\def\cL{{\cal L}}

\def\cS{{\cal S}}

\def\part{\partial}

\def\dag{\dagger}

\input epsf
\def\caption#1{\centerline{\vbox{\hsize 12truecm \noindent #1}}}

\def\journal#1, #2, #3, 1#4#5#6{               
    {\sl #1~}{\bf #2}, (1#4#5#6) #3}           
\def\cit#1{[\cite{#1}]}
\def\refto#1{[\cite{#1}]}
\def\Ref#1{Ref. [\cite{#1}]}
\def\ref#1{Ref. [\cite{#1}]}
\def\head#1{
\vskip 10mm
\noindent {\bf #1}
\vskip 5mm
\noindent
}
\def\subhead#1{
\vskip 6mm
\noindent {\it #1}
\vskip 3mm
\noindent
}
\def\La{K}
\def\tt{{\bf t}}
\def\ss{{\bf s}}
\def\ll{{\bf l}}
\def\\{\hfil\break}

\singlespace
\hsize 16truecm
\vsize 24truecm
\noindent
{\bf 
\uppercase{ Topological orders and Edge excitations in  FQH states}
}
\vskip 7mm
\noindent
Xiao-Gang Wen
\vskip 5mm

\noindent
Department of Physics, MIT, 77 Massachusetts Avenue, Cambridge, MA 02139, USA
\vskip 5.5mm
\noindent
{\bf Abstract:}
Fractional quantum Hall (FQH) liquids contain extremely rich internal structures
which represent a whole new kind of ordering. We discuss the 
characterization and
classification of the new orders (which is called topological orders).
We also discuss the edge excitations in FQH liquids, which form
the so-called chiral Luttinger liquids. The chiral Luttinger liquids
at the edges also have very rich structures as a reflection of the rich
topological orders in the bulk. Thus, edge excitations provide us
a practical way to measure topological orders in experiments.

\vskip 7mm
\noindent
{\bf Table of contents}

\item{1.} Introduction
\item\item{1.1} {\sl Characterization of topological orders}
\item\item{1.2} {\sl Classification of topological orders}
\item\item{1.3} {\sl Edge excitations --
a practical way to measure topological orders in experiments}
\item{2.}  Effective theory of FQH liquids
\item\item{2.1} {\sl Effective theory of the hierarchical FQH states}
\item\item{2.2} {\sl Effective theory of simple multi-layer FQH states}
\item{3.} Edge excitations in abelian FQH liquids
\item\item{3.1} {\sl Hydrodynamical approach to the edge excitations
 -- 1/m Laughlin state}
\item\item{3.2} {\sl Hydrodynamical approach to the edge excitations
 -- 2/5 and 2/3 states}
\item\item{3.3}  {\sl Bulk effective theory and the edge states}
\item\item{3.4} {\sl Charged excitations and electron propagator on the edges
of generic FQH states}
\item\item{3.5} {\sl Some simple phenomenological consequences of chiral Luttinger
liquids at FQH edges}
\item{4.} Shift and spin vectors -- New topological quantum numbers
for FQH liquids
\item\item{4.1} {\sl Quantum Hall states on a sphere and the shift}
\item\item{4.2} {\sl Spin vectors in QH liquids}
\item\item{4.3} {\sl Wave functions and $(K,\tt,\ss)$ of FQH liquids}
\item{5.}  Classification of Abelian Hall States
\item{6.} Algebraic approach and non-abelian FQH liquids
\item\item{6.1} {\sl Ideal Hamiltonians, zero energy states, 
and edge excitations}
\item\item{6.2} {\sl Edge excitations and current algebra}
\item\item{6.3} {\sl Non-abelian FQH liquids and their edge excitations}
\item{7.} Remarks and acknowledgments

\head{1. Introduction}
\taghead{1.}
The fractional quantum Hall (FQH) effect appears
in two-dimensional electron systems in a strong magnetic field. Since its
discovery in 1982 \refto{TSG,Lfqh}, experiments on FQH systems have 
continued to reveal many new phenomena and surprises.
These, together with the observed rich
hierarchical structures \refto{HH}, 
indicate that electron systems that demonstrate a
fractional quantum Hall effect (those systems are called FQH liquids) contain
extremely rich internal structures.  In fact FQH liquids 
represent a whole new
state of matter. One needs to develop new concepts and new techniques
to understand this new kind of states.

The first attempt to characterize the internal structures of FQH liquids
was proposed in \Ref{GL0}, where it was shown that a primary FQH liquid
(with an inverse filling fraction $\nu^{-1}=$integer) contains a
off-diagonal-long-range-order in a non-local operator.
Such an observation lead to a description of FQH liquids in terms of
Ginzburg-Landau-Chern-Simons effective theory\refto{GL}.
Those developments have lead to many interesting and important
results and deeper understanding of FQH liquids.
However, in this article I will try to describe the internal structures of FQH
liquids from a more general point of view. It appears that
some internal structures of FQH liquids (especially those in the so-call
non-abelian FQH liquids) cannot be described by the 
Ginzburg-Landau-Chern-Simons effective theory and the associated
off-diagonal-long-range-orders. Thus we need to develop
more general concepts and formulations for
the internal structures in FQH liquids.
Let us start with an general discussion of correlated systems.

Gases are one of simplest systems. The motion of a molecule in a gas hardly
depends on the positions and motion of other molecules. Thus gases are
weakly correlated systems. 
As we lower the temperature, the motions of molecules become more
and more correlated. Below a certain temperature the
motions of molecules are completely correlated and the molecules form a
crystal which is a strongly correlated system. 
In a crystal, an individual atom can hardly move by itself.
Excitations in a crystal always correspond to collective motions of many
atoms (which are called phonons).

FQH Hall liquids are another type of strongly correlated structure. Since
electrons are much lighter than atoms, they have much stronger quantum
motions. These quantum fluctuations prevent electrons in FQH systems 
from forming 
a crystal. Nevertheless, the motions of electrons in FQH liquids are
highly correlated.
For example let us look at Laughlin's wave function
for a filling fraction 1/3 FQH liquid \refto{Lfqh}
$$
\Psi_0=\prod_{i<j} (z_i-z_j)^3 \exp{(-{1\o 4} \sum_i |z_i|^2)}
\eqno(1.1)$$
where $z_i=x_i+iy_i$ is the coordinate of the $i^{th}$ electron.
We see from Laughlin's wave function that
\item{1)} all electrons are in the first Landau level and doing their own
cyclotron motion; and 
\item{2)} the wave function has a third order zero as any pair of electrons 
approach each other,
hence electrons try to stay away from each other
as much as possible.

The Laughlin wave function represents a particular
state in which all electrons
do their own cyclotron motion in the first Landau level
and {\it at same time} avoid collision with each
other. This requires 
all the electrons in the Laughlin state to dance collectively
following a strict global dancing pattern. This collective dancing
minimizes the ground state
energy and gives rise to a very stable correlated state. 
The strict collective
dancing pattern makes FQH liquids highly organized systems which contain
well defined internal orders.
The highly correlated nature of FQH liquids also makes them 
incompressible liquids. A compression of FQH liquids
create first order zeros in the Laughlin state and
breaks the global dancing pattern of electrons, hence costing finite
energies.

We would like to remark that above picture of collective dancing is quite
vague and general. Collective motions of electrons exist in all correlated
systems. To have a more concrete description of the internal structures
in FQH liquids, one need to find measurable quantum numbers that reflect
the internal correlations in FQH liquids. As we will see later that
the correlations between electrons in FQH liquids are very special
that we need a completely new set of quantum numbers to describe them.

Laughlin states represent only the simplest dancing pattern.
Different FQH liquids contain different dancing patterns.
Rich families of
FQH liquids observed in experiments indicate that the possible dancing
patterns are very rich. The internal orders (\ie the dancing patterns)
of FQH liquids are very different from the internal orders in other
correlated systems, such as crystal, superfluids, {\it etc}. The internal orders
in the latter systems can be described by order parameters associated with
broken symmetries. As a result, the ordered states can be described by 
the Ginzburg-Landau
effective theory. The internal order in FQH liquids
is a new kind of ordering which cannot be described by order parameters
associated with broken symmetries. Thus it would be very interesting to
see what kind of effective theory describe the low energy physics of
FQH liquids.

Despite the absence  
of broken symmetries and associated order parameters,
the internal orders in FQH liquids
are universal in the sense that they are robust against {\it
arbitrary} perturbations. These (universal) internal orders
characterize a phase in phase diagram.
This makes FQH liquids a new state of matter.
A concept of ``topological order" was introduced to describe this new kind of
ordering in FQH liquids \refto{Wtop,WN}.
Later we will give a more precise definition of the topological
orders. We will introduce measurable quantities to characterize
different topological orders. The universality of the topological orders
means that these characterization quantities are independent of 
various perturbations.

It is also useful to introduce the notion of topological fluids to describe
quantum liquids with non-trivial topological orders. FQH liquids are not
the only kind of topological fluid. Other examples of topological fluids
include anyon superfluids \refto{anyon}, chiral spin states
\refto{cspin,Wtop}, and short-ranged RVB
states for spin systems \refto{srvb}. 
A general but brief discussion of topological
order can be found in \Ref{topcs}.

It is instructive to compare FQH liquids with crystals.
FQH liquids are similar to crystals in the sense that they both
contain rich internal patterns (or internal orders).
The main difference is that the patterns in the crystals are static
related to the positions of atoms, while the patterns in FQH liquids
are dynamic associated with the ways that electrons ``dance" around each other.
However, many of the same questions for crystal orders can also be asked 
and should be addressed for 
topological orders. One important question is: How do we
characterize and
classify the topological orders? Another is: How do we
 experimentally measure the topological orders?

\subhead{1.1 Characterization of topological orders}
In the above, the concept of topological order (the dancing pattern)
is introduced through
the ground state wave function. This is not quite correct because
the ground state wave function is not universal.
To establish a new concept, such as topological order, one needs to find
physical characterizations or measurements of topological orders.
Or, in other words, one needs to find universal quantum numbers which are
independent of details of interactions, effective mass, \etc, 
but can take different values
for different classes of FQH liquids. The existence of such quantum numbers 
implies
the existence of topological orders.

One way to show the existence of the topological orders
in FQH liquids is to study their ground state degeneracies
(in the thermodynamical limit).
FQH liquids have a very special property. Their  ground state degeneracy
depends on the topology of space \refto{sphere,torus,WN}.
For example,
the $\nu={1\o q}$ Laughlin state has $q^g$ degenerate ground states on
a Riemann surface of genus $g$ \refto{WN}. 
The ground state degeneracy in FQH liquids
is {\it not} a consequence of symmetry of the Hamiltonian. 
The ground state degeneracy is
robust against arbitrary perturbations (even impurities that break all the 
symmetries in the Hamiltonian)\refto{WN}.
The robustness of  ground state degeneracy indicates that the internal
structures that give rise to ground state degeneracy is universal, hence
demonstrating the  existence of universal internal orders -- topological orders.

The ground state degeneracy can be regarded as a new quantum number
that (partially) characterize the topological orders in FQH states.
In particular some hierarchical states with the same filling fraction can have
different ground state degeneracies, indicating different topological
orders in those hierarchical states.
Thus the ground state degeneracies provide new information, in addition to 
the filling fraction, about the internal orders of FQH liquids.
According to this picture we would like to suggest that
the different hierarchical FQH states are characterized by the different
topological orders instead of by their symmetry properties.

In compact space, the low-energy physics of FQH liquids is very unique.
There is only a finite number of low-energy excitations ({\ie} the 
degenerate ground
states), yet the low-energy dynamics is non-trivial since the ground
state degeneracy depends on the topology of the space. Such special 
low-energy dynamics, which depend only on the topology of the space,
is described by the so called topological field theory which was under
intensive study recently in the high-energy physics community\refto{NAB,TOP}.
Topological
theories are effective theories for FQH liquids just as the Ginzburg-Landau
theory for superfluids (or other symmetry broken phases).

The dependence of the ground-state degeneracy on the topology of the space
indicates the existence of some kind of
long range order (the global dancing pattern
mentioned above) in FQH liquids, despite the absence of 
long-range correlations for all local physical operators.
In some sense we may say that FQH liquids contain hidden long range orders.

\subhead{1.2 Classification of topological orders}
A long standing problem has been
how to label and classify the rich topological orders in FQH liquids.
We are able to classify all crystal orders because we know the crystal orders
are described by symmetry group. But our
understanding of topological orders in FQH liquids is very poor, and
the mathematical structure behind the topological orders is unclear.

Nevertheless, we have been able to find a simple and
unified treatment for a class of FQH liquids -- abelian FQH liquids.
It has been shown that the internal structures (the topological orders)
in such fluids can
be characterized by a symmetric integer
matrix $K$, a charge vector $\tt$ and a
spin vector $\ss$ \refto{WZR,R,BW,FZ,class,spinv}.
Various physical quantities can be determined from $(K,\tt,\ss)$
that include ground state degeneracy,
quasiparticle quantum numbers, structure of the edge excitations,
exponents of electron and quasiparticle propagators along the edge, \etc
 All quasiparticle excitations in this class of
FQH liquids have abelian statistics, which leads to the name abelian FQH
liquids. The low energy effective theory of abelian FQH liquids is a
topological Chern-Simons theory with several $U(1)$ gauge 
fields $a_{I\mu}$ \refto{WZR,WN,BW,FZ,class}
$$\cL ={1\o 4\pi} K_{IJ} a_{I\mu}\part_\nu a_{J\la}\veps^{\mu\nu\la}
-eA_\mu t_I \part_\nu a_{I\la}\veps^{\mu\nu\la}
\eqno(1.1a)$$
The above effective theory is the dual form of the Ginzburg-Landau-Chern-Simons
effective theory discovered earlier \refto{GL}.
Almost all FQH liquids observed in
experiments were believed to be in this class.

We know that there are many different schemes to construct FQH states at
filling fractions different from $1/m$ \refto{HH}.
 Different constructions in general
lead to different wave functions, even when they have
the same filling fraction. 
From the construction schemes themselves, it is usually
difficult to see whether 
or not the 
FQH states described by those
different wave functions belong to the same universality class.
The above characterization (using $K$, $\tt$, and $\ss$) of the topological 
orders 
has an advantage in that it is independent of specific 
construction schemes and provides
a universal description of the universality classes of
abelian FQH liquids. One can usually derive an effective theory 
for the constructed
FQH states from the construction
schemes \refto{BW,R,FZ,class,spinv}. 
If two different constructions lead to the same $K$-matrix, 
charge vector $\tt$, and the spin vector $\ss$ 
(up to a field-redefinition of the gauge fields), then
we can say that
the two constructions generate the same FQH liquid. This approach
allows us to determine\refto{BW,R}
that the two 2/5 states obtained by hierarchical construction and 
Jain's construction\refto{HH} belong to the same universality class.

The physical meaning of the $K$-matrix is quite transparent in a class
of multi-layer FQH states obtained by generalizing Laughlin's
construction for the 1/m state \refto{Hmult}. The wave function
of those  multi-layer FQH states have the form
$$\prod_{a,b,i,j} (z_{a i} -z_{b j})^{{1\o 2} k_{ab}} \exp{(-\sum |z_{a i}|^2)}
\eqno(1.2)$$
where $z_{a i}$ is the coordinate of $i^{th}$ electron in the $a^{th}$ layer,
and $k_{ab}$ are integers satisfying $k_{ab}=k_{ba}$ and $k_{aa}=$odd.
The $K$-matrix that describes the multi-layer FQH state is nothing but
the matrix formed by the integer exponents in \(1.2): $K=(k_{ab})$.
Thus, $K$ describes the pattern of zeros in the wave function which
determines the way that electrons dance around each other.
The zeros in the wave function can also be interpreted as 
a certain number of flux quanta
being attached to each electron.
For example the third order
zero in the Laughlin wave function in \(1.1) implies that
three flux quanta are attached to  the electrons, since the phase
of the wave function changes by $3\times 2\pi$ as one electron moves 
around another. In this sense $K$ also describes the way we attach flux 
quanta to electrons. It is obvious that the multi-layer FQH states
naturally admit the $K$-matrix description. However, it is less obvious that
the hierarchical states and the many states obtained in Jain construction
also admit the $K$-matrix description.

The above classification of FQH liquids is not complete. Not every
FQH state is described by $K$-matrices.
In the last few years a new class of FQH states -- 
non-abelian FQH  states -- was
proposed \refto{MR,NABW,WWH}. A non-abelian FQH state contains
quasiparticles with non-abelian statistics.
The observed
filling fraction 5/2 FQH state \refto{5/2}
is very likely to be one such state \refto{HR}.
Recent studies in Refs. \cit{MR,NABBW,WWH} reveal that the
topological orders in some non-abelian FQH
states can be described by conformal field theories. 
However, we are still
quite far from a complete classification of all possible topological
orders in non-abelian states.

\subhead{1.3 Edge excitations -- 
a practical way to measure topological orders in experiments}
FQH liquids as incompressible liquids have a finite energy gap for all
their bulk excitations. However, FQH liquids of finite size always contain
one-dimensional gapless edge excitations, which is another unique property
of FQH fluids. The structures of edge excitations are extremely
rich which reflect the rich bulk topological orders. Different bulk
topological orders lead to different structures of edge excitations.
Thus we can study
and measure the bulk topological orders by studying structures of edge
excitations.

The edge excitations of integer quantum Hall (IQH) 
liquids was first studied by Halperin \refto{H}.
 He found that
edge excitations are described by 1D Fermi liquids. In the last few
years we have begun to understand the edge excitations of FQH 
liquids. We found, due to the
non-trivial bulk topological order, that electrons at the edges of (abelian)
FQH liquids
form a new kind of correlated state -- chiral Luttinger 
liquids \refto{WCL,LW,WH,BW,edgere}. The
electron propagator in chiral Luttinger liquids develops an anomalous
exponent: $\<c^\dag (t,x) c(0)\>\propto (x-vt)^{-g}$, $g\neq 1$. (For Fermi
liquids $g=1$). The exponent $g$, in many cases,
is a topological quantum number which
does not depend on detailed properties of edges. Thus, $g$ is a new quantum
number that can be used to characterize the topological orders in FQH
liquids. Recently a Maryland-IBM group successfully measured the exponent $g$
through the temperature dependence of tunneling conductance between two
edges, \refto{Web} which was predicted 
to have the form $\si \propto T^{2g-2}$\refto{WT}. This
experiment demonstrates the existence of new chiral
Luttinger liquids and opens the door to experimental study of the rich
internal and edge structures of FQH liquids.

The edge states of non-abelian FQH liquids form more exotic 1D correlated
systems which are not yet named. These edge states were found to be
closely related to conformal field theories at 1+1 dimensions \refto{WWH}.
\vskip .2 in
\epsfysize=2.0truein
\centerline{ \epsffile{toprevf1.eps} }
\caption{
Fig. 1.1: A 1D crystal passing an impurity will generate a narrow
band noise in the voltage drop.
}
\vskip .2 in

We know that crystal orders can be measured by X-ray diffraction
experiments. In the following we would like to indicate that the
topological orders in FQH liquids can be measured (in principle)
through noise spectrum in an edge transport experiment. Let us first
consider a 1D crystal driven through an impurity (see Fig. 1.1a).
Because of the crystal order, 
the voltage across the impurity has a narrow band noise at frequency $f=I/e$
if each unit cell has only one charged particle.
More precisely the noise spectrum has a singularity:
$S(f) \sim A\de(f-{I\o e})$.  If each unit cell contains two charged
particles (see Fig. 1.1b), 
we will see an additional narrow band noise at $f=I/2e$:
$S(f)\sim B\de(f-{I\o 2e}) +A\de(f-{I\o e})$. In this example we see that
the noise spectrum allows us to measure crystal orders in 1 dimension.
A similar experiment can also be used to measure topological orders
in FQH liquids. Let us consider a FQH sample with a narrow constriction
(see Fig. 1.2). The constriction induces a back scattering 
through quasiparticle tunneling between the two edges.
The back scattering causes a noise in the voltage
across the constriction. In the weak back scattering limit,
the noise spectrum contains singularities at certain frequencies which
allows us to measure the topological orders in the FQH 
liquids.\footnote{*}{The discussion presented here applies only to the
FQH states whose edge excitations all propagate in the same direction.
This requires, for abelian states, 
all the eigenvalues of $K$ to have the same sign.} 
 To be more specific, the singularities in the noise spectrum have the form,
$$S(f) \sim \sum_a C_a |f-f_a|^{\ga_a}
\eqno(1.3)$$
The frequencies and the exponents of the singularities $(f_a, \ga_a)$
are determined by the topological orders. For the abelian state
characterized by the matrix $K$ and charge vector $\tt$, the allowed
values of the pair  $(f_a, \ga_a)$ are given by
$$
f_a={I\o e\nu} \tt^T K^{-1}\ll,\ \ \ \ \ \ \
\ga_a=2 \ll^T K^{-1}\ll -1
\eqno(1.4)$$
where $\ll^T=(l_1,l_2,...)$ is an arbitrary integer vector, and
$\nu=\tt^T K^{-1}\tt$ is the filling fraction.
The singularities in the noise spectrum are caused by quasiparticle
tunneling between the two edges. The frequency of the
singularity $f_a$ is determined by the electric charge of the tunneling 
quasiparticle $Q_q$: $f_a={I\o e}{Q_q \o e\nu}$. The exponent $\ga_a$
is determined by the statistics of the tunneling quasiparticle
$\th_q$: $\ga=2{|\th_q|\o \pi}-1$.
Thus, the noise spectrum measures the charge and the statistics of 
the allowed quasiparticles.
\vskip .2 in
\epsfysize=2.0truein
\centerline{ \epsffile{toprevf2.eps} }
\caption{
Fig. 1.2:  A FQH fluid passing through a constriction will generate
narrow band noises due to the back scattering of the quasiparticles.
}
\vskip .2 in

\subhead{1.4 Organization}
In Chapter 2 we will construct the effective theory of FQH liquids, through 
which we will demonstrate that the abelian FQH liquids are characterized
by $K$-matrices and charge vectors. In Chapter 3 we will derive
the low energy effective theory for the 
gapless edge excitations in FQH liquids from the
bulk effective theory, and demonstrate the close connection between the
bulk topological orders and the structures of edge states.
We will also study some experimental consequence of the edge excitations.
In particular we will discuss the
effects of long range Coulomb interaction, smooth edge
potential, and impurities on the structures of FQH edges.
Chapters 4 and 5 contain some further study of topological orders, where we 
will introduce spin vectors and discuss classification of 
the abelian states. In Chapter 6, we make an attempt to describe 
non-abelian FQH states and their edge excitations using an operator
algebraic approach. 

I have tried to make each chapter more or less
self-contained. Because of this, there are some
overlaps between different chapters.

\head{2. Effective theory of FQH liquids}
\taghead{2.}
One way to understand the topological orders is to construct a low-energy
effective theory for FQH liquids. The effective theory should
capture the universal properties of FQH liquids and provide hints on how
to characterize and label different topological orders in FQH liquids.
In the following we will review a way to construct
effective theories that is closely related to the hierarchical
construction proposed by Haldane and Halperin \refto{HH}.
The first effective field theory for the Laughlin states was written
in the form of a Ginzburg-Landau theory with a Chern-Simons term \refto{GL}.
In this chapter we will adopt a different form of the effective theory,
which contains only pure Chern-Simons terms. 
The pure Chern-Simons form is more compact and more convenient for studying
hierarchical FQH states.
The two forms of effective theory is related by duality 
transformations \refto{dual}.

\subhead{2.1 Effective theory of the hierarchical FQH states}
In this section we will consider only single-layer spin-polarized
quantum Hall (HQ)
systems. To construct the effective theory for the hierarchical
states, we will start with the effective theory (in the pure Chern-Simons
form) of the Laughlin state. Then we will use the hierarchical construction
to obtain the effective theory for the hierarchical states.

First let us review the effective theory for the Laughlin
state \refto{WN,WZR,GL}.
Consider a charged boson or fermion system in a magnetic field,
\def\KE{ {\rm Kinetic~Energy}}
$$ \cL=-e{\bf A}\cdot {\bf J} +\KE
\eqno(4.1)
$$
where 
$$\eqalign{
{\bf J} = &\sum_i {\bf v_i} \de({\bf x} -{\bf x_i}) \cr
J^0 = &\sum_i \de({\bf x} -{\bf x_i}) \cr}
\eqno(4.1a)$$
are the current and the density of the particles and $({\bf x_i},{\bf v_i})$
are the position and the velocity of the i$^{th}$ particle.
We have also assumed that the charge of each particle is $-e$.

At a filling fraction $\nu=1/m$, where
$m$ is an even integer for boson and an odd integer for fermion, 
the ground state of the the above system is given by the Laughlin wave
function\refto{Lfqh}
$$\left[ \prod (z_i-z_j)^{m} \right] e^{-{1\o 4} \sum |z_i|^2}
\eqno(4.2)$$
To construct the effective theory of such a state, we note that the state
\(4.2) is an incompressible fluid and that the particle number current $J_\mu$
has the following response to a change of
electromagnetic fields:
\footnote{$^\dagger$}
{Here we use the convention that the Greek letters $\mu,\nu$ represent
space-time indices
$0,1,2$ and the Rome letter $i,j$ represent spatial indices $1,2$.}
$$ -e \de J_\nu= 
\si_{xy}\veps^{\mu\nu\lambda} \part_\nu \de A_\la
={\nu e^2\o 2\pi} \veps^{\mu\nu\lambda} \part_\nu \de A_\la
\eqno(4.2a)$$
as a result of finite Hall conductance $\si_{xy}={\nu e^2\o h}$
(in this paper we always choose $\hbar=1$).
In a hydrodynamic approach to the incompressible Hall liquid, we can
write the effective theory in terms of the current. We choose the
Lagrangian in such a way that it produces the equation
of motion \(4.2a). It is convenient to introduce a 
$U(1)$ gauge field $a_\mu$ to describe the conserved
particle number current:
$$J_{\mu}=
{1 \over 2 \pi}  \partial_\nu a_{ \lambda}~\veps^{\mu\nu\lambda}
\eqno(4.4)$$
The current defined this way automatically satisfies the conservation law.
Then the effective Lagrangian that produces \(4.2a) takes the following form
$$
\cL =
\left[ -m{1 \over 4 \pi}
a_{ \mu} \partial_\nu a_{ \lambda}~\veps^{\mu\nu\lambda}
-{e \over 2 \pi} A_{ \mu} \partial_\nu a_{ \lambda}~\veps^{\mu\nu\lambda}
\right] 
\eqno(4.3a)$$

\(4.3a) describes only the linear response of the ground state to the
external electromagnetic fields. To have a more complete description
of the topological fluid such as QH liquid, we need to introduce the 
boson or fermion excitations in our effective theory. 

In the effective theory \(4.3a), inserting the following source term
which carries a $a_\mu$-charge $q$,
$$ q a_0 \de({\bf x} -{\bf x_0})
\eqno(4.3aa)$$
will create an excitation of charge $Q=-q e/m$.
This can be seen  from the equation of motion ${\de \cL\o \de a_0}=0$,
$$  j_0={1\o 2\pi} \veps_{ij} \part_i a_j=- {e\o 2\pi m} B + {q\o m}
\de({\bf x} -{\bf x_0})
\eqno(4.3ab)$$
The first term indicates that
the filling fraction 
$\nu\equiv 2\pi {j_0\o -eB}$ is indeed $\nu=1/m$, and the second term 
corresponds to the increase of the 
particle density associated with the excitation.

We also see that the excitation created by the source term
\(4.3aa) is associated with $q/m$ unit of the $a_\mu$-flux.
Thus, if we have two excitations carrying $a_\mu$-charge $q_1$ and 
$q_2$, moving one excitation around the other will induce a phase
$2\pi \times $(number of $a_\mu$-flux quanta)$\times a_\mu$-charge,
$$2\pi \times {q_1 \o m}\times q_2
\eqno(4.3ac)$$
If $q_1=q_2\equiv q$, 
the two excitations will be identical. Interchanging them will
induce half of the phase in \(4.3ac),
$$\th=\pi {q^2\o m}
\eqno(4.3ad)$$
Here $\th$ is nothing but the statistical angle of the excitation that carries 
$q$ unit of the $a_\mu$-charge.

Our bosons or fermions carry a charge of $-e$. From the above discussion, we
see that a charge $-e$ excitation
can be created by inserting a source term of $m$ units of the $a_\mu$-charge.
Such an excitation has a statistical angle $\th=\pi m$ (see \(4.3ad)); thus it
is a boson if $m$ is even and a fermion if $m$ is odd. Therefore, we can
identify the excitations of $m$ units of the $a_\mu$-charge with the
particles (the bosons or the fermions) that form the QH liquid.
We would like to stress that the identification of the fundamental
particles (the bosons or the fermions) in the effective theory is very 
important. It is this identification, together with
the effective Lagrangian, that provides a complete description of the
topological properties of the QH liquid. We will see below that this
identification allows us to determine the fractional charge and the
fractional statistics of the quasiparticle excitations. 

A quasihole excitation at a position described by a complex number
$\xi=x_1+i x_2$ 
is created by multiplying
$\prod_i (\xi-z_i)$ to the ground state wave function \(4.2).
Note that the phase of the wave function changes by $2\pi$ as
a boson or a fermion
going around the quasihole. Thus the quasihole also 
behaves like a vortex in the condensate of the bosons or the fermions.

Now let us try to create an excitation by inserting a source
term of $q$ units of the $a_\mu$-charge. Moving a boson or a fermion around such
an excitation will induce a phase $2\pi q$ (see \(4.3ac)). The single-value
property
of the boson or fermion wave function requires such a phase to be multiples of
$2\pi$ and hence $q$ must be quantized as an integer for allowed 
excitations. From the charge of the excitations, we find that
$q=-1$ corresponds to the fundamental quasihole excitation
described above,
while $q=1$ corresponds to the fundamental quasiparticle excitation.
The quasiparticle excitation carries an electric charge $-e/m$ and the
quasihole $e/m$. Both have statistics
$\th=\pi/m$, as one can see from \(4.3ad). 
We see that the effective theory reproduces
the well known results for quasiparticles in Laughlin states \refto{ASW}.
The full effective theory with 
quasiparticle excitations is given by
$$\eqalign{
\cL =&
\left[ -m{1 \over 4 \pi}
a_{ \mu} \partial_\nu a_{ \lambda}~\veps^{\mu\nu\lambda}
-{e \over 2 \pi} A_{ \mu} \partial_\nu a_{ \lambda}~\veps^{\mu\nu\lambda}
\right] \cr
&+ l a_\mu j_\mu + \KE  \cr}
\eqno(4.3)$$
where $j_\mu$ is the current of the quasiparticles which has the
form in \(4.1a).
For fundamental quasiparticles the integer $l$ in \(4.3) has a value
of $l=1$, and for fundamental quasiholes
$l=-1$. $l$ takes other integer values for composite quasiparticles.
\(4.3), together with the quantization condition on $l$,
is a complete low energy
effective theory which captures the topological properties
of the $1/m$ Laughlin state. It can be shown that the effective theory \(4.3)
is simply the dual form \refto{dual,WN} of the
Ginzburg-Landau-Chern-Simons effective theory discovered 
earlier \refto{GL}.

Now consider a $1/m$ FQH state formed by 
electrons, which corresponds to the fermion case
discussed above. Let us increase the filling fraction by
creating the fundamental quasiparticles, which
are labeled by $l=1$. \(4.3) with $l=1$ describes the $1/m$ state
in the presence of these quasiparticles.
Now, two equivalent pictures emerge:
\item{a)} In a mean-field-theory approach,
we may view the gauge field $a_\mu$ in \(4.3) 
as a fixed background and do not allow $a_\mu$ to respond to the inserted source 
term $j_\mu$. In this case the quasiparticle gas
behaves like bosons in the ``magnetic" field $b=\part_i a_j \veps_{ij}$,
as one can see from the second term in \(4.3). These bosons do not carry
any electric charge since the quasiparticle number current $j_\mu$ does not
directly couple with the electromagnetic gauge potential $A_\mu$.
When the boson density satisfies
$$j_0={1\o p_2 }{b\o 2\pi}
\eqno(4.6)$$
where $p_2$ is even, the bosons have a filling fraction ${1\o p_2}$. The ground
state of the bosons can again be 
described by  a Laughlin state. The final electronic
state that we obtained
is just a second level hierarchical FQH state constructed by 
Haldane \refto{HH}.
\item{b)} If we let $a_\mu$ respond to the insertion $j_\mu$, then 
quasiparticles will be dressed by the $a_\mu$ flux. 
The dressed quasiparticles carry an electric charge of $e/m$
and a statistics of $\th=\pi/m$. When the quasiparticles have the density
$$j_0={1\o (p_2 +{\th\o \pi})}{eB\o 2\pi m}
\eqno(4.6)$$
where $p_2$ is even, the quasiparticle will have a filling fraction 
${1\o (p_2+{\th\o \pi})}$. In this case the quasiparticle 
system can form a Laughlin state
described by the wave function
$$\prod_{i<j} (z_i -z_j)^{p_2+{\th\o \pi}} 
\eqno(4.6a)$$
The final electronic
state obtained this way
is again a second level hierarchical FQH state.
This construction  was first proposed by
Halperin \refto{HH}. The two constructions in a) and b) lead to the same
hierarchical state and are equivalent.

In the following we will follow Haldane's hierarchical construction to
derive the effective theory of hierarchical FQH states. 
Notice that under the assumption a),
 the boson Lagrangian (the second term of \(4.3) with $l=1$) is just
\(4.1) with an 
external electromagnetic field $-eA_\mu$ replaced by $a_\mu$. Thus,
we can follow the same steps from \(4.2) to \(4.3)
to construct the effective theory of the
boson Laughlin state. Introducing a new $U(1)$ gauge field $\t a_\mu$
to describe the boson current, we find that the boson effective theory takes
the form
$$
\cL=-{p_2 \over 4 \pi}
\t a_{ \mu} \partial_\nu \t a_{ \lambda}~\veps^{\mu\nu\lambda}+
{1 \over 2 \pi} a_{ \mu} \partial_\nu \t a_{ \lambda}~\veps^{\mu\nu\lambda}
\eqno(4.6a)$$
In \(4.6a) 
the new gauge field $\t a_\mu$ describes the density $j^0$ and the current
$j^i$ of the bosons
and is given by
$$j^{\mu}=
{1 \over 2 \pi}  \partial_\nu \t a_{ \lambda}~\veps^{\mu\nu\lambda}
\eqno(4.6b)$$
This reduces the coupling between $a_\mu$ and the boson current, 
$a_\mu j_\mu$, to a Chern-Simons term between $a_\mu$ and $\t a_\mu$
(which becomes the second term in \(4.6a)).
The total effective theory (including the original electron condensate)
has the form
$$
\cL = \left[
-{p_1 \over 4 \pi}
a_{ \mu} \partial_\nu a_{ \lambda}~\veps^{\mu\nu\lambda}
-{e \over 2 \pi} A_{ \mu} \partial_\nu a_{ \lambda}~\veps^{\mu\nu\lambda} \right]
+\left[ -{p_2 \over 4 \pi}
\t a_{ \mu} \partial_\nu \t a_{ \lambda}~\veps^{\mu\nu\lambda}
+{1 \over 2 \pi} a_{ \mu} \partial_\nu \t a_{ \lambda}~\veps^{\mu\nu\lambda}
\right]
\eqno(4.7)$$
where $p_1=m$ is an odd integer.
\(4.7) is the effective theory of a second level hierarchical FQH state.

The second level hierarchical FQH state contains 
two kinds of quasiparticles. One 
is quasihole (or vortex) in the original electron condensate,
and the other
is quasihole (or vortex) in the new boson condensate. 
The two kinds of the quasiholes are created by inserting the 
source terms $-j^\mu a_\mu$ and $-\t j^\mu \t a_\mu$, respectively,
where $\t j^\mu$ and $j^\mu$ have a similar form as in \(4.1a).
The first kind of quasihole is created by 
multiplying $\prod_i (\xi -z_i)$ to the electron   wave function,
while the second kind is created by multiplying
$\prod_i (\eta -\xi_i)$ to the boson Laughlin wave function (here
$\xi_i$ are the complex coordinates of the boson and $\eta$ is the
position of the quasihole).
The effective theory for the quasiholes in the second level
hierarchical states has the form
$$ \t j^\mu \t a_\mu + j^\mu a_\mu +\KE
\eqno(4.8)$$
We can use the effective theory in \(4.7) and \(4.8) to calculate the quantum 
numbers of the quasiholes.

The total filling fraction can be
determined from the equation of motion ${\de \cL\o \de a_0}={\de \cL\o
\de \t a_0}=0$,
$$-eB=p_1 b-\t b, \ \ b=p_2\t b
\eqno(4.9)$$
We find
$$\nu={b\o -eB}={1\o p_1-{1\o p_2}}.
\eqno(4.9a)$$
\(4.7) can be written in a more
compact form by introducing $(a_{1\mu}, a_{2\mu})=(a_{\mu}, \t a_{\mu})$,
$$\cL=-\sum_{I,I^\prime} {1 \over 4 \pi}\La_{II^\prime}
a_{I \mu} \partial_\nu a_{I^\prime \lambda}~\veps^{\mu\nu\lambda} -
{e \over 2 \pi} A_{ \mu} \partial_\nu t_I a_{I \lambda}\veps^{\mu\nu\lambda}
\eqno(4.10)$$
where  the matrix $\La$ has integer elements,
$$\La=\pmatrix{p_1&-1 \cr
                -1& p_2 \cr}
\eqno(4.11)$$
and $\tt^T=(t_1,t_2)=(1,0)$ will be called the charge vector.
The filling fraction \(4.9a) can be rewritten as $\nu= \tt^T K^{-1} \tt$.

A generic quasiparticle is labeled by two integers that
consist of $l_1$ number of quasiparticles of the first kind and
$l_2$ number of quasiparticles of the second kind. 
Such a quasiparticle carries  $l_1$
units of the $a_{1\mu}$ charge
 and $l_2$ units of the $a_{2\mu}$ charge and is described by
$$ (l_1 a_{1\mu} +l_2 a_{2\mu}) j^\mu
\eqno(4.11aa)$$
After integrating out the
 gauge fields, we find that
such a quasiparticle carries $\sum_J K_{I J} l_J$ units of the
$a_{I\mu}$-flux. Hence 
the statistics of such a quasiparticle are given by
$$\th=\pi \ll^T\La^{-1} \ll={1\o p_2p_1-1}(p_2l_1^2+p_1 l_2^2+2l_1l_2)
\eqno(4.12)$$
and the electric charge of the quasiparticle is
$$Q_{q}=-e\tt^T\La^{-1}\ll=-e{p_2l_1+l_2\o p_2p_1-1}
\eqno(4.13)$$

\def\od{\o\displaystyle}
The above construction can be easily generalized to the level-$3$ 
hierarchical FQH states with the filling fraction
$$\nu={1\od p_1 -{1\od p_2 -{1\od p_3}}}
\eqno(4.14aa)$$
by allowing the boson (described by $\t j$ current)  in \(4.8) 
to condense into a $1/p_3$ 
Laughlin state. The effective theory of this third level hierarchical state 
still has the form in \(4.10), but now $I,I'$ run from 1 to 3.
The new $K$-matrix has the form
$$K=\pmatrix{p_1& -1  & 0 \cr
              -1& p_2 &-1 \cr
               0& -1  &p_3\cr}
\eqno(4.11a)$$
and the charge vector $\tt^T=(1,0,0)$.
One can go further to obtain the effective theories of the $n^{th}$ level
hierarchical states.
In the hierarchical
 construction one always assumes that the quasiparticles from the last
 condensate
 ``condense" to get the next level hierarchical state. 
Therefore the $K$-matrix $K$ for
a hierarchical FQH state has a tri-diagonal form
$$\La_{II'}=p_I\de_{I,I'}-\de_{I,I'-1}-\de_{I,I'+1}
\eqno(4.15)$$
with the charge vector given by
$$t_I=\de_{1I}
\eqno(4.15a)$$
In \(4.15) $p_1$ is odd and $p_i|_{i>1}$ are even.
The filling fraction of such a state is given by \refto{BW}
$$\nu=\tt^T K^{-1}\tt={1\od p_1 -{1\od p_2 -{1\od\dots -{1\od p_n}}}}
\eqno(4.14)$$

We can also construct FQH states which are more general than those
obtained from the standard hierarchical construction.
The effective theory for these more general FQH states
still has the form in \(4.10) but now $I$ runs from 1 to an integer $n$
($n$ will be called the level of the FQH state).
To obtain the form of the matrix $\La$, let us assume
that at level $n-1$ the
effective theory is given by eq. \(4.10) with $a_{I\mu},\ I=1,\dots, n-1$ and
$\La=\La^{(n-1)}$. The quasiparticles
carry integer charges of the $a_{I\mu}$ gauge fields. (This can always
be achieved by 
properly normalizing the gauge fields.) 
Now consider an $n^{th}$ level hierarchical state which is 
obtained by ``condensation" of quasiparticles
with the $a_{I\mu}$ charge $l_I|_{I=1,..,n-1}$. The effective theory
of this level $n$ hierarchical state
will be given by eq.\(4.10)
 with $n$ gauge fields. The $n$-th gauge field $a_{n\mu}$
comes from the new condensate.
The matrix $\La$ is given by
$$\La^{(n)}=\pmatrix{\La^{(n-1)}& -l \cr
                            -l^T & p_n\cr}
\eqno(4.14a)$$
with $p_n=$even. The charge vector $\tt$ is still given by \(4.15a).
By iteration, we see that the generalized hierarchical states are
always described by integer symmetric matrices, with $K_{II}$=even
except $K_{11}$=odd.
The new  condensate gives rise to a new kind of quasiparticle which again
carries integer charge of the new gauge field $a_{n\mu}$.
Hence, a  generic quasiparticle always carries integral charges of the
$a_{I\mu}$ field. Assume those charges are $l_I$. Then
the electric charge and the statistics of the quasiparticle are given by the
following general formulae
$$\th=\pi \ll^T\La^{-1} \ll ,\ \ \
Q_q=-e\tt^T \La^{-1}\ll
\eqno(4.16)$$
The filling fraction is given by
$$\nu = \tt^T K^{-1} \tt
\eqno(4.16a)$$

From the above discussion we see that more general hierarchical states,
in principle, can be obtained
by assuming condensation of different types of  quasiparticles.
In the standard Haldane-Halperin hierarchical scheme 
one assumes the quasiparticles from the last condensate condense to
generate next level hierarchical states.
This choice of condensing quasiparticles 
is valid if such quasiparticles have the smallest energy gap. 
However, in principle one can not exclude the
possibility that a different type of quasiparticles 
(rather than the one from the last
condensate) might have the smallest energy gap.
In this case,  we will obtain a new hierarchical state from the condensation
of different quasiparticles.
Certainly for the FQH systems used in real experiments, there are good
reasons to believe that the quasiparticles from the last
condensate do
have the smallest energy gap and it is those quasiparticles
that generate the next level hierarchical states.
It would be interesting to find out under which conditions other types
of quasiparticles may have smaller energy gaps.

From the generic effective theory in \(4.10) we obtain one of the important
results in this paper; that 
the generalized hierarchical states can be labeled by an integer
valued $K$-matrix and a charge vector $\tt$. 
Now we would like to ask a question: Do different $(K,\tt)$s 
describe different FQH states?
Notice that, through a redefinition of the gauge fields $a_{I\mu}$,
one can always diagonalize the $K$-matrix into one with $\pm 1$ as
the diagonal elements. Thus, it seems that
all $K$-matrices with the same signature describe the same FQH states,
since they lead to the same effective theory after a proper
redefinition of the gauge fields. Certainly this conclusion
is incorrect. We would like to stress that the effective Lagrangian \(4.10)
alone does not provide a proper description of the topological orders
in the hierarchical states. It is the effective Lagrangian \(4.10)
together with the quantization condition of the $a_{I\mu}$ charges
that characterize the topological order. A $U(1)$ gauge theory
equipped with a quantization condition on the allowed $U(1)$ charges
is called compact $U(1)$ theory. Our effective theory \(4.10) is
actually a compact $U(1)$ theory with all $U(1)$ charges quantized
as integers. Thus the allowed $U(1)$ charges form an $n$-dimensional
cubic lattice which will be called the charge lattice. Therefore,
when one considers the equivalence of two different $K$-matrices, one
can use only the field redefinitions that keep the charge
quantization condition unchanged (\ie keep the charge lattice unchanged).
The transformations that map the charge lattice onto itself belong
to the group $SL(n,Z)$ (a group of integer matrices with a unit determinate)
$$ a_{I\mu} \to W_{IJ} a_{J\mu}, \ \ \ \ \ W\in SL(n,Z)
\eqno(4.16c)$$
From the above discussion we see that the two FQH states
described by 
$(K_1,\tt_1)$ and $(K_2,\tt_2)$ are equivalent (\ie they belong to the
same universality class) if there exists a $W \in SL(n,Z)$ such that
$$\tt_2=W\tt_1,\ \ \ \ \ \ K_2=WK_1 W^T
\eqno(4.16d)$$
This is because, under \(4.16d),  an effective theory described 
by $(K_1,\tt_1)$ 
simply changes into another effective theory described by $(K_2,\tt_2)$.

We would like to remark that
in the above discussion we have ignored another topological
quantum number -- spin vector. Because of this, the equivalence 
condition in \(4.16d) does not apply to clean systems.
However \(4.16d) does apply to disordered FQH systems because
the angular momentum is not conserved in disordered systems and
the spin vector is not well defined. A more detailed discussion
of the spin vectors will be given in chapter 4.

In the following we list the $K$-matrix, the charge vector $\tt$, and the
spin vector $\ss$ for  some common single-layer spin-polarized FQH states:
$$\eqalign{
\nu=1/m, \ \ \ \ & \tt=(1),\ \ \ K=(m),\ \ \ \ \ \ \ \ 
\ \ \ \ \ \ \ \ \ \ss=(m/2) \cr
\nu=1-1/m, \ \ \ \ & \tt=\pmatrix{1 \cr
                                0 \cr},\ 
                     K=\pmatrix{1 & 1      \cr
                                1 & -(m-1) \cr},\
                     \ss=\pmatrix{1/2     \cr
                                (1-m)/2 \cr} \cr
\nu=2/5, \ \ \ \   & \tt=\pmatrix{1 \cr
                                0 \cr},\ 
                     K=\pmatrix{3 & -1 \cr
                                 -1 & 2 \cr},\ \ \ \ \  \ \ \
                     \ss=\pmatrix{1/2 \cr
                                1   \cr} \cr
\nu=3/7, \ \ \ \   & \tt=\pmatrix{1 \cr
                                0 \cr 
                                0 \cr},\ 
                     K=\pmatrix{ 3&-1& 0\cr
                                -1& 2&-1\cr
                                 0&-1& 2 \cr},\
                     \ss=\pmatrix{1/2 \cr
                                1 \cr
                                1 \cr}\cr
}
\eqno(4.16b)$$

\subhead{2.2 Effective theory of simple multi-layer FQH states}
The same approach used to construct the effective theory of the hierarchical 
states can also be used to construct the effective theory for the
multi-layer FQH states. The connection between the FQH wave function and the
$K$-matrix becomes very transparent for the multi-layer FQH states.
In this section we will concentrate on double-layer FQH states.
However, the generalization to the $n$-layer FQH sates is straight forward.

We would like to construct an effective theory for the following
simple double-layer FQH state,
$$\prod_{i<j}(z_{1i}-z_{1j})^l
  \prod_{i<j}(z_{2i}-z_{2j})^m
  \prod_{i,j}(z_{1i}-z_{2j})^n 
  e^{-{1\o 4} (\sum_i|z_{1i}|^2 + \sum_j|z_{2j}|^2)}  ,
\eqno(2.2.1)$$
where $z_{Ii}$ is the complex coordinate of the i$^{th}$ electron in the
I$^{th}$ layer. Here $l$ and $m$ are odd integers so that the wave function
is consistent with the
Fermi statistics of the electrons, while $n$ can be any non-negative
integer. The above wave function was first suggested by Halperin as a
generalization of the Laughlin wave function \refto{Hmult}. It appears that
these wave functions can explain some of the dominant FQH filling fractions
observed in double layer FQH systems.

We start with a single layer FQH state in the first layer,
$$
\prod_{i<j}(z_{1i}-z_{1j})^l e^{-{1\o 4} \sum_i|z_{1i}|^2 }
\eqno(2.2.2)$$
which is a $1/l$ Laughlin state and is described by the effective theory
$$
\cL =
\left[ -l{1 \over 4 \pi}
a_{1 \mu} \partial_\nu a_{1 \lambda}~\veps^{\mu\nu\lambda}
-{e \over 2 \pi} A_{ \mu} \partial_\nu a_{ \lambda}~\veps^{\mu\nu\lambda}
\right]
\eqno(2.2.2a)$$
where $a_{1\mu}$ is the gauge field that describes the electron density and 
current in the first layer.

Examining the wave function in \(2.2.1) we see that an electron in the second
layer is bounded to a quasihole excitation in the first layer.
Such a quasihole excitation is formed by $n$ fundamental quasihole
excitations and carries an $a_{1\mu}$-charge of $-n$. A gas of the quasiholes
is described by the following effective theory
$$ \cL= -n a_{1\mu} j_\mu + \KE
\eqno(2.2.3)$$
where $j_\mu$ has the form in \(4.1a).
As we mentioned before, in the mean-field theory, if we ignore the response
of the $a_{1\mu}$ field, \(2.2.3) simply describe a gas of bosons in a
magnetic field $-n b_1$ where $b_1 = \veps_{ij} \part_i a_{1j}$.
Now we would like to attach an electron (in second layer) to each quasiholes
in \(2.2.3). Such an operation has two effects: a) the bound
state of the quasiholes and the electrons can directly couple with the
electromagnetic field $A_\mu$ since the electron carries the charge $-e$;
and b)
the bound state behaves like a fermion.
The effective theory for the bound states has the form
$$ \cL= (-e A_\mu -n a_{1\mu}) j_\mu + \KE
\eqno(2.2.4)$$
which now describes a gas of fermions (in the mean field theory).
These fermions see an effective magnetic field $-eB-nb_1$.

When the electrons (\ie the fermions in \(2.2.4))
in the second layer have a density 
${1\o m} {-eB-nb_1 \o 2\pi}$ (\ie have an effective filling fraction $1/m$),
they can form a $1/m$ Laughlin state, which corresponds
to the $ \prod_{i<j}(z_{2i}-z_{2j})^m$
part of the wave function. (Note here $eB <0$.) 
Introducing a new gauge field
$j_{\mu}=
{1 \over 2 \pi}  \partial_\nu a_{ \lambda}~\veps^{\mu\nu\lambda}$
to describe the fermion current $j_\mu$ in \(2.2.4), the effective theory
of the $1/m$ state in the second layer has the form
$$
\cL =
 -m{1 \over 4 \pi}
a_{2 \mu} \partial_\nu a_{2 \lambda}~\veps^{\mu\nu\lambda}
\eqno(2.2.5)$$
Putting \(2.2.2a), \(2.2.4) and \(2.2.5) together we obtain the total
effective theory of the double layer state which has the form in \(4.10)
with the $K$-matrix and the charge vector \tt\ given by
$$ K=\pmatrix{ l & n \cr
                   n & m \cr}, \ \ \ \
   \tt = \pmatrix{ 1 \cr
                   1 \cr}
\eqno(2.2.6)$$
We see that the $K$-matrix is nothing but the exponents in the wave function.
The filling fraction of the FQH state is still given by \(4.16a).

There are two kinds of (fundamental) quasihole excitations in the double
layer state.  The first kind is created by multiplying 
$\prod_i (\xi - z_{1i})$ to the ground state wave function and the second kind
is created by $\prod (\xi - z_{2i})$. 
As the vortices in the two electron condensates in the first and
the second layers,
a first kind of quasihole is
created by the source term $-a_{1\mu} j^\mu$, and the second kind
of quasihole by $-a_{2\mu} j^\mu$.
Thus, a generic quasiparticle in the double layer state is a bound state
of several quasiholes of the first and the second kind and is described by
\(4.11aa). The quantum numbers of such quasiparticles are still given by
\(4.16).

In general, 
a multi-layer FQH state of type \(2.2.1) is described by a $K$-matrix
whose elements are integers and whose diagonal elements are odd integers.
The charge vector has the form $\tt^T=(1,1,...,1)$.

People usually label the double layer FQH state \(2.2.1) by $(l,m,n)$.
In the following we 
list the $K$-matrix, the charge vector $\tt$, and the
spin vector $\ss$ for  some simple double layer states:
$$\eqalign{
\nu=1/m, \ \ \ \ & \tt=\pmatrix{1 \cr
                              1 \cr},\ \
                     K=\pmatrix{m & m \cr
                                m & m \cr},\ \
                     \ss=\pmatrix{m/2 \cr
                                m/2 \cr} \cr
\nu=2/5, \ \ \ \   & \tt=\pmatrix{1 \cr
                                1 \cr},\ \
                     K=\pmatrix{3 & 2 \cr
                                2 & 3 \cr},\ \
                     \ss=\pmatrix{3/2 \cr
                                3/2 \cr} \cr
\nu=1/2, \ \ \ \   & \tt=\pmatrix{1 \cr
                                1 \cr},\ \
                     K=\pmatrix{3 & 1 \cr
                                1 & 3 \cr},\ \
                     \ss=\pmatrix{3/2 \cr
                                3/2 \cr} \cr
\nu=2/3, \ \ \ \   & \tt=\pmatrix{1 \cr
                                1 \cr},\ \
                     K=\pmatrix{3 & 0 \cr
                                0 & 3 \cr},\ \
                     \ss=\pmatrix{3/2 \cr
                                3/2 \cr} \cr
\nu=2/3, \ \ \ \   & \tt=\pmatrix{1 \cr
                                1 \cr},\ \
                     K=\pmatrix{1 & 2 \cr
                                2 & 1 \cr},\ \
                     \ss=\pmatrix{1/2 \cr
                                1/2 \cr} \cr
}
\eqno(2.2.7)$$

From the above we see that the (332) double layer state has the filling 
fraction 2/5 -- a filling fraction that also appears in single layer
hierarchical states. Now a question arises: Do the double layer
2/5 state and the single layer 2/5 belong to the same universality class?
This question has experimental consequences.
We can imagine the following experiment. We start with a (332) double
layer state in a system with very weak interlayer tunneling. As we make
the interlayer tunneling stronger and stronger, while keeping the filling
fraction fixed, the double layer state will eventually become a single
layer 2/5 state. The question is whether the transition between
the double layer (332) state and
the single layer 2/5 is a smooth crossover or a phase transition.
If we ignore the spin vector we see that the $K$-matrices and the charge
vectors of the two 2/5 states are equivalent since they are related by
a $SL(2,Z)$ transformation. Therefore, in the presence of disorders
(in which case the spin vector is not well defined) the two
2/5 states can change into each other smoothly.
However, the cleaner samples may have a smaller
activation gap for the DC conductance in the crossover region.
When we include the spin vector, the two 2/5 
states are not equivalent and, for pure systems, they 
are separated by a first order phase transition.

The transition between the above two 2/5 states should be similar to the
transition in the following system. Let us consider an electron system
with a strange kinetic energy such that the first Landau level has an energy
$E_1$ and the second level has $E_2$. Originally $E_1<E_2$. Assume that
by changing a
certain parameter we can make $E_1 > E_2$. This will cause a transition
between the $\nu=1$ state in the first Landau level and the the 
$\nu=1$ state in the second Landau level. For pure systems, such a transition
must be first order since the two $\nu=1$ states have different spin vectors.
For disordered systems the transition can be a smooth crossover.
Certainlyi, it is always possible that some other states may appear between
the two $\nu=1$ (or $\nu=2/5$) states.
The bottom line is that the two $\nu=1$ states cannot be smoothly
connected in a clean sample.

From \(2.2.7) we also see that there are two different 2/3 double layer states.
When the intra-layer interaction is much 
stronger than the interlayer interaction
(a situation in real samples), the ground state wave function prefers
to have higher order zeros between electrons in the same layers. Thus,
the (330) state should have lower energy than the (112) state.
The $K$-matrix and the charge vector of the single layer
2/3 state is equivalent
to those of the (112) state, and is not equivalent to those of the
(330) state. Thus, to change a double layer 2/3 state (\ie the (330) state)
into the single layer 2/3 state,
one must go through a phase transition for both random and
pure systems.

After replacing the layer index by the $S_z$ spin index, 
all the results  obtained in this section can be directly
applied to describe spin unpolarized FQH states of spin-1/2 electron systems.
In particular, one can show that the $\nu=2/5$ and the second $\nu=2/3$
states in \(2.2.7) describe two spin singlet FQH states \refto{spin}.
Due to the different
spin vectors, both states are inequivalent to their spin polarized
counterpart -- the $\nu=2/5$ and the $\nu=2/3$ FQH states in \(4.16b).
Thus for pure system, the above spin singlet FQH states and the spin
polarized FQH states are separated by first order transitions.
However for dirty system the first order transition may be
smeared into a smooth crossover.

The double layer $(mmm)$ state is an interesting state since det$(K)=0$.
It turns out that the $(mmm)$ state (in the absence of interlayer tunneling)
spontaneously break a $U(1)$ symmetry and contain 
a superfluid mode.\refto{WZmmm}
More detailed discussions (including their experimental implications)
can be found in \Ref{WZmmm,mmm}.

\def\r{\rho}
\def\sgn{{\rm sgn}}

\head{3. Edge excitations in abelian FQH liquids}
\taghead{3.}
Due to the repulsive interaction and strong
correlation between the electrons, a FQH liquid is an incompressible state
despite the fact that the
first Landau level is only partially filled. All the bulk excitations in 
FQH states have finite energy gaps. FQH states and insulators are very
similar
in the sense that both states have finite energy gaps and  short ranged
electron
propagators. Because of this similarity people were puzzled by the
fact that 
FQH systems apparently have very different transport properties than
ordinary
insulators. Halperin first pointed out that the integral quantum Hall (IQH)
states contain gapless edge excitations \refto{H}.
 Although the electronic states in the
bulk are localized, the electronic states at the edge of the sample are
 extended
(\ie  the electron propagator along  the edge is long-ranged) \refto{niu}. 
Therefore,
the nontrivial transport properties of the IQH states come from the gapless
edge excitations \refto{H,EdgeT}
\eg a two probe measurement of a QH sample can result in a finite
resistance only when the source and the drain are connected by the edges.
If the source and the drain are not connected by any edge, the two probe
measurement will
yield an infinite resistance at zero temperature, a result very similar to
the insulators.
The edge transport picture has been supported by many
experiments \refto{EdgeE}.
 Halperin also studied
the dynamical properties of the edge excitations of the IQH states and 
found that
the edge excitations are described by a chiral 1D Fermi liquid theory.

Using the gauge argument in \ref{LG,H,WG}, one can easily
show that  FQH states also support gapless edge excitations.
Thus it is natural to conjecture that the transport in  FQH states is
also governed by the edge excitations \refto{B,M}. However, 
since  FQH states are
intrinsically many-body states, the edge excitations in the
FQH states cannot be constructed from a single-body theory. Or in other
words, the edge excitations of  FQH states should not be described by a
Fermi liquid.
In this case we need completely new approaches to understand the dynamical
properties of the edge states of FQH liquids.
Recent advances in fabrication of
small devices make it possible to study in detail the dynamical properties
of the edge states in FQH liquids. 
Thus it is important to develop a quantitative
theory for FQH edge states to explain new experimental
data.

There is another motivation to study the edge states in  FQHE.
We know that different FQH states were generally labeled by their filling
fractions. However, it becomes clear that  FQH states contain extremely
rich internal structures that the filling fraction alone is not enough to
classify
all the different universality classes of  FQH states \refto{Wtop,WN}.
 One can easily
construct different FQH states with the same filling
fraction \refto{HH,BW,R,FZ,class}.
From the last chapter we see that (generalized) hierarchical states and
simple double layer states can be labeled by $K$-matrices and charge vectors
$\tt$. But in the last chapter, $K$ and $\tt$ merely appear as parameters
in some theoretical effective theory. An important question is that can one
find experiments that measure $K$ and $\tt$. Certainly a combination of $K$
and $\tt$ can be 
measured through the filling fraction, and the determinant of $K$
can be measured through the ground state degeneracy of the FQH liquid on a
torus \refto{torus,WN}.
Experimentally, however one can never put
an HQ state on a torus. Thus, results in \ref{Wtop,WN,torus}
can be checked only in numerical calculations.
In the following we will see that
edge excitations in FQH states provide an important (probably the
only practical) probe to detect the topological 
orders in the bulk FQH states. Through tunneling experiments between
FQH edges one can measure many different combinations of $K$ and $\tt$.
 Using
the edge excitations we also can tell whether a FQH state is an abelian
FQH state or a non-abelian FQH state.
Thus, the edge states provide us with  a practical window through
which we can look into the internal structures in FQH states.
The measurements of the edge states can provide us new quantum numbers,
in addition to the filling fractions, to characterize different quantum Hall
states.

\subhead{3.1 Hydrodynamical approach to the edge excitations
 -- 1/m Laughlin state}
The simplest (but not complete) way to understand the dynamics of 
edge excitations is to use the hydrodynamical approach. In this approach,
one uses the fact that QH (IQH or FQH) states are incompressible ir-rotational
liquids
that contain no low-energy bulk excitations. Therefore, the only low lying
excitations (below the bulk energy gap) are surface waves on a HQ droplet.
These surface waves are identified as
edge excitations of the HQ state \refto{HE,S,LW}.

In the hydrodynamical approach
we first study the classical theory of the surface wave on the HQ droplet.
Then we quantize the classical theory to obtain the quantum description
of the edge excitations. It is amazing that the simple quantum description
obtained from the classical theory provides a complete description of the
edge excitations at low energies and allows us to calculate
the electron and the quasiparticle propagators along the edges.

Consider a QH droplet with a filling fraction $\nu$ confined by a 
potential well.
Due to the non-zero conductance, the electric field of the
potential well generates a persistent current flowing along the edge,
$${\bf j}=\si_{xy}\hat z\times {\bf E},\ \ \ \   \si_{xy}=\nu{e^2\o h}
\eqno(2.1)$$
This implies that the electrons near the edge drift with a velocity
$$ v={E\o B} c
\eqno(2.2)$$
where $c$ is the velocity of the light. Thus, the edge
wave also propagates with the velocity $v$.
Let us use one dimensional density $\r(x)=nh(x)$ to describe the edge wave,
where $h(x)$ is the displacement of the edge, $x$ is the coordinate along
the edge, and $n={\nu\o 2\pi l_B^2}$ is the two dimensional electron density
in the bulk.
We see that the propagation of the
edge waves are described by the following wave equation,
$$\part_t\r -v\part_x \r=0
\eqno(2.3)$$
Notice that the edge waves always propagate in one direction, there are no
waves that propagate in the opposite direction.

The Hamiltonian (\ie the energy) of the edge waves is given by
$$H=\int dx \12 eh\r E=\int dx\ \pi {v\o \nu} \r^2
\eqno(2.4)$$
In the momentum space \(2.3) and \(2.4) can be rewritten as
$$\eqalign{
\dot \r_k=& i vk\r_k  \cr
H=& 2\pi {v\o \nu} \sum_{k>0} \r_k\r_{-k}  \cr}
\eqno(2.5)$$
where $\r_k=\int dx {1\o \sqrt{L}} e^{ikx}\r(x)$, and $L$ is the length
of the edge.
Comparing \(2.5) with the standard Hamiltonian equation,
$$\dot q= {\part H\o \part p}, \ \ \ \ \dot p=-{\part H\o \part q}
\eqno(2.6)$$
we find that if we identify $\r_k|_{k>0}$ as the ``coordinates", then
the corresponding canonical ``momenta" can be identified
as $p_k=i2\pi \r_{-k}/\nu k$.
We would like to stress that because the edge waves are chiral, the
 displacement
$h(x)$ contains both the ``coordinates" and the ``momenta".

Knowing the canonical coordinates and momenta, it is easy to quantize
the classical theory. We simply view $\r_k$ and $p_k$ as operators that
satisfy $[p_k, \r_{k'}]=i\de_{kk'}$. Thus after quantization we have
$$\eqalign{
[\r_k, \r_{k'}]=& {\nu\o 2\pi}k\de_{k+k'} \cr
k,k^\prime=&{\rm integer}\times {2\pi\o L}  \cr
[H,\r_k]=& v\r_k  \cr}
\eqno(2.7)$$
The above algebra is called the ($U(1)$) Kac-Moody (K-M) algebra \refto{KM}.
A similar algebra has also appeared in the Tomonaga model \refto{T}.
Notice that \(2.7) simply describes a collection of decoupled harmonic
oscillators (generated by $(\r_k, \r_{-k})$ ). Thus \(2.7) is an one
  dimensional free phonon theory
(with only a single branch) and is exactly soluble.
We will show later that \(2.7) provides
a complete description of the low lying edge excitations of the HQ state.

To summarize, we find that the edge excitations in the QH states are described
by a free (chiral) phonon theory at low energies. We not only
show the existence of the gapless edge excitations, we also obtain the density
of states of the edge excitations. The specific heat (per unit length) of the
edge excitations
is found to be ${\pi\o 6}{T\o v}$. The edge
excitations considered here do not change the total charge of the system
and hence are neutral. In the following, we will discuss the charged
excitations and calculate the electron propagator from the K-M algebra \(2.7).

The low lying charge excitations obviously correspond to adding (removing)
 electrons to (from) the edge.
Those charged excitations carry integer charges and are created by electron
operators $\Psi^\dag$.  The above
 theory of the edge excitations is formulated in
terms of a 1D density operator $\r(x)$. So the central question is to write
the electron operator in terms of the density operator.
The electron operator on the edge creates a localized
charge and should satisfy
$$[\rho(x), \Psi^\dag(x')]= \de(x-x')\Psi^\dag(x')
\eqno(3.1)$$
Since $\rho$ satisfies the Kac-Moody algebra \(2.7),
one can show that the operators that satisfy \(3.1)
are given by\refto{WCL}
$$
\Psi\prop e^{ i{1\o \nu}\phi}
\eqno(3.2)$$
where $\phi$ is given by
$\rho={1\o 2\pi}\part_x \phi$.

\(3.1) implies only that the operator $\Psi$ carries 
the charge $e$. In order to
identify $\Psi$ as an electron operator we need to show that $\Psi$ is a
fermionic operator. Using the K-M algebra \(2.7) we find that
$$\Psi(x)\Psi(x')=(-)^{1/\nu}\Psi(x')\Psi(x)
\eqno(3.2a)$$
We see that the electron operator $\Psi$ in \(3.2) is fermionic
only when $1/\nu=m$ is an odd integer, in which case the QH state is a
Laughlin state \refto{WCL,FK}.

In the above discussion we have made an assumption
that is not generally true. We have assumed that the incompressible QH liquid
contains only {\it one} component of incompressible fluid which leads to
one branch of edge excitations.  The above result implies that,
when $\nu\neq 1/m$, the edge theory with only one branch does not contain
the electron operators and is not self-consistent.
Therefore we conclude that the FQH states with $\nu\neq 1/m$ must contain more
than one branch of edge excitations.
(Here we have ignored
the possibility of the pairing between the electrons.)
Later (in section 6.1) we will see that the one-branch assumption is true
only for a simple Laughlin state with filling fraction $\nu=1/m$. For
hierarchical FQH states, there are several condensates corresponding to
several components of incompressible fluid. Each
component gives rise to a branch of the edge excitations. Thus a generic
QH state may contain many branches of the edge excitations,\refto{M,WH}
even when electrons are all in the first Landau level (see section 3.2 and 3.3).

Now let us calculate electron propagator along the edge of the Laughlin
states with $\nu=1/m$. In this case the above simple edge theory is valid.
Because $\phi$ is a free phonon field with a propagator
$$\<\phi(x,t) \phi(0)\>=-\nu \ln (x-vt) +\hbox{const.}
\eqno(3.3)$$
the electron propagator can be easily calculated as\refto{WCL}
$$G(x,t)=\<T(\Psi^\dag(x,t) \Psi(0))\>=\exp[{1\o \nu^2}\<\phi(x,t) \phi(0)\>]
\prop {1\o (x-vt)^{m} }
\eqno(3.4)$$

The first thing we see is that the electron propagator on the edge of a FQH
state acquires a non-trivial exponent $m=1/\nu$ that is not equal to one.
This implies that the electrons
on the edge of the FQH state are strongly correlated and cannot be described
by Fermi liquid theory. We will call this type of an electron state chiral
Luttinger liquid.

The K-M algebra \(2.7) and the electron operator \(3.2) provided a complete
description of both neutral and charged edge excitations at low energies.
We would like to remark that the propagator \(3.4) is correct only for large
$x$ and $t$. At a 
short distance the form of the propagator depends on the details
of the electron interactions and the edge potentials. We would
also like to
emphasize
that the exponent $m$ of the edge propagator is determined by the bulk state.
Such an exponent is a topological number that is independent of electron
interactions, edge potential, \etc.
The quantization of the exponent is directly related to the fact that the
exponent is linked to the statistics of the electrons (see \(3.2a)).
Thus the exponent can be regarded as a
quantum number that characterizes the topological orders in the bulk FQH
states.

In the momentum space the electron propagator has the form
$$G(k,\om)\prop {(vk+\om)^{m-1}\o vk-\om}
\eqno(3.5)$$
The anomalous exponent $m$ can be measured in tunneling experiments. The
tunneling density of states of electron is given by
$$N(\om)\prop |\om|^{m-1}
\eqno(3.6)$$
This implies that deferential conductance has the form ${dI\o dV} \prop V^{m-1}$
for a metal-insulator-FQH junction.

\subhead{3.2 Hydrodynamical approach to the edge excitations
 -- 2/5 and 2/3 states}
In this section we will use the hydrodynamical approach discussed above
to study edge structures of second level hierarchical states. We will
concentrate on the 2/5 and 2/3 states as examples. In particular we
will study the structures of the electron and the quasiparticle
operators on the edges of the hierarchical states.
We will also see that the 2/3 state contains two edge modes that
propagate in the opposite directions, which is quite counter intuitive.

First let us consider the $\nu={2\o 5}$ FQH state. According to
the hierarchical theory, the $\nu={2\o 5}$ FQH state is generated by the
condensation of quasiparticles on top of the $\nu={1\o 3}$ FQH state. 
Thus the 2/5 state contains two components of incompressible fluids.
To be
definite let us consider a special edge potential such that
the FQH state consists of two droplets, one is the
electron condensate with a filling fraction ${1\o 3}$ and radius $r_1$,
and the other is the quasiparticle condensate 
(on top of the 1/3 state) with a filling fraction
${1\o 15}$ ( note ${1\o 3} +{1\o 15}={2\o 5}$) and radius $r_2< r_1$.

When
$r_1-r_2 \gg l_B$, the two edges are independent. Generalizing the
hydrodynamical approach in section 3.1, we can show that there are two
branches of the edge excitations whose low energy dynamics are described
by
$$\eqalign{
[\r_{Ik},\r_{Jk'}]=& {\nu_I\o 2\pi} k \de_{IJ} \de_{k+k'}  \cr
H=&2\pi \sum_{I,k>0} {v_I \o \nu_I} \r_{Ik}\r_{I-k}  \cr}
\eqno(2.3a.1)$$
where $I=1,2$ labels the two branches, $(\nu_1,\nu_2)=({1\o 3},{1\o
15})$ are filling fractions of the electron condensate and the
quasiparticle condensate, and $v_I$ are the velocity of the
edge excitations. $\r_I$ in \(2.3a.1) are the 1D electron densities given by
$\r_I=h_I \nu_I{1\o 2\pi l_B^2}$ where $h_I$ are the amplitude of the
edge waves on the two droplets. 

Because the electrons are interacting with each other, the edge
velocities are determined by $v_I=c{E^*_I/B}$ where $E_I^*$ are the
effective electric fields that include contributions from both the edge
potential and the electrons. In order for the Hamiltonian to be bounded
from below, we require $\nu_Iv_I>0$. We find that the stability of the
$\nu ={2\o 5}$ FQH state requires both $v_I$ to be positive.

Generalizing the discussion in section 3.1, the electron
operators on the two edges are found to be
$$\Psi_I=e^{i{1\o \nu_I}\phi_I(x)}
\ \ \ \ \ I=1,2
\eqno(2.3a.1a)$$
with $\part_x\phi_I= {1\o 2\pi}\r_I$. The electron propagators have the form
$$\<T(\Psi_I(x,t)\Psi^\dag_I(0))\>= e^{ik_Ix} {1\o (x-v_I t)^{-1/|\nu_I|} }
\ \ \ \ \ I=1,2
\eqno(2.3a.1b)$$
where $k_I={r_I\o 2l_B^2}$.

According to the hierarchical picture, the $\nu={2\o 3}$ FQH state
is also formed by two condensates, an electron condensate with a filling
fraction $1$ and a hole condensate with a filling fraction $-{1\o 3}$. Thus,
the above discussion can also be applied to the $\nu = {2\o 3}$ FQH
state by choosing $(\nu_1,\nu_2)=(1,-{1\o 3})$. Again there are two
branches of the edge excitations but now with {\it opposite} velocities
if the Hamiltonian is positive definite.
This result, although surprising, is not difficult to understand.
The stability of both the electron droplet and the hole droplet requires
$E_1^*$ and $E^*_2$ to have opposite signs.

As we bring the 
two edges together ($r_1-r_2 \sim l_B$) the interaction between
the two branches of the edge excitations can no longer be ignored. In this
case the Hamiltonian has the form
$$
H=2\pi \sum_{I,J,k>0} V_{IJ} \r_{Ik}\r_{J-k}
\eqno(2.3a.2)$$
(The Hamiltonian may also contain terms that describe the electron
hopping between edges. But
those terms are irrelevant at low energies due to the chiral property of
the edge excitations. For example one can show that those terms can
never open an energy gap\refto{WG})

The Hamiltonian \(2.3a.2)
can still be diagonalized. For $\nu_1\nu_2 >0$, we may choose
$$\eqalign{
\t \r_{1k}&=\cos(\th){1\o \sqrt{|\nu_1|}}\r_{1k}
          +\sin(\th){1\o \sqrt{|\nu_2|}}\r_{2k} \cr
\t \r_{2k}&=\cos(\th){1\o \sqrt{|\nu_2|}}\r_{2k}
          -\sin(\th){1\o \sqrt{|\nu_1|}}\r_{1k} \cr
\tan(2\th)&=2{\sqrt{|\nu_1\nu_2|} V_{12} \o |\nu_1|V_{11}-|\nu_2|V_{22} } \cr}
\eqno(2.3a.3)$$
One can check that $\t \r$s satisfy
$$\eqalign{
[\t\r_{Ik},\t\r_{Jk'}]=&  {\sgn(\nu_I)\o 2\pi} k \de_{IJ} \de_{k+k'}  \cr
H=&2\pi \sum_{I,k>0} \sgn(\nu_I) \t v_I  \t\r_{Ik}\t\r_{I-k}  \cr}
\eqno(2.3a.4)$$
where the new velocities of the edge excitations $\t v_I$ are given by
$$\eqalign{
\sgn(\nu_1) \t v_1=& {\cos^2(\th) \o \cos(2\th)}|\nu_1|V_{11}
-{\sin^2(\th) \o \cos(2\th)}|\nu_2|V_{22}  \cr
\sgn(\nu_2) \t v_2=& {\cos^2(\th) \o \cos(2\th)}|\nu_2|V_{22}
-{\sin^2(\th) \o \cos(2\th)}|\nu_1|V_{11}  \cr  }
\eqno(2.3a.4a)$$
We see that there are still two
branches of the edge excitations. However in this case
the edge excitations with
a definite velocity are mixtures of those on the inner edge and the
outer edge. One can also show that as long as the Hamiltonian \(2.3a.2)
is bounded from below, the velocities of the two branches $\t v_I$ are always
positive.
This result has been
confirmed by numerical calculations \refto{JM}.

After rewriting the electron operator $\Psi_I$ in terms of $\t \r_I$
by inverting \(2.3a.3),
we can calculate their propagators using \(2.3a.4a)
$$\<T(\Psi_I(x,t)\Psi^\dag_I(0))\>= e^{ik_Ix}
    {1\o (x-\t v_1 t)^{\a_I }}
    {1\o (x-\t v_2 t)^{\be_I } }
\eqno(2.3a.5)$$
where
$$(\a_1,\a_2)=({1\o |\nu_1|}\cos^2\th,{1\o |\nu_2|}\sin^2\th),\ \ \ \
 (\be_1,\be_2)=({1\o |\nu_1|}\sin^2\th,{1\o |\nu_2|}\cos^2\th)
\eqno(2.3a.6)$$

However,
when the two edges are close to each other within the magnetic length,
the $\Psi_I$ are no longer the most general electron operators on the
edge. The generic electron operator may contain charge transfers between the
two edges. For the $\nu=2/5$ FQH state, the inner edge and the outer edge
are separated by the $\nu={1\o 3}$ Laughlin state. Thus,
the elementary charge transfer operator is given by
$$\eta(x)=e^{i(\phi_1-{\nu_1\o \nu_2}\phi_2)}
=(\Psi_1\Psi_2^\dag)^{\nu_1}
\eqno(2.3a.7)$$
which transfers a $\nu_1 e=e/3$ charge from the outer edge to the inner edge.
The generic electron operator then takes the form
$$\eqalign{
  \Psi(x)=& \sum_{n=-\infty}^{+\infty} c_n\psi_n(x) \cr
  \psi_n(x)=&\Psi_1(x) \eta^n(x)  \cr}
\eqno(2.3a.8)$$
 To understand this result, we notice
that each operator $\psi_n$ always creates a unit localized charge
and is a fermionic operator regardless of the value of the integer $n$.
Therefore, each $\psi_n$
is a candidate for the electron operator on the edge. For a generic interacting
system the electron operator on the edge is expected to be a superposition
of different $\psi_n$s as represented in \(2.3a.8).
Note $\Psi_2=\psi_{-{1\o \nu_1} }$.
The propagator of $\psi_n$ can be
calculated in a similar way as outlined above and is given by
$$\<T(\psi_n(x,t)\psi^\dag_m(0))\>\propto
    \de_{n,m}e^{i[k_1+n\nu_1(k_2-k_1)]x}
    \prod_I (x-\t v_I t)^{-\ga_{In} }
\eqno(2.3a.9)$$
where $\ga_{In}$ are
$$\eqalign{
\ga_{1n}=& \left[(n+{1\o |\nu_1|})\sqrt{|\nu_1|}\cos\th
-{n\nu_1\o \nu_2}\sqrt{|\nu_2|} \sin\th\right]^2  \cr
\ga_{2n}=& \left[(n+{1\o |\nu_1|})\sqrt{|\nu_1|}\sin\th
+{n\nu_1\o \nu_2}\sqrt{|\nu_2|} \cos\th\right]^2  \cr   }
\eqno(2.3a.10)$$
From \(2.3a.8) and \(2.3a.9) we see that the electron
propagator has singularities at discrete momenta $k=k_1+n\nu_1(k_2-k_1)$.
They are analogs to the $k_F, 3k_F,...$ singularities of the electron propagator
in the interacting 1D electron systems.

For the $\nu={2\o 3}$ FQH state, $\nu_1\nu_2<0$. In this case
we need to choose
$$\eqalign{
\t \r_{1k}=&\ch(\th){1\o \sqrt{|\nu_1|}}\r_{1k}
          +\sh(\th){1\o \sqrt{|\nu_2}|}\r_{2k} \cr
\t \r_{2k}=&\ch(\th){1\o \sqrt{|\nu_2|}}\r_{2k}
          +\sh(\th){1\o \sqrt{|\nu_1|}}\r_{1k} \cr
\hbox{th}(2\th)=&
2{\sqrt{|\nu_1\nu_2|} V_{12} \o |\nu_1|V_{11}+|\nu_2|V_{22} } \cr}
\eqno(2.3a.12)$$
to diagonalize the Hamiltonian. One can check that $\t \r_I$ also satisfies
the K-M algebra \(2.3a.4) but now
$$\eqalign{
\sgn(\nu_1) \t v_1=& {\ch^2(\th) \o \ch(2\th)}|\nu_1|V_{11}
-{\sh^2(\th) \o \ch(2\th)}|\nu_2|V_{22}  \cr
\sgn(\nu_2) \t v_2=& {\ch^2(\th) \o \ch(2\th)}|\nu_2|V_{22}
-{\sh^2(\th) \o \ch(2\th)}|\nu_1|V_{11}  \cr  }
\eqno(2.3a.12a)$$
Again as long as
the Hamiltonian $H$ is positive definite, the velocities of
the edge
excitations $\t v_I$ always have opposite signs.
The electron operator still has the form
\(2.3a.8) with $\eta=(\Psi_1\Psi_2^\dag)^{\nu_1}$. The propagator of
$\psi_n$ is still given by  \(2.3a.9) with
$$\eqalign{
\ga_{1n}=& \left[(n+{1\o |\nu_1|})\sqrt{|\nu_1|}\ch\th
+{n\nu_1\o \nu_2}\sqrt{|\nu_2|} \sh\th\right]^2  \cr
\ga_{2n}=& \left[(n+{1\o |\nu_1|})\sqrt{|\nu_1|}\sh\th
+{n\nu_1\o \nu_2}\sqrt{|\nu_2|} \ch\th\right]^2  \cr   }
\eqno(2.3a.13)$$

From \(2.3a.10) and \(2.3a.13), we see that exponents in the electron
propagator depend on the inter-edge interactions. However, we can show that
the exponents satisfy a sum rule:
$$\sum_I \sgn(\nu_I) \ga_{In} \equiv \la_n=(n+{1\o |\nu_1|})^2\nu_1
+{n^2\nu_1^2\o \nu_2}
\eqno(2.3a.14)$$
$\la_n$ always take odd-integer values and are independent of the details of
the electron system.
The quantization of $\la_n$ is again due to the fact that
$\la_n$ are
directly linked to the statistics of the electrons:
$$\psi_n(x)\psi_n(x')=(-)^{\la_n} \psi_n(x')\psi_n(x)
\eqno(2.3a.15)$$
We would like to point out that the values of $\la_n$ are determined by
the internal correlations (topological orders) of the bulk FQH state.
$\la_n$ can be changed only by changing the bulk topological orders through
a {\it two dimensional } phase transition, despite the fact that
they are  properties
of 1D edge excitations.
Therefore $\la_n$ are topological quantum numbers that can be used
to characterize and to experimentally measure the 2D bulk topological
orders.

The total exponents $g_e^{(n)}$ of the electron operators
$$ g_e^{(n)} =\sum_I \ga_{In}
\eqno(2.3a.15a)$$
are every important quantities. For the 2/5 state,
the minimum value of the exponents
$$g_e\equiv {\rm Min}(g_e^{(n)})=g_e^{(0)}=g_e^{(-1)}=3
\eqno(2.3a.15b)$$
controls the scaling properties of tunneling of electrons between
two edges. For example the tunneling conductance scales at finite temperatures
as \refto{WT}
$$\si \propto T^{2g_e -2}
\eqno(2.3a.15c)$$

For the 2/5 state the sum rule \(2.3a.14)
implies that $g_e^{(n)}$ and $g_e$ are
topological quantum numbers that are independent of details of
electron interactions and edge potential. However for 2/3 state
the sum rule cannot fix the value of $g_e$, and $g_e$ is not
universal. Later in section 3.5 we will see that in the presence of long range
Coulomb interaction and/or in the presence of edge impurity, $g_e$ for the 2/3
state will take a universal value $g_e=2$ \refto{KFP}.

\subhead{3.3  Bulk effective theory and the edge states}
In this section we will directly derive the macroscopic theory of the edge
excitations from the effective theory of the bulk FQH 
states \refto{BW,FK,class}.
 In this approach we do not rely on a specify construction of the FQH states.
The relation between the bulk topological orders and edge states becomes
quite clear.

We know a hierarchical (or generalized hierarchical) FQH state
contains many different condensates, 
electron condensate forms the Laughlin states, and additional
quasiparticle condensates
on top of that give rise to a hierarchical state. Each condensate corresponds to one
component of the incompressible fluid. The idea is to generalize the
hydrodynamical approach in section 3 to multi-component fluids
and to obtain the low energy effective theory of the edge excitations.
To accomplish this, we would first like to write down the low energy effective
theory of the bulk FQH state. The effective theory contains 
information about the internal structures ( the topological orders) 
in the bulk states, such as
the number of the condensates and how different condensates interact
with each other.

The different condensates in FQH states are not independent. The particles
in one condensate behave like a flux tube to the particles in other condensates.
To describe such a coupling, it is convenient to use $U(1)$ gauge fields to
describe the density and the current of each condensate. In this case
the couplings between different condensates are described by the
Chern-Simons term
of the gauge fields (see chapter 2) \refto{BW,FZ,class}.
The most
general abelian FQH states of the electrons are classified by integer
valued symmetry matrices $K$ and described by
 the following effective theory
$$\cL=-{1\o 4\pi} K_{IJ}a_{I\mu}\part_\nu a_{J\la}\veps^{\mu\nu\la} 
- {e\o 2\pi} t_I A_{\mu}\part_\nu a_{I\la}\veps^{\mu\nu\la}
\eqno(2.4.1)$$
In the following we will choose the so called symmetric basis with
the charge vector $\tt^T=(1,1,...,1)$. In this case 
the diagonal elements of $K$ must be odd 
integers. We can always change the hierarchical basis with $\tt=(1,0,...,0)$
to the symmetric basis with $\tt=(1,1,...,1)$ by a field-redefinition of the 
$U(1)$ gauge fields. Let $\ka=$dim$(K)$. Then
the FQH state described by \(2.4.1)
contains $\ka$ different condensates and there are $\ka$ kinds of
different quasiparticle
excitations. 

The quasiparticle excitations can be viewed as vortices in different 
condensates.
A generic quasiparticle is labeled by $\ka$ integers $l_I|_I=1,..,\ka$, and can
be generated by a source term
$$\cL= l_I a_{I\mu} j^\mu
\eqno(2.4.1a)$$
Such a quasiparticle will be denoted as $\psi_\ll$.
$j_{\mu}$ in \(2.4.1a) and has the form 
$$\eqalign{
{\bf j}(x) = & {\bf v} \de({\bf x} -{\bf x_0}) \cr
j^0(x) = &\de({\bf x} -{\bf x_0}) \cr}
\eqno(2.4.1b)$$
which create a quasiparticle at ${\bf x_0}$ with velocity ${\bf v}$.
The density and the current of
the $I^{th}$ component of incompressible fluid (\ie the $I^{th}$ condensate)
are given by
$$J_{I\mu}={1\o 2\pi}\veps_{\mu\a\be} \part_\a a_{I\be} 
\eqno(2.4.2a)$$

As we create a quasiparticle $\psi_{\ll}$, it will induce a change
in the
density of all the condensates, $\de J_{I0}$.
 From the equation of motion of \(2.4.1) and \(2.4.1a),
we find that $\de J_{I0}$ satisfies
$$ \int d^2x \de J_{I0}=l_J (K^{-1})_{J I}\int d^2x j_{ 0}=l_J (K^{-1})_{J I}
\eqno(2.4.1aa)$$
The charge and the statistics of the
quasiparticle $\psi_\ll$ are given by
$$\th_{\ll}=\pi \ll^TK^{-1}\ll,\ \ \ \ Q_{\ll}=-e t_I  \int d^2x \de J_{I0}
=-e \ll^T K^{-1} \tt
\eqno(2.4.2)$$
A generic electron excitation can be viewed as a special kind of 
(generic) quasiparticles, $\psi_e=\psi_{\ll_e}$, where the components
of $\ll_e$ are given by
$$
l_{eI}=K_{IJ}L_J,\ \ \
L_I={\rm integers}, \ \  \
\sum_I L_I=1
\eqno(2.4.2b)$$
We can show
that these electron excitations satisfy the following properties:
a) they carry a unit
charge (see \(2.4.2)); b) they have the fermionic statistics; c) moving an
electron excitation $\psi_e=\psi_L$
around any quasiparticle excitations $\psi_\ll$ always
induces a phase of multiple of $2\pi$; and d) the excitations defined in
\(2.4.2b) are all the excitations satisfying the above three conditions.

We would like to point out that the effective theory \(2.4.1) not only
applies to the standard
QH system in which all electrons are spin polarized and are in the first
Landau level, it also applies to the QH system in which electrons may
occupy several Landau levels and/or occupy several layers and/or
carry different spins. In this case the index $I$ may label the condensates
in different Landau levels, in different layers and/or with different
spins. 

To understand the relation between the effective theory and the edge states,
let us first consider the simplest FQH state of
the  filling fraction $\nu=1/q$ and try to re-derive the results in section 3.1
from the bulk effective theory.
Such a FQH state is described by the $U(1)$
Chern-Simons theory with the action\refto{WN,FK}
$$ S={q\o 4\pi} \int a_\mu \part_\nu a_\la \veps^{\mu\nu\la} d^3x
\eqno (2.4.4)$$
 Suppose that our sample has a
boundary. For the simplicity we shall assume that
the  boundary is the $x$-axis and the  sample covers the  lower half-plane.

There is one problem with the effective action \(2.4.4) for FQH liquids with
boundaries. It is not invariant
under gauge transformations $a_\mu\to a_\mu+\part_\mu f$ due to the
presence of the boundary: 
$\De S = {q\o 4\pi} \int_{y=0} dx dt f (\part_0 a_1-\part_1 a_0)$. 
To solve this problem we will restrict the gauge transformations
to be zero on the boundary $f(x,y=0,t)=0$. Due to this restriction some degrees
of freedom of $a_\mu$ on the boundary become dynamical.

We know that the effective theory
\(2.4.4) is derived only for a bulk FQH state without boundary. Here 
we will take \(2.4.4)
with the restricted gauge transformation as the definition of the effective
theory for a FQH state with boundary. Such a definition is definitely 
self-consistent. 
In the following we will show that such a definition reproduces
the results obtained in section 3.1.

One way to study the dynamics of gauge theory is to choose the gauge
 condition 
$a_0=0$ and regard the equation of motion for $a_0$ as a
constraint. For the Chern-Simons theory such a constraint becomes
$f_{ij}=0$. Under this constraint, we may
write $a_i$ as $a_i=\part_i \phi$. Plugging this into
\(2.4.4), one
 obtains\refto{TOP} an effective 1D theory
on the edge with an action
$$ S_{edge}={m\o 4\pi} \int \part_t\phi \part_x \phi dxdt\eqno (2.4.5)$$
This approach,
however, has a problem. It is easy to see that a Hamiltonian associated
with the action \(2.4.5) is zero and the boundary excitations described by
\(2.4.5) have zero velocity. Therefore, this action cannot
be used to describe
any physical edge excitations in real FQH samples. The edge excitations in
FQH states always have finite velocities.

 The appearance of finite velocities of edge excitations is a boundary effect.
 The bulk effective theory defined
by Eq. \(2.4.1) does not contain the information about the
velocities of the edge excitations.
 To determine the dynamics of the edge excitations
from the effective theory we must find a way to input the information
about the
edge velocity. The edge velocities must be
treated as the external parameters that are not contained in the bulk
effective theory.  The problem
is how to put these parameters in the theory.

\par Let us now note  that the condition $a_0=0$ is not the only choice
for the gauge fixing condition.
A more general gauge fixing condition has the form
$$ a_\tau =a_0+v a_x=0\eqno (2.4.6)$$
Here $a_x$ is the component
of the vector potential parallel to the
boundary of the sample and $v$ is a parameter that has a dimension of
velocity.

It is convenient to choose new coordinates that satisfy
$$\eqalign{&\t x =  x-vt \cr
& \t t =   t,  \ \ \ \t y= y \cr}
\eqno (2.4.7)$$
In these coordinates the components of the gauge field are given by
$$\eqalign{
\t a_{\t t}=& a_t +v a_x \cr
\t a_{\t x}=& a_x \cr
\t a_{\t y}=& a_y \cr}
\eqno(2.4.8)$$
The gauge fixing condition becomes the one discussed before in the new coordinates.
It is easy to see that  the
form of the Chern-Simons action  is preserved
under the transformation \(2.4.7) and \(2.4.8):
$$S={q\o 4\pi} \int d^3x \  a_\mu \part_\nu a_\la \veps^{\mu\nu\la}=
{q\o 4\pi} \int d^3x\  \t a_{\t\mu} \part_{\t \nu}
\t a_{\t \la} \veps^{\t\mu\t\nu\t\la}
\eqno (2.4.9)$$
Repeating the previous derivation, we find that the edge action is given by
$$S={q\o 4\pi}\int d\t t d\t x \part_{\t t}\phi \part_{\t x}\phi
\eqno(2.4.10)$$
In terms of the original physical coordinates the above action acquires
the form
$$S={q\o 4\pi}\int dt dx (\part_{t}+v \part_x)\phi \part_x\phi
\eqno(2.4.11)$$
which is a chiral boson theory \refto{CB}.
It is easy to see that the edge excitations described by \(2.4.11) have a
non-zero velocity.
The quantization of chiral boson theory has been discussed in detail
in \Ref{CB}.
The canonical momentum $\pi (x)$ is equal to $\pi =
{\delta L\o \delta \phi_t}={q\o 4\pi}\part_x\phi$.
The ``coordinate'' $\phi$ and ``momentum'' $\pi$ obey the  commutation
relations:
$$\eqalign{
&[\pi (x), \phi (y)]={1\o 2}
\delta (x-y)\cr
&[\phi (x),\phi (y)]= {\pi \o q}
 \sgn (x-y)\cr}\eqno (2.4.12)$$

\par The Hamiltonian of the theory \(2.4.11) is given by
$$H=-{qv\o 4\pi}\int dx  \part_x\phi \part_ x\phi
\eqno(2.4.13)$$
 The Hilbert space contains only left-moving degrees of freedom
(or right moving degrees of freedom if $v<0$).
 The theory \(2.4.12) and \(2.4.13)
describes  free left (or right) moving phonons
(\ie the edge density waves). One can easily show that \(2.4.12)
and \(2.4.13) are
equivalent to the K-M algebra \(2.7) by identifying $\r={1\o 2\pi}\part_x \phi$.

In the following we would like to show that $\r={1\o 2\pi}\part_x \phi$
can actually be interpreted as the 1D electron density on the edge. First we
notice that the coupling between the electrons and the external electromagnetic
potential is given by $\int A_\mu J_\mu d^3 x=\int {1\o 2\pi} A\part ad^3 x$
(see \(2.4.2a)). But for a finite sample such a coupling should be
written as
$$\int A_\mu J_\mu d^3 x \sim \int {1\o 2\pi} a_\mu\part_\nu 
A_\la \veps^{\mu\nu\la}d^3 x
\eqno(2.413aa)$$
which is invariant under the gauge transformation of the electromagnetic 
field $A_\mu$ and under the restricted gauge transformation of the gauge field
$a_\mu$.
 Assuming that $A_\mu$ is independent
of $y$ and $\t y$ and $A_y=A_{\t y}=0$, we see that,
from $\t a_{\t i}=\part_{\t i}\phi$,
$$\eqalign{
  & \int {1\o 2\pi} a_{\t \mu}\part_{\t \nu} A_{\t \la} 
    \veps^{\t \mu\t \nu\t \la}d^3 \t x \cr
= &-\int d\t x d\t y d\t t {1\o 2\pi} \part_{\t y}\phi 
   (\part_{\t x} A_{\t t} -\part_{\t t} A_{\t x} )  \cr
= &-\int d\t x d\t t {1\o 2\pi} \phi 
   (\part_{\t x} A_{\t t} -\part_{\t t} A_{\t x} )  \big |_{\t y=0}\cr 
= & \int dxdt {1\o 2\pi} (A_t+vA_x)\part_x\phi  \big |_{\t y=0}\cr}
\eqno(2.4.13a)$$
where we have used the equation of 
motion $(\part_t +v \part_x)\phi=\part_{\t t} \phi=0$ and
the transformations \(2.4.7) and \(2.4.8). \(2.4.13a) clearly indicates
that the 1D edge electron density is given by ${1\o 2\pi}\part_x \phi=\r$.
This identification, together with the algebra \(2.4.12) and \(2.4.13),
completes our proof that our treatment of the effective theory 
\(2.4.4) and \(2.4.6) reproduces the edge theory obtained in section 3.1.

The velocity of the edge excitations $v$
 enters our theory through the gauge fixing condition. Notice
that  under the restricted gauge transformations
the gauge fixing conditions \(2.4.6) with different $v$ cannot be transformed
into each other. They are physically inequivalent. 
This agrees with our assumption that 
that $v$ in the gauge fixing condition
is physical and  actually determines the velocity of
the edge excitations.

The Hamiltonian \(2.4.13) is bounded from below only when $vq<0$.
The consistency of our theory requires $v$ and $q$ to have opposite signs.
 Therefore the sign
of the velocity (the chirality) of the edge excitations is determined
by the sign of the coefficient in
front of the Chern-Simons terms.

The above results can be easily generalized to the generic FQH states
described
by \(2.4.1) because the matrix $K$ can be diagonalized. The resulting
effective edge theory has the form
$$S_{edge}={1\o 4\pi}\int dt\ dx [K_{IJ}\part_t \phi_I \part_x \phi_J
-V_{IJ}\part_x \phi_I \part_x \phi_J]
\eqno(2.4.14)$$
The Hamiltonian is given by
$$H_{edge}={1\o 4\pi}\int dt\ dx V_{IJ}\part_x \phi_I \part_x \phi_J
\eqno(2.4.15)$$
Therefore $V$ must be a positive definite matrix. Using this result
one can show that a positive eigenvalue of $K$ corresponds to a left moving
branch and a negative eigenvalue corresponds to a right moving one.

The effective theory of the $\nu=2/5$ FQH state is given by
$$K=\pmatrix{3 &  2 \cr
             2 &  3 \cr}
\eqno(2.4.16)$$
Since $K$ has two positive eigenvalues, the edge excitations of the $\nu=2/5$
FQH state have  two branches moving in the same direction. The
$\nu=1-{1\o n}$ FQH state is described by the effective theory with
$$K=\pmatrix{1 &  0 \cr
             0 &  -n \cr}
\eqno(2.4.17)$$
The two eigenvalues of $K$ now have opposite signs, hence, the two branches of
the edge excitations move in opposite directions. 

\subhead{3.4 Charged excitations and electron propagator on the edges 
of generic FQH states}
In the last two sections we studied the dynamics of edge excitations 
in FQH
liquids at low energies. We found that the low lying edge
excitations are described
by a  free phonon theory that is exactly soluble. In this section we will
concentrate on the generic charge excitations. In particular we will
calculate the propagators of the electrons and the quasiparticles for the
most general (abelian) FQH states \refto{class}. (See chapter 5)
The key point again is to write the electron or the
quasiparticle operators in terms of the phonon operator $\r_I$.
Once we do so, the propagators can be easily calculated because the phonons
are free (at low energies and long wave length).

We know that for
the FQH state described by \(2.4.1), the edge states are described by the action
\(2.4.14). The Hilbert space of the edge excitations forms a representation
of K-M algebra
$$\eqalign{
[\rho_{Ik}, \rho_{Jk^\prime}]=&
(K^{-1})_{IJ} {1\o 2\pi}k\de_{k+k^\prime}  \cr
k,k^\prime=&{\rm integer}\times {2\pi\o L}  \cr }
\eqno(3.7)$$
where $\r_I={1\o 2\pi}\part_x \phi_I$ is 
the edge density of the $I^{th}$ condensate
in the FQH state, $I,J=1,...,\ka$, and $\ka$ is the dimension of $K$.
The electron density on the edge is given by
$$\r_e=-e\sum_I \r_I
\eqno(3.7a)$$
The dynamics of the edge excitations are described by the Hamiltonian:
$$H=2\pi \sum_{IJ} V_{IJ}\r_{I,k}\r_{J,-k}
\eqno(3.7b)$$
where $V_{IJ}$ is a positive definite matrix.

Let us first try to write down the quasiparticle operator
$\Psi_\ll$ on the edge which creates a quasiparticle labeled by $l_I$. 
We know that inserting the quasiparticle on the edge will
cause a change $\de\r_I$ in the edge density of the $I^{th}$
condensate (see \(2.4.1aa))
that satisfies
$$\int dx \de \r_I=l_J(K^{-1})_{JI}.
\eqno(3.7c)$$
Because $\Psi_\ll$ is a local operator that only causes a local change
of the density, we have
$$[\rho_I(x), \Psi_\ll(x')]= l_J(K^{-1})_{JI}\de(x-x')\Psi_\ll(x')
\eqno(3.8)$$
Using the Kac-Moody algebra \(3.7)
one can show that the quasiparticle operators that satisfy \(3.8)
are given by
$$
\Psi_\ll \prop e^{ i\phi_I l_I}
\eqno(3.9)$$
The charge of the quasiparticle $\Psi_\ll$ is determined from the commutator
$[\r_e, \Psi_\ll]$ and is found to be
$$Q_\ll
=-e\sum_{IJ} l_J (K^{-1})_{IJ}
\eqno(3.10)$$

From \(2.4.2b) and \(3.9), we see that the electron operator can be
written as
$$\eqalign{
\Psi_{e,L}\prop & \prod_I \Psi_I^{l_I}
\prop e^{ i\sum_{I}l_I \phi_I}   \cr
l_I=& \sum_J K_{IJ}L_J,\ \ \  \sum_I L_I= 1 \cr}
 \eqno(3.11)$$
The above operators carry a unit charge as one can see from \(3.10).
The commutation of $\Psi_{e,L}$ can be calculated as
$$\eqalign{
  &\Psi_{e,L}(x)\Psi_{e,L}(x')=(-)^{\la} \Psi_{e,L}(x')\Psi_{e,L}(x)   \cr
  &\la=\sum_{IJ}L_IK_{IJ}L_J  \cr}
\eqno(3.11b)$$
Because the diagonal elements of $K$ are odd integers, we can show that
$(-)^\la=-1$. The electron operators defined in \(3.11) are indeed fermion
operators.

Since all the operators $\Psi_{e,L}$ for different choices of $L_I$
carry a unit charge and are fermionic,
each $\Psi_{e,L}$ can be a candidate for the electron operators.
In general the true electron operator is a superposition of $\Psi_{e,L}$s
$$\Psi_e=\sum_L C_L\Psi_{e,L}
\eqno(3.11c)$$
In this paper, when we say there are many different electron operators on the
edge, we really mean that the true physical electron operator is a
superposition of the those operators.

Using the K-M algebra \(3.7)  and the Hamiltonian \(3.7b), we can calculate
the propagators of a generic quasiparticle operator
$$\Psi_{\ll}
\prop e^{ i\sum_{I}l_I \phi_I}
 \eqno(3.11a)$$
(which includes the electron operators for suitable choices of $\ll$).
First we notice that after a suitable
redefinition of $\r_I$:
$$ \t \r_I=\sum_J U_{IJ}\r_J
\eqno(3.12)$$
$K$ and $V$ can be simultaneously diagonalized,  \ie in terms of $\t \r_I$
\(3.7) and \(3.7b) become
$$\eqalign{
[\t\rho_{Ik}, \t\rho_{Jk^\prime}]=&  \si_I\de_{IJ}
{1\o 2\pi}k\de_{k+k^\prime} \cr
 H=& 2\pi\sum_{I} |v_I| \t\r_{I,k}\t\r_{I,-k}  \cr}
\eqno(3.13)$$
where $\si_I=\pm 1$ is the sign of the eigenvalues of $K$.  The velocity
of the edge excitations created by $\t \r_I$ is given by
$ v_I=\si_I | v_I|$.

To prove the above result
we first redefine $\r_I$ to transform $V$ into the identity matrix:
$V\to U_1 VU_1^T=1$.  This is
possible because $V$ is a positive definite symmetric matrix. Now $K$ becomes
a new symmetric matrix $K_1=U_1KU_1^T$ whose eigenvalues have the same sign
as the
eigenvalues of $K$ (although the absolute values may differ). Than we make an
orthogonal transformation to diagonalize $K_1$: $(K_1)_{IJ}\to \si_I|v_I|
\de_{IJ}$. After a trivial rescaling of the densities, we obtain \(3.13).
In terms of $\t \r_I$ the operator $\Psi_\ll$ has the form
$$\eqalign{
\Psi_\ll& \prop e^{ i\sum_{I}\t l_I\t \phi_I} \cr
\t l_J=&\sum_I l_I U^{-1}_{IJ}  \cr}
\eqno(3.14)$$

From \(3.13) and \(3.14)
we see that the propagator of $\Psi_\ll$ has the following general form
$$\<\Psi^\dag_\ll(x,t)\Psi_\ll(0)\>\prop e^{il_Ik_I x}
 \prod_I (x-v_It+i\si_I \de)^{-\ga_{I}},\ \ \ \ \ \ga_I=\t l_I^2
\eqno(3.15)$$
where $v_I=\si_I |v_I|$ is the velocities of the edge excitations.
$\ga_{I}$ in \(3.15)
satisfy the sum rule
$$\sum_I \si_I \ga_I\equiv \la_\ll=\sum l_IK^{-1}_{IJ}l_J
\eqno(3.16)$$
In order to prove  the above sum rule, we have used the relation
$$(UKU^T)_{IJ}=\si_I \de_{IJ}
\eqno(3.17)$$
From \(3.11b) and \(3.16) we see that the sum rule is directly related to the
statistics  of $\Psi_\ll$ and $\la_\ll$ is a topological quantum number.
If $\Psi_\ll$ represents the electron
operator, $\la_\ll$ will be an odd integer.
From \(3.15) we also see that the operator $\Psi_\ll$ creates an excitation with
momentum near $\sum_I l_Ik_I$.

As an application of the formalism developed above,
let us discuss the hierarchical states with the filling fractions
$\nu={p\o pq+1}$ ($q=$even)
in more detail. Those states include $\nu=2/5,3/7,2/9,...$
FQH states. The hierarchical states with $\nu={p\o pq+1}$
is described by the 
$p$ by $p$ matrices $K=1+qC$, where $C$ is the pseudo identity
matrix: $C_{IJ}=1$, $I,J=1,...,p$. $K^{-1}=1-{q\o pq+1}C$.
Because all the edge excitations move in the same direction, we have
$$\<\Psi_\ll^\dag(x=0,t)\Psi_\ll(0)\>\prop t^{-\la_\ll}
\eqno(3.18)$$
where $\la_\ll$ is given by \(3.16). The fundamental quasiparticle is given
by $\ll^T=(1,0,...,0)$ and carries the charge ${1\o pq+1}$. The exponent in its
propagator is $\la_\ll=1-{q\o pq+1}$. The quasiparticle with the smallest
exponent is given by $\ll^T=(1,...,1)$ and carries the charge ${p \o pq+1}$.
The exponent is $\la_\ll={p\o pq+1}$ that is less than $1-{q\o pq+1}$ (note
we have $q\ge 2$ and $p\ge 1$). 
Such a quasiparticle (with the charge ${p \o pq+1}$)
dominates the tunneling between two edges of the {\it same}
FQH fluid at low energies.

The electron operators are given by $\Psi_{e,L}=\psi_\ll$ with $\ll$ satisfying
$\sum_I l_I=pq+1$. The exponent in the propagator is given by
$\la_\ll=\sum l_I^2 -q(pq+1)$.
The electron operator with a minimum exponent in its propagator
is given by $\ll^T=(q,...,q,q+1)$. The value of the minimum exponent is
$\la_\ll=q+1$. Such an electron operator dominates the tunneling
between edges of two {\it different} FQH fluid at low energies.

\subhead{3.5 Some simple phenomenological consequences of chiral Luttinger
liquids at FQH edges}
In the last four sections 
we have shown that the electrons at the edges of a FQH liquid
form a chiral Luttinger liquid. As one of the 
characteristic properties of chiral
Luttinger liquids, the electron and the quasiparticle propagators
obtain anomalous exponents:
$$
\< \Psi_e^\dag(t,x=0) \Psi_e(0)\> \sim t^{-g_e},\ \ \ \ \ 
\< \Psi_q^\dag(t,x=0) \Psi_q(0)\> \sim t^{-g_q}
\eqno(3.5.1)$$
Those anomalous exponents can be directly measured by tunneling experiments
between the FQH edges. 

Two situations need to be considered separately: I) two edges of FQH liquids
are separated by an insulator; and II) two edges are separated by a FQH liquid.
In case I, only electron can tunnel between the edges, and in case II
the quasiparticles supported by the FQH liquid that separates the two edges
can tunnel. The tunneling operator $A$ is given by 
$A\propto \Psi_{e1}\Psi_{e2}^\dag$ for case I, where $\Psi_{e 1}$
and $\Psi_{e 2}$
are the electron operators on the two edges. 
The tunneling operator $A$ has the form
$A\propto \Psi_{q1}\Psi_{q2}^\dag$ for case II, where 
$\Psi_{q1}$ and $\Psi_{q 2}$ 
are the quasiparticle operators on the two edges. 
The physical properties of the tunneling can be calculated from the correlation
of the tunneling operator,
 which in turn can be expressed as a product
of the electron or quasiparticle propagators on the two edges.

Let $g=g_e$ for case I and $g=g_q$ for case II. Then at zero temperature
one can show that\refto{WT}
the anomalous exponents of the electron and quasiparticle operators
lead to non-linear tunneling $I$-$V$ curve,
$$I \propto V^{2g-1}
\eqno(3.5.1a)$$
The noise spectrum of the tunneling current
also contains a singularity at the frequency $f= Q V/h$,
$$ S(f) \sim |f -{Q V\o h}|^{2g-1}
\eqno(3.5.2)$$
where $V$ is the voltage difference between the two edges and $Q$ is the 
electric charge of the tunneling electron or the tunneling quasiparticle.
At finite temperature $T$, the zero bias conductance also has
a power law dependence:
$$\si \propto T^{2g-2}
\eqno(3.5.3)$$
We see that the anomalous exponents can be easily measured by the tunneling
experiments. The noise spectrum further reveals the charges of the tunneling
particles.

The exponents $g_e$ and $g_q$ are calculated in section 3.4. To summarize the
results in a simple way, it is convenient to divide the FQH edges into
three classes. A) all edge excitations move in one direction --
the direction of the normal drift velocity along the edge. Such a direction
will called the normal direction. B)
One branch of edge excitations propagates in the normal direction, and
all other branches propagate in the opposite direction.
C) There are two or more branches propagating in the normal
direction and one or more branches propagating in the opposite
direction.

For the class A edges, 
the exponents $g_e$ and $g_q$ are directly related to the
statistics of the electrons and quasiparticles. In this case $g_e$ and $g_q$ are
universal and independent of details of electron interactions and 
edge potentials. 
The following table lists some FQH states
which support the class A edge states, as well as
the corresponding values of the
exponents $g_{e,q}$ and the electric charge of the associated particles
(note only the minimum values of $g_e$ and $g_q$ are listed):
$$
\matrix{
\hbox{FQH states}:&
\nu=1/m & \nu =2/5 & \nu={n\o 2n+1} & (331)_{\nu=1/2} & (332)_{\nu=2/5} \cr
g_e:&
m & 3 & 3 & 3 & 3 \cr
g_q: &
1/m & 2/5 & {n\o 2n+1} & 3/8 & 2/5 \cr
\hbox{charge }Q_q/e: &
1/m & 2/5 & {n\o 2n+1} & 1/4 & 2/5 \cr
}
$$

We also see from section 3.4 that, due to the presence of the modes propagating
in both directions, $g_e$ and $g_q$ are not universal for the
class B and class C edges. Their values depend on the electron interactions
and the edge potentials. However the electrons in the real FQH sample interact
through a quite special interaction -- the long range Coulomb interaction.
In this case the edge Hamiltonian has a special form,
$$H = \sum [V_{ij} \r_i \r_j + V_c (\sum t_i \r_i)^2 ]
\eqno(3.5.5)$$
with $V_c \gg V_{ij}$ (in fact $V_c \sim -\ln k$ at long wave lengths).
In \(3.5.5) $t_i$ is the charge vector, and $\sum t_i \r_i$ is the total charge
density on the edge. 
In the limit $V_c \gg V_{ij}$ the edge has a special structure.
One edge mode  propagates at a large velocity $v\sim V_c$ in the normal
direction.
All other modes propagate at smaller velocity $v \sim V_{ij}$. What
is more important is that only the fast mode carries charge and all the
other slow modes are neutral. The separation between the charge and the 
neutral modes in the presence of the long range Coulomb interaction
make it very difficult to detect more than one edge branch 
in the edge magnetoplasma
experiments \refto{mp}. In general, the
edge magnetoplasma experiments can see only 
a single charge mode. The neutral modes are hard to detect due to
the weak coupling. Thus, the edge magnetoplasma experiment in \Ref{mp2/3}
does not contradict our picture that the 2/3 FQH state contains edge
modes that propagate in both directions.

In addition to separating the charge and the neutral modes, the long
range Coulomb interaction also makes the exponents $g_e$ and $g_q$
universal for the class B edges. The exponent $g_\ll$ in the propagator
of a quasiparticle labeled by an integer vector $\ll$ can be calculated
as follows. The calculation applies
only to the class B edges in the limit $V_c \gg V_{ij}$. We first
separate $\ll$ into a charge part and a neutral part,
$$\ll = \ll_c +\ll_n
\eqno(3.5.6)$$
where the neutral part $\ll_n$ satisfies
$$ \tt^T K^{-1} \ll_n =0
\eqno(3.5.7)$$
Therefore, we have
$$
\ll_n= \ll - \tt{\tt^T K^{-1} \ll \o \tt^T K^{-1} \tt},\ \ \ \ \ \
\ll_c= \tt{\tt^T K^{-1} \ll \o \tt^T K^{-1} \tt}
\eqno(3.5.8)$$
One can show that the operator $\sum_i l_{n,i} \r_i$ generates the slow
neutral modes and $\sum_i l_{c,i} \r_i$ generates the fast charge mode.
Following the calculation outlined in section 3.4, one can show that 
$$\eqalign{
g_\ll =  & |\ll_c^T K^{-1} \ll_c| + |\ll_n^T K^{-1} \ll_n| \cr
=& | \ll^T M \ll | \cr}
\eqno(3.5.9)$$
where
$$M =K^{-1} -{2\o \tt^T K^{-1} \tt} K^{-1} \tt \tt^T K^{-1}
\eqno(3.5.9a)$$
The simple result \(3.5.9) is due to two facts: a) the charge and the neutral
modes separate, and b) all the neutral modes propagate in one direction.
From \(3.5.9) we can calculate the minimum values of $g_e$ and $g_q$.
The following table lists some FQH states
which support the class B edge states, and
the corresponding values of the
exponents $g_{e,q}$ 
(again only the minimum values of $g_e$ and $g_q$ are listed).
$$
\matrix{
\hbox{FQH states}:&
\nu=2/3 &\nu=1-1/m & \nu =3/5 & \nu={n+1 \o 2n+1} & \nu= 2/7 \cr
g_e:&
2& {m+1\o m-1} & 7/3 & 3-{2\o n+1} & 4  \cr
g_q: &
2/3& {1\o m}{m+1\o m-1} & 3/5 & {n+1\o 2n+1} & 2/7 \cr
\hbox{charge }Q_q/e: &
1/3,2/3& 1/m & 3/5 & {n+1\o 2n+1} & 1/7,2/7 \cr
}
$$

In section 3.3 we argued that the edge excitations of the a FQH state
characterized by a matrix $K$ are described by $U(1)$ K-M algebras
characterized by the same matrix (see \(3.7)). This result is correct
only for sharp edges where the electron density drops to zero in a range
of order magnetic length near the edge. 
However, the edge potentials in real QH samples
are very smooth, and the electron density gradually drops to zero 
in a range of a few
thousands angstrom in order to minimize the electrostatic energy \refto{elesta}.
Thus the edge structures for smooth and sharp edge
potentials are quite different. 
It was shown in \Ref{CW} that as the edge potential
changes from a sharp one to a smooth one, the QH edge will undergo a so
called edge reconstruction. One or more pairs of edge branches are generated
after the edge reconstruction. Each pair contains two branches that propagate
in opposite directions, and behaves like the usual non-chiral Luttinger
liquid in one dimension. After the edge reconstruction, even a $\nu=1$ IQH state
can have several edge branches.
The edge excitations on a smooth edge
are generally described by the following matrix:\refto{CW}
$$K_{edge}=K_{bulk}\oplus \pmatrix{K'&0 \cr 0&-K' \cr}
$$
where $K_{bulk}$ is the matrix characterizing the bulk FQH state and $K'$
is another integer matrix that depends on the edge potential. 
For example, the edge excitations on a smooth 2/3 edge could be described by
$$K_{edge}=\pmatrix{-3&  &  &   \cr
                      &1 &  &   \cr
                      &  &-1&   \cr
                      &  &  &1  \cr}
$$

In an ideal FQH sample where electrons and/or quasiparticles are not allowed
to be scattered between different edge branches, a bar of the FQH liquid has
a quantized two-terminal conductance only when the FQH liquid has 
a class A edge state. However, in the presence of the long
range Coulomb interaction, any FQH liquids will have a (approximately) quantized
two-terminal conductance $\si=\nu e^2/h$, due to the separation of the charge 
and the neutral modes in the presence of the long range interaction.

In the real sample, there are many mechanisms that allow electrons and
quasiparticles to scatter between different edge branches. 
These mechanisms include
inelastic scattering of phonons and elastic scattering of impurities.
These scatterings equilibrate different edge branches.
Using the sum rule obtained in \Ref{WG,M}, one can show that 
equilibrated edge always gives quantized two-terminal conductance.
Thus in addition to the 
long range Coulomb interaction, there are many mechanisms to
cause the two-terminal conductance to be quantized at $\si=\nu e^2/h$,
even for the class B and C edges.

In the following we will briefly discuss some effects of impurities on the
properties of FQH edge states. We know that the electrons in the non-chiral
Luttinger liquids are localized in the presence of impurities. 
Because of this, the additional pairs of the edge branches generated by the 
smooth edge will be localized by the impurities.
Thus it is possible that the dirty smooth edges may have very similar
physical properties to the clean sharp edges at low enough energies.
It is also obvious that
the impurities cannot localize the electrons in the class A edge due to
the lack of back scattering. 

The situation becomes  very interesting for the
edge state of the $\nu=2/3$ FQH liquid, which contains two branches propagating
in opposite directions. In this case the impurities do cause back 
scattering. However the back scattering will not cause localization
of the electrons on the edge.
Mobile edge excitations are required to cancel the gauge anomaly from
the bulk Chern-Simons term, so that the effective
action for a finite QH system
will be invariant under the gauge transformations of the electromagnetic
gauge potential \refto{WG}. A detailed dynamical theory
for the disordered 2/3 edge has been proposed in \Ref{KFP}, where it was
shown that even at the strong-disordered fixed point the electrons on the edge
remain delocalized. It was further argued that the strong-disordered fixed
point has a very special dynamical property -- the charge mode and the
neutral mode separate even in the absence of long range interaction. The
separation of the charge and the neutral modes leads to the universal
$g_{e,q}$ whose values coincide with those calculated above for the
long range interaction. \Ref{KFP} also studied in detail some experimental 
consequences of the disordered edges. 
However, a different point of view was raised in \Ref{Hedge}.

\head{4. Shift and spin vectors -- New topological quantum numbers
for FQH liquids}
\taghead{4.}
In Chapter 2 we studied the effective theory of (abelian) FQH liquids. We found 
that the effective theory depends on the $K$-matrix and the charge vector
\tt. This result seems to suggest that the (abelian) FQH liquids are completely
characterized by $K$ and $\tt$. However, in this chapter we will see that
there
is a new kind of topological quantum numbers for FQH liquids. This new
type of quantum numbers 
does not directly appear in the effective theory on the plane.
However they can still be measured by experiments
in planar systems, due to a new set of selection rules induced by
these quantum numbers.

\subhead{4.1 Quantum Hall states on a sphere and the shift}
\def\S{{su(2)}}
To understand the appearance of new quantum numbers -- spin vectors, let us
first consider the $\nu=1$ IQH state on a sphere. Assume that 
the magnetic field
is uniform on the sphere and let $N_\phi$ be the total number of the
flux quanta passing through the sphere. The Hamiltonian has an $su(2)$
symmetry associated with the rotation of the sphere. Thus, the degenerate
eigenstates form irreducible representations of the $su(2)$ group.
The ground states form a representation with a $\S$-spin $S^\S=N_\phi /2$
and thus have $2S^\S+1=N_\phi+1$ folds degeneracy \refto{sphere}. 
These $N_\phi+1$ states
form the first Landau level on the sphere. The next energy level has
$N_\phi+3$ degeneracy and forms the $S^\S={N_\phi \o 2}+1$ representation of 
the $\S$, which corresponds to the second Landau level. In general the
states in the n$^{th}$
Landau level form the  $S^\S={N_\phi \o 2} +n-1$ representation of the $\S$
rotation of the sphere.

The electron wave functions on the sphere can be expressed in a simple form
through the following spinor coordinates\refto{sphere}
$$(u, v)=( \cos(\th) e^{i\phi/2}, \sin(\th) e^{-i\phi/2})
\eqno(4.1.1)$$
The $N_\phi+1$ states in the first Landau level
are described by the wave functions $u^m v^n$, $m=N_\phi-n=0,...,N_\phi$,
which form the $S^\S=m+n$ representation with $S^\S_z=m$. 

We see that the $\nu=1$ IQH state with a filled first Landau level contains
$N_e=N_\phi+1$ number of electrons. Thus $N_e/N_\phi$ is not exactly
equal to one. We will call $\cS$ in the relation
$$N_\phi=\nu^{-1} N_e -\cS
\eqno(4.1.2)$$
the shift. The shift of a QH state depends on the topology of the space.
The $\nu=1$ IQH state has a shift $\cS=1$ on a sphere, and a shift $\cS=0$
on a torus \refto{torus}.

We can construct another $\nu=1$ IQH state with filled {\it second} 
Landau level. Such a  state contains $N_e=N_\phi+3$ electrons and has
a shift $\cS=3$ on the sphere. On the torus the shift still vanishes.

Both the above two IQH states are described by the same effective theory
with $K=1$ and $\tt=1$ on the plane. This indicates that the topological
quantum number $\cS$ cannot be determined from  $K$ and $\tt$. Thus the
$K$-matrix and the charge vector do not provide a complete description
of the QH liquids. We need to find some additional quantum numbers
which will at least enable us to determine the shift \refto{spinv,cmhi,EG}.

To gain some understanding of the origin of the new quantum numbers, let us
consider in more detail the above two $\nu=1$ IQH states.
The two IQH states differ only by the different cyclotron motions of the
electrons. We know that as the Landau level index increases, the orbital
angular momentum carried by the cyclotron motion also increases.
In the following we will call such orbital angular momentum -- orbital spin.
Note here that the orbital spin is associated with the $U(1)$ rotation
of the plane and should be distinguished from the ordinary spin quantum
number of the electron and the $su(2)$ quantum $S^\S$ discussed above.
In this chapter we will consider only spinless electrons (which correspond
to the spin polarized electrons in experiments).

As an object with no-zero orbital spin moves along a loop $C$
on a curved 2D space, a non-zero
Berry's phase will be induced. The Berry's phase is given by
$$\phi =S_{ob} \oint_C d x^i \om_i =S_{ob} \int_{S_C} d^2 x R
\eqno(4.1.3)$$
where $S_C$ is the area enclosed by the loop $C$. Here $R$ is the
curvature and $\om_i$ the connection whose curl gives rise to the curvature.
The relation between $\om_i$ and $R=\veps_{ij}\part_i \om_j$ 
is the same as relation between the
gauge potential $a_i$ and the ``magnetic field" $b=\veps_{ij}\part_i a_j$.
When a 2D space is embedded in a 3D space, the integral
$\oint_C d x^i \om_i=\int_{S_C} d^2 x R$ has a geometric meaning. The norm 
vector of the 2D surface spans a solid angle $\Om$ as it moves along the
loop $C$. The above integral is simply determined by this solid angle:
$$\oint_C d x^i \om_i=\int_{S_C} d^2 x R = \Om 
\eqno(4.1.4)$$
Thus the total curvature of a sphere is $\int_{sphere} d^2 x R=4\pi$.

From the above discussion we see that because electrons carry a non-zero
orbital spin, the total flux seen by the electron is the sum of the 
magnetic flux and the Berry's phase induced through the coupling of the
orbital spin to the curvature of the space. It is this Berry's phase
that causes the shift. Thus, the electrons with different orbital spin
will cause different shifts. According to this picture we also
obtain the
well known fact that the shift always vanishes on torus \refto{torus},
since the curvature on torus is zero.

According to the semiclassical picture of the cyclotron motion,
the electrons in the n$^{th}$
Landau level have an orbital spin of $S_{ob}=n+$const. Thus, the orbital spins
of the electrons in the first and the second Landau levels differ by one,
and the induced Berry's phase differ by two flux quanta on the sphere.
Therefore, the shifts of the $\nu=1$ state in the first landau level and the
second Landau level differ by two.

To determine the absolute value of the orbital spin for each Landau level,
let us move an electron in the n$^{th}$ Landau level around a loop $C$
on the sphere. There are two ways to calculate the phase $\Phi$ induced by such
a motion. The phase $\Phi$ can be obtained as the sum of the phase induced
by the magnetic field and the Berry's phase induced by the orbital spin
$$\Phi =\oint_C (dx^i -eA_i + S_{ob} \om_i)
       = 2\pi (N_\phi + 2 S_{ob}) {\Om \o 4\pi}
\eqno(4.1.5)$$
where $\Om$ is the solid angle spanned by the loop $C$ on the sphere.
The phase $\Phi$ can also be calculated from the fact that the
electrons in the n$^{th}$ Landau level form an $S^\S ={N_\phi \o 2} +n-1$
representation of the $su(2)$. An electron in the n$^{th}$ Landau level
and at position $(\th, \phi)$ on the sphere is an eigenstate of
${\bf n} \cdot {\bf S}^\S$ with an eigenvalue ${N_\phi \o 2} +n-1$, where
${\bf n}$ is the unit vector in the direction $(\th, \phi)$ and
${\bf S}^\S$ are the three generators of the $su(2)$ rotation of the sphere.
As ${\bf n}$ traces out the loop $C$, a Berry's phase will be induced due to
the non-zero $su(2)$ spin $S^\S$. The induced phase is
$$\Phi=S^\S \Om = [N_\phi +2(n-1) ]{\Om \o 2}
\eqno(4.1.6)$$
Comparing \(4.1.5) with \(4.1.6) we find the orbital spin
for the n$^{th}$ Landau level to be
$$S_{ob}=n-1
\eqno(4.1.7a)$$
The orbital spin for first Landau level vanishes.
\(4.1.5), \(4.1.6) and \(4.1.7a) reveal a direct relation between the following quantities:
a) the orbital spin of electrons; b) the number of effective flux quanta,
$N^*_\phi =N_\phi + 2S_{ob}$, seen by the electrons;
c) the $S^\S$ quantum number of the electrons, $S^\S =N^*_\phi/2$,
d) the number of states in the corresponding Landau level,
$N=2S^\S +1 =N^*_\phi +1$, and e) the value of the shift,
$\cS = N -N_\phi =2 S_{ob} +1$, for the $\nu=1$ state.

\subhead{4.2 Spin vectors in QH liquids}
In the last section we saw that the shift is closely related to the coupling
between the orbital spin and the curvature of the space. In this section we will
attempt to include such coupling in the effective theory so that we will
be able to calculate the shift from the effective theory. In the process of
building such an effective theory we find that it is necessary to introduce
a new topological quantum number -- spin vector -- in our description
of QH liquids. 

First let us consider a system of bosons or fermions on a sphere with a uniform
magnetic field. 
We also assume that the bosons or fermions carry an orbital spin $S_{ob}$.
Those particles are described by
$$\cL =-et_1 A_\mu J^\mu+S_{ob} \om_i J^i +\KE
\eqno(4.2.1a)$$
Here in addition to the orbital spin, we have 
also included two other complications.
The first one is that we allow the magnetic field $B$ to have either positive
or negative values. The second one is that we include $t_1$ in \(4.2.1a).
$t_1$ is equal to 1 if the particles carry a charge of $-e$,
and is equal to $-1$ if the particles carry a charge $e$.

Let $N_\phi=-eB/\Phi_0$ be the number of the flux quanta. If $N_\phi t_1>0$,
then
a filling-fraction-$1/m$ Laughlin state of the boson or the fermion
is described by the wave function
$$\Phi=\prod_{i<j} (v_iu_i-v_ju_j)^m
\eqno(4.2.1)$$
where $(u,v)$ is the spinor coordinate (see \(4.1.1)). 
If $t_1 N_\phi<0$, the $1/m$ Laughlin state will has a form
$$\Phi=\prod_{i<j} (v_i^* u_i^* -v_j^* u_j^*)^m
\eqno(4.2.1aa)$$
Here we always assume $m>0$.
For a given $i$,
$\Phi$ is a homogeneous
polynomial of $(u_i,v_i)$ (or $(u_i^*,v_i^*)$)
of degree $m (N_p-1)$ where
$N_p$ is the number of the bosons or fermions. Thus the single-particle
states form an $S^\S=m (N_p-1)/2$ representation of the $su(2)$ rotation
of the sphere. Knowing that all particles are in the first Landau level
we conclude that the total number of the effective flux quanta
seen by the particles is given by
$N^*_\phi=2S^\S=m (N_p-1) =|t_1N_\phi +2 S_{ob}|$. This implies that 
$|N_\phi|=mN_p -[m+2 S_{ob}{\rm sgn}(t_1 N_\phi) ]$.
The shift is defined through $|N_\phi|=mN_p -\cS$ when $N_\phi$ can be negative. 
So the shift of our $1/m$ state is given by
$$\cS=2 S_{ob}{\rm sgn}(t_1 N_\phi) +m
\eqno(4.2.2)$$
\(4.2.2) is valid for both positive and negative $t_1$ and $N_\phi$.
We also would like mention that the number of the particles $N_p$ is
assumed to be positive in the above discussion.

Now the question is how to reproduce the shift \(4.2.2) from the effective 
theory. Here we propose the Chern-Simons effective theory of the $1/m$ state
on the sphere to have the form
$$
\cL =
\left[ -m{\rm sgn}(t_1 N_\phi){1 \over 4 \pi}
a_{ \mu} \partial_\nu a_{ \lambda}~\veps^{\mu\nu\lambda}
-{e t_1\over 2 \pi} A_{ \mu} \partial_\nu a_{ \lambda}~\veps^{\mu\nu\lambda}
+s  \om_i \partial_\nu a_{ \lambda}~\veps^{i\nu\lambda}
\right]
\eqno(4.2.3)$$
where the third term describes the coupling of the curvature.
The new quantum number $s$ will be called the spin vector (here it has
only one component). From the
equation of motion ${\part \cL\o \part a_0}=0$, the total number of
the particles can be shown to be
$$ \eqalign{
N_p=&\int d^2 x {1\o 2\pi} \veps_{ij} \part_i a_j
=\int d^2 x {1\o 2m\pi{\rm sgn}(t_1 N_\phi)} 
\veps_{ij} \part_i (-t_1 eA_j + 2\pi s\om_j) \cr
=&{1\o m{\rm sgn}(t_1 N_\phi)} (t_1 N_\phi+2s) \cr}
\eqno(4.2.4)$$
We see that the inclusion of sgn($t_1 N_\phi$) in \(4.2.3)
ensure that $N_p >0$.
\(4.2.4) implies that $N_p={1\o m} [|N_\phi|+2{\rm sgn}(t_1 N_\phi)s]$.
Thus the shift is simply given by $\cS=2{\rm sgn}(t_1 N_\phi)s$ 
and we find the spin vector
for the above $1/m$ state to be (by comparing with \(4.2.2) )
$$s=S_{ob}+{m\o 2}{\rm sgn}(t_1 N_\phi)
\eqno(4.2.5a)$$

It is interesting to see that the spin vector receives two contributions.
The first contribution, coming from the 
orbital spin $S_{ob}$, is easy to understand.
One way to understand the second contribution
is the following. We know that the
$1/m$ state can be viewed as a boson condensation of composite bosons.
A composite boson is a bound state of the boson or the fermion with
$m$ units of flux quanta. The binding of the flux not only changes the 
statistics of the particle, it also changes the 
orbital spin of the boson or the fermion to a new value, and $s$ in \(4.2.5a)
can be regarded as the orbital spin of the composite bosons.

We would like to mention that if the finite orbital spin $S_{ob}$ is due
to the bosons or fermions occupying the $n^{th}$ Landau level, then
$$S_{ob}=(n-1) {\rm sgn}(t_1 N_\phi)
\eqno(4.1.7)$$
which generalizes \(4.1.7a) to include the possibilty that $t_1$ and $N_\phi$
may be negative.
Thus \(4.2.5a) can be rewritten as
$$s=(n-1+{m\o 2}){\rm sgn}(t_1 N_\phi)
\eqno(4.2.5b)$$

\(4.2.5b) only applies to the Laguangian
$$
\cL =
\left[ -K_{11}{1 \over 4 \pi}
a_{ \mu} \partial_\nu a_{ \lambda}~\veps^{\mu\nu\lambda}
-{e t_1\over 2 \pi} A_{ \mu} \partial_\nu a_{ \lambda}~\veps^{\mu\nu\lambda}
+s  \om_i \partial_\nu a_{ \lambda}~\veps^{i\nu\lambda}
\right]
\eqno(4.2.3b)$$ 
with $K_{11}$ and $t_1 N_\phi$ to have the same sign. In this case
$\eps_{ij}\part_i a_j/2\pi$ is positive and can be regarded as the
density of the particles. When $K_{11}$ and $t_1 N_\phi$ have the
opposite signs, we should view $-\eps_{ij}\part_i a_j/2\pi$ as the
density of the particles. Or we can make a transformation
 $a_\mu\to -a_\mu$ and $t_1\to -t_1$ in \(4.2.3b) to make
$K_{11}$ and $t_1 N_\phi$ to have the same sign.
Thus \(4.2.5b) can be generalized to
$$s=(n-1+{|K_{11}|\o 2}){\rm sgn}(t_1 N_\phi)
\eqno(4.2.5)$$
which applies to the 1/m state described by \(4.2.3b) where
$K_{11}$ can have any signs.  Here we have assumed that 
the condensed particles occupy the $n^{th}$ Landau level.

Applying the above results to the electron systems we see that the $1/m$ state
in the first Landau level has a spin vector $s=m/2$ 
while a
$1/m$ state in the n$^{th}$ Landau level has a spin vector 
$s=n-1+{m\o 2}$
(Here we have assumed $t_1>0$ and $N_\phi >0$).

Now, knowing the effective theory of the first level hierarchical states, we
can include the quasiparticle excitations
(or electron excitations in the next layer), and 
construct the effective theory of the second level hierarchical states
(or double layer FQH states).

Consider a $\ka^{th}$ level hierarchical state (or a $\ka$-layer FQH
state) of an electron system on a sphere.
Motivated by the effective theory of the Laughlin state \(4.2.3b),
let us assume the effective theory for the FQH state
to have a form
$$\cL=-{1 \over 4 \pi}\La_{II^\prime}
a_{I \mu} \partial_\nu a_{I^\prime \lambda}~\veps^{\mu\nu\lambda} 
-{e \over 2 \pi} t_I A_{ \mu} \part_\nu a_{I \la}\veps^{\mu\nu\la}
+s_I \om_i \part_\nu a_{I \la}\veps^{i\nu\la}
\eqno(4.2.6)$$
where the curvature couples 
with all the condensates through the spin vector $\ss^T=(s_1,s_2,...)$.  
From the
equation of motion ${\part \cL\o \part a_0}=0$, we obtain the number of 
electrons
$$ N_e =\tt^T K^{-1} \tt N_\phi + 2 \tt^TK^{-1} \ss
\eqno(4.2.7a)$$
For an electron system 
$\tt^T K^{-1} \tt N_\phi=\nu N_\phi$ 
is always positive, and $N_e$ is
positive as expected. However for a hole system (where the particles
carry a charge of $e$),
$\tt^T K^{-1} \tt N_\phi=\nu N_\phi$ 
is always negative. To include this possibility \(4.2.7a) should be
generalized to
$$ N_e =|\tt^T K^{-1} \tt N_\phi + 2 \tt^TK^{-1} \ss|
\eqno(4.2.7)$$
Now the positive $N_e$ can be regarded as
the number of electrons or holes.
We can see that the shift, for electron or hole systems, is given by
$$\cS = {2 \tt^TK^{-1} \ss \o \tt^T K^{-1} \tt }{\rm sgn}(N_\phi)
\eqno(4.2.8)$$
on the sphere. We see that the spin vector determines the shift
of the QH state.

Now the question is how to determine the value of the spin vector.
The idea is that the FQH state described by \(4.2.6) contain $\ka$
componants, and each componant is desribed by a Laughlin state.
Thus we may use the result \(4.2.5) to calculate the spin vector $\ss$.

Let us first concentrate on the first componant described by $a_{1\mu}$.
\(4.2.6) can be rewritten as
$$\cL= -{1 \over 4 \pi} K_{11} a_{1 \mu} \part_\nu a_{1 \la}\veps^{\mu\nu\la}
+(-\sum_{I>1} {1 \over 2 \pi}\La_{I1}
a_{I \mu} 
-{e \over 2 \pi} t_1 A_{ \mu}) \part_\nu a_{1 \la}\veps^{\mu\nu\la}
+s_1 \om_i \part_\nu a_{1 \la}\veps^{i\nu\la} + ...
\eqno(4.2.6a)$$    
where ``...'' represents terms that do not contain $a_{1\mu}$.
\(4.2.6a) and \(4.2.3b) have the same form except $-et_1 A_\mu$ in \(4.2.3b)
is replaced by $-et_1 A_\mu -\sum_{I>1} \La_{I1} a_{I \mu}$.
Thus from \(4.2.5) we see that
$$s_1={|K_{11}|\o 2} {\rm sgn}(t_1 N_\phi -\sum_{I>1} K_{I1} {b_I\o 2\pi})
\eqno(4.2.6b)$$    
if we assume the condensate occupies the first Landau level.
Since ${b_I\o 2\pi}=(K^{-1})_{IJ}t_J N_\phi$, \(4.2.6b) can be rewritten as
$$\eqalign{
s_1=&{|K_{11}|\o 2} 
{\rm sgn}((t_1 -\sum_{I=2,J=1}^\ka K_{I1}(K^{-1})_{IJ}t_J) N_\phi ) \cr
=& {K_{11}\o 2} {\rm sgn}( \sum_J (K^{-1})_{1J}t_J  N_\phi ) \cr}
\eqno(4.2.6c)$$ 
Generaling the above to a general condensate, we get
$$s_I={K_{II}\o 2}{\rm sgn}(\sum_J  (K^{-1})_{IJ}t_J  N_\phi )
\eqno(4.2.9c)$$ 
Note $t_I\sum_J  (K^{-1})_{IJ}t_J  N_\phi$ is the number of charges
(in the unit of  electron charge $-e$)
in the $I^{th}$ condensate.

To obtain \(4.2.9c) we have assumed that the each condensate forms a
Laughlin state in the first
Landau level, which is a part of assumptions in the hierarchical construction.
To construct more general states we may assume (although it may not be
energetically favorable) that the $I^{th}$ condesate forms a Laughlin
state in the $n_I^{th}$ Landau level.
In this case we should replace \(4.2.9c) by
$$s_I={K_{II}\o 2}{\rm sgn}(\sum_J  (K^{-1})_{IJ}t_J  N_\phi )
+(n_I-1) {\rm sgn}(K_{II} \sum_J  (K^{-1})_{IJ}t_J  N_\phi )
\eqno(4.2.9d)$$ 
where $n_I$ is an integer.

An important consequence of \(4.2.9d) is the quantization
of the spin vector. We see that two times the spin vector, 
$2\ss$, is always an integer vector.

We would like to point out that \(4.2.9c) and 
\(4.2.9d) apply to both the hierarchical
FQH states and the multi-layer 
FQH state. In general we expect the stable states to have $n_I=1$.
For the multi-layer states with $t_I=1$ and
$K_{IJ}$s all positive, 
$(K^{-1})_{IJ}t_J  N_\phi$ is the number of electrons in the $I^{th}$
layer which is positive. Therefore
$$s_I={K_{II}\o 2}
\eqno(4.2.9m)$$
In particular for the $(lmn)$ state
$$\ss=\pmatrix{ l/2 \cr
              m/2 \cr}
\eqno(4.2.9e)$$
For the hierarchical states with $\tt^T=(1,0,0...)$,
\(4.2.9c) reduces to
$$s_I={K_{II}\o 2}{\rm sgn}( (K^{-1})_{I1} N_\phi )
\eqno(4.2.9h)$$ 
which allows us to calculate the spin vectors for the
hierarchical states listed in \(4.16b)

Combining \(4.2.9m) and \(4.2.9h) with \(4.2.8), we can calculate the shift
for the hierarchical or the multi-layer states. Notice that the shift
$\cS$ is independent of sgn$(N_\phi)$ due to a cancelation, which is expected.

In the following we will calculate the orbital spin quantum number of 
quasiparticle excitations\refto{spinv,cmhi,EG} from the new effective theory
\(4.2.6) that contains the spin vector.
Let us create a generic quasiparticle
labeled by $l_I$ in the FQH state \(4.2.6) on the sphere.
The charge and the statistics of the quasiparticle are given by
\(4.16). In the presence of the quasiparticle the relation between
the number of  electrons and the number of the flux quanta is no longer given
by \(4.2.7). There is an additional shift due to the quasiparticle.
Again from the equation of motion ${\part (\cL+\cL_q)\o \part a_0}=0$,
we find
$$N_e =\tt^T K^{-1} \tt N_\phi + 2 \tt^TK^{-1} \ss + \tt^T K^{-1} \ll
\eqno(4.2.10)$$
To calculate the orbital spin of the quasiparticle, we move
the quasiparticle around a loop $C$ which spans a solid angle $\Om$.
There are two ways to calculate the induced phase $\Phi$. First the phase
$\Phi$ can be obtained as a 
sum of the phase induced by a magnetic field and the phase induced
by the orbital spin, 
$$\Phi={Q_q\o e} N_\phi {\Om\o 2} + 2 S^q_{ob} {\Om\o 2}
\eqno(4.2.11)$$
where $Q_q=e\tt^T K^{-1} \ll$ is the charge and
$S^q_{ob}$ is the orbital spin of the quasiparticle.
In fact, \(4.2.11) can be viewed as the definition of the
orbital spin $S^q_{ob}$ of the quasiparticle.
Note $N_\phi$ in \(4.2.11) is the number of the flux quanta
in the presence of the quasiparticle and is given by \(4.2.10).
The same phase can also be calculated from the coupling 
$$ l_I a_{I\mu} j_\mu
\eqno(4.2.9)$$
First let us assume that the densities of each condensate (\ie the field
strength of $a_{I\mu}$) is uniform on the sphere even in the
presence of the quasiparticle. This can be achieved by including
the term 
$$ \cL_K=g (f^{I}_{\mu\nu})^2,\ \ \ \
 f^{I}_{\mu\nu} = \part_\mu a_{I\nu}-\part_\nu a_{I\mu}
\eqno(4.2.12)$$
in the effective theory \(4.2.6) and assuming $g$ is very very large.
In this limit, the phase induced by the coupling \(4.2.9) can be
easily calculated. From the equation of motion we find,
$$ \int d^2x b_I 
= \int d^2 x\veps_{ij}\part_i a_{Ij}
= 2\pi K^{-1}_{IJ} (t_J N_\phi +2 s_J +l_J)
\eqno(4.2.13)$$
where we integrate over the whole sphere.
Since $b_I$ are constants on the sphere, we have
$$\Phi=l_I {\Om \o 4 \pi} \int d^2x b_I
=l_I  K^{-1}_{IJ} (t_J N_\phi +2 s_J +l_J){\Om \o 2}
\eqno(4.2.14)$$

Although $\Phi$ in \(4.2.14) is calculated in the large $g$ limit, in the
following we would like to show that $\Phi$ is topological and is
independent of $g$. Notice that the solid angle $\Om$ spanned by the
loop $C$ is defined only up to a multiple of $4\pi$. The consistency of
the theory requires that the ambiguity of $\Om$ should only cause an ambiguity
of multiples of $2\pi$ in the phase $\Phi$. Thus the
coefficient in front of $\Om /2$ must be quantized as an integer for
any value of $g$,
$$\ll^T  K^{-1}\tt N_\phi +2 \ll^T  K^{-1}\ss +\ll^T K^{-1}\ll ={\rm integer}
\eqno(4.2.15)$$
This implies that
the coefficient in front of $\Om /2$ is independent of $g$ and that \(4.2.14)
is valid for all values of $g$.
Comparing \(4.2.11) and \(4.2.14) we find
$$ S_{ob} ={1\o 2} \ll^T K^{-1}\ll + \ll^T  K^{-1}\ss
\eqno(4.2.16)$$
It is interesting to see that the orbital spin of the quasiparticle receives
two contributions. The first term ${1\o 2} \ll^T K^{-1}\ll ={\th\o 2\pi}$
is directly related to the statistics of the quasiparticle $\th$
and satisfies the spin-statistics theorem. The second term is due
to the spin vector of the condensates, which violate the
spin-statistics theorem \refto{spinv,cmhi,EG}.

We would like to remark that for a given $K$-matrix, \(4.2.15)
sometimes cannot be satisfied for any integer choice of $N_\phi$,
if $\ll$ takes certain values. This implies that the
quasiparticle labeled by these $\ll$ cannot be created individually
on a sphere \refto{ELi}. For example, one cannot create 
a single charge-1/4 quasiparticle on top of 
the $\nu=1/2$ (331) FQH state on a sphere, although one can create two of them
by inserting a unit flux.
This selection rule, however, does not invalidate the above calculation
of $S_{ob}$, because
mathematically we can always choose a non-integer value for $N_\phi$.

From the construction discussed above we can easily obtain the spin vectors
for the FQH states listed in \(4.16b) and \(2.2.7). 
Using \(4.2.16) we can calculate the orbital spin quantum numbers
of various excitations. In particular, one can show that some
neutral plasma modes carry non-trivial orbital spin quantum numbers.
The conservation of the angular momentum implies that to
excite these plasma modes through, \eg Raman scattering, the photon
must transfer a certain amount of angular momentum to the plasma mode.
By performing angle-resolved circular-polarized Raman scattering
we can measure the amount of transferred angular momentum, which in turn
measures
the orbital spin of the plasma mode. Such measurement
allows us to obtain
information about the spin vector of the FQH state \refto{cmhi}.

\subhead{4.3 Wave functions and $(K,\tt,\ss)$ of FQH liquids}
In this section we will study the relation between 
$(K,\tt,\ss)$ (the $K$-matrix, charge vector $\tt$,  and the
spin vector $\ss$) and the wave functions of FQH liquids.
The results obtained here allow us to easily calculate
$(K,\tt,\ss)$ for various FQH liquids obtained in various
construction schemes.

Let $\Psi_{K_1\tt_1\ss_1}(\{z_i\})$ 
and $\Psi_{K_2\tt_2\ss_2}(\{w_i\})$ be two wave
functions of two FQH liquids labeled by $(K_1,\tt_1,\ss_1)$ 
and $(K_2,\tt_2,\ss_2)$.
Putting them together we obtain a new FQH liquid labeled
by $(K_3,\tt_3,\ss_3)$
$$\Psi_{K_1\tt_1\ss_1}(\{z_i\}) \Psi_{K_2\tt_2\ss_2}(\{w_i\})
=\Psi_{K_3\tt_3\ss_3}(\{z_i\},\{w_i\})
\eqno(4.3.1)$$
Obviously
$$\eqalign{
  K_3=&\pmatrix{ K_1& 0\cr
                0  &K_2 \cr}=K_1\oplus K_2 \cr
  \tt_3=&\pmatrix{ \tt_1\cr
                \tt_2\cr}=\tt_1\oplus \tt_2 \cr
  \ss_3=&\pmatrix{ \ss_1\cr
                \ss_2\cr}=\ss_1\oplus \ss_2 \cr}
\eqno(4.3.2)$$

We also know that $\Psi_{K\tt\ss}^\dag (\{z_i\})$ 
is a wave function for the holes
(with charge $e$) which will be called the charge conjugate state of
the $\Psi_{K\tt\ss}$ state. The Hall conductance of the 
charge conjugate state has an opposite sign to that of the original state.
Notice that the following transformation in the effective
theory \(4.2.6):
$$ \tt \to -\tt, \ \ \ \ss\to -\ss,\ \ \ K\to -K,\ \ \ a_{I\mu}\to a_{I\mu}
\eqno(4.3.3)$$
leaves the equation motion unchanged. However, the transformation
changes the sign of the Hall conductance and the sign of electric
current $J_e^\mu=-{e\o 2\pi}t_I \part_\nu a_{I\la} \veps^{\mu\nu\la}$.
Thus,
$$\Psi_{K\tt\ss}^\dag =\Psi_{-K,-\tt,-\ss}
\eqno(4.3.4)$$
\ie the charge conjugation changes the signs of $\tt$, $\ss$, and $K$.

Given a spin-polarized FQH state in a single layer at a filling fraction
$\nu_{K\tt}$ and labeled by $(K,\tt,\ss)$, there is a particle-hole conjugate
state at the filling fraction $\nu=1-\nu_{K\tt}$. 
This conjugate state is obtained simply by putting the $\nu=1$ IQH state
and the charge conjugate of the $(K,\tt,\ss)$ state together.
Thus the label of the $1-\nu_{K\tt}$ state is given by
$$K_{1-\nu_{K\tt}} =1_{1\times 1}\oplus (-K),\ \ \
  \tt_{1-\nu_{K\tt}} =1 \oplus (-\tt),\ \ \
  \ss_{1-\nu_{K\tt}} ={1\o 2} \oplus (-\ss)
\eqno(4.3.5)$$
Similarly, given a spin-polarized FQH state 
in double layers at a filling fraction
$\nu_{K\tt}$ and labeled by $(K,\tt,\ss)$, there is a particle-hole conjugate
state at the filling fraction $\nu=2-\nu_{K\tt}$. Such a conjugate state is
labeled by
$$  K_{2-\nu_{K\tt}} =\pmatrix{1&0\cr 0&1\cr} \oplus (-K),\ \ \
  \tt_{2-\nu_{K\tt}} =\pmatrix{1\cr 1\cr} \oplus (-\tt),\ \ \
  \ss_{2-\nu_{K\tt}} =\pmatrix{1/2\cr 1/2\cr} \oplus (-\ss)
\eqno(4.3.6)$$

From the two FQH states $\Psi_{K_1\tt_1\ss_1}(\{z_i\})$ 
and $\Psi_{K_2\tt_2\ss_2}(\{w_i\})$,
we can construct a new FQH state by multiplying the two wave functions 
together and setting $z_i=w_i$:
$$\Psi_{K_1\tt_1\ss_1}(\{z_i\}) \Psi_{K_2\tt_2\ss_2}(\{w_i\})|_{z_i=w_i}
=\Psi_{K_3\tt_3\ss_3}(\{z_i\})
\eqno(4.3.7)$$
To calculate $(K_3,\tt_3,\ss_3)$ it is convenient to use the $SL(n,Z)$ 
transformation 
$$ K\to WKW^T,\ \ \ \tt\to W\tt, \ \ \ \ss\to W\ss
\eqno(4.3.8)$$
to transform $\tt_1$ and $\tt_2$ 
into $\tt_{1,2}^T=(1,0,0,...)$. This is possible
if the elements in $\tt_{1,2}$ do not contain common integer factors
(\ie the largest common divisor is 1).
In the new basis $K_{1,2}$ have the form
$$ K_{1,2}=\pmatrix{ m_{1,2}& k_{1,2}^T \cr
                     k_{1,2}& \t K_{1,2}\cr}
\eqno(4.3.9)$$
where $m_{1,2}$ are odd integers, $k_{1,2}$ are integer vectors in 
${\rm dim}K_{1,2}-1$
dimensions, and $\t K_{1,2}$ are 
$({\rm dim}K_{1,2}-1)\times ({\rm dim}K_{1,2}-1)$
integer matrices with even diagonal elements (see chapter 5).
 Accordingly the spin vectors
have the form,
$$\ss_{1,2}=\pmatrix{ s^1_{1,2} \cr
                    \t \ss_{1,2}\cr}
\eqno(4.3.9a)$$
where $s^1_{1,2}$ are half odd integers and $\t \ss_{1,2}$ are integer vectors
in ${\rm dim}K_{1,2}-1$ dimensions. 

Using the $(K, \tt, \ss)$ in
the new basis, one can show that $(K_3,\tt_3,\ss_3)$ are given by
$$\eqalign{
  K_3=&\pmatrix{m_1+m_2 & k_1^T & k_2^T  \cr 
                k_1     & \t K_1& 0      \cr
                k_2     & 0     & \t K_2 \cr} \cr
  \tt_3^T=&(1,0,0,...) \cr
  \ss_3=&\pmatrix{s^1_1+s^1_2 \cr
                \t \ss_1      \cr
                \t \ss_2      \cr} \cr}
\eqno(4.3.10)$$
The results in \(4.3.10) are obtained from the following consideration.
Before imposing the constraint $z_i=w_i$, the effective theory
only describes the two independent FQH states and has the form
$$\eqalign{
\cL=&
-{1 \over 4 \pi}K_{1IJ} a^1_{I \mu} \part_\nu a^1_{J \la}~\veps^{\mu\nu\la}
+ s_{1I} \om_i \part_\nu  a^1_{I \la}\veps^{i\nu\la} \cr
&-{1 \over 4 \pi}K_{2IJ} a^2_{I \mu} \part_\nu a^2_{J \la}~\veps^{\mu\nu\la}
+ s_{2I} \om_i \part_\nu  a^2_{I \la}\veps^{i\nu\la}  \cr}
\eqno(4.3.11)$$
After imposing the constraint $z_i=w_i$, the charge currents in the two
FQH states $j^\mu_{1,2}={1\o 2\pi} \veps^{\mu\nu\la} \part_\nu a^{1,2}_{1\la}$
are always equal to each other $j^\mu_1=j^\mu_2$ and the relative
fluctuations $j^\mu_1-j^\mu_2$ are not allowed. Thus
the constraint  $z_i=w_i$ can be realized in the effective theory
by setting $a^{1}_{1\mu}=a^{2}_{1\mu}$. This changes the effective theory
\(4.3.11) into one with $(K,\ss)$ given by \(4.3.10). The charge vector
can be obtained by realizing that $A_\mu$ only couples with the charge current
$j^\mu_1=j^\mu_2$.

We would like to remark that  \(4.3.10) is not always valid.
Let $\chi_n$ be the wave function of the
$\nu=n$ integer QH state with first $n$ Landau levels filled.
Then the FQH state described by
$\chi_1(\{z_i\})\left(\chi_2(\{z_i\})\right)^2$
is in fact a $\nu=1/2$ non-abelian state rather than an abelian
state as implied by  \(4.3.10). 
More detailed discussions can be found in \Ref{hnab}.

However  \(4.3.10) does apply to the following wave function,
$(\chi_1)^p \Psi_{K\tt\ss}$. Notice that $(\chi_1)^p$ is just the wave function
of the $1/p$ Laughlin state. From \(4.3.10) one can show
that multiplying $(\chi_1)^p$ to a FQH
wave function $\Psi_{K\tt\ss}$ changes $(K,\tt,\ss)$ to
$$ K\to K+p \tt \tt^T, \ \ \ \tt\to \tt,\ \ \ \ss\to \ss+{p\o 2}\tt
\eqno(4.3.12)$$

A sequence of FQH states, $\nu=1/3, 2/5, 3/7,...,{n\o 2 n+1},..$, 
under Jain's construction, are described by the wave functions 
$\chi_1^2 \chi_n$. The $(K,\tt,\ss)$ for the IQH state $\chi_n$ are given by
$$K=1_{n\times n},\ \ \ \tt^T=(1,...,1),\ \ \ \ss^T=(1/2,3/2,5/2,...,(2n-1)/2)
\eqno(4.3.13)$$
Thus the $(K,\tt,\ss)$ for the $\nu={n\o 2 n+1}$ state are
$$K=1_{n\times n}+2C,\ \ \ \tt^T=(1,...,1),\ \ \ 
\ss^T=(3/2,5/2,7/2,...,(2n+1)/2)
\eqno(4.3.14)$$
where $C$ is a matrix with all its elements equal to 1.
The hierarchical construction can also generate a sequence
of FQH states with filling fraction $\nu={n\o 2 n+1}$. 
The $K$-matrix obtained from the hierarchical construction
is a $n\times n$
tri-diagonal matrix with off diagonal elements equal to $-1$
and diagonal elements equal to $2$, except $K_{11}=3$.
The charge and the spin vectors are given by
$\tt^T=(1,0,...)$ and $\ss^T=(3/2,1,1,...)$. Although the two sets of
$(K,\tt,\ss)$ obtained from Jain's construction and the hierarchical 
construction
are very different, one can show that they are equivalent under a $SL(n,Z)$
transformation. In fact the matrix
$W_{IJ}=\de_{IJ} -\de_{I-1,J}$ transforms the $(K,\tt,\ss)$ in \(4.3.14)
into the ones obtained from the hierarchical construction
through \(4.3.8). Therefore the $\nu={n\o 2 n+1}$ states obtained
from the Jain's construction and the hierarchical construction
belong to the same universality class.

Another interesting example is that the following three sets of
$(K,\tt,\ss)$s describe the same $\nu=2/3$ FQH state:
$$\eqalign{
K=&\pmatrix{1&2\cr 2&1\cr}, \ \ \ \tt^T=(1,1),\ \ \ \ss^T=(1/2,-1/2)\cr
K=&\pmatrix{1&1\cr 1&-2\cr}, \ \ \ \tt^T=(1,0),\ \ \ \ss^T=(1/2,-1)\cr
K=&\pmatrix{1&0\cr 0&-3\cr}, \ \ \ \tt^T=(1,-1),\ \ \ \ss^T=(1/2,-3/2)\cr}
\eqno(4.3.15)$$
In fact $W=\pmatrix{1&0\cr -1&1\cr}$  and $W=\pmatrix{1&0\cr -2& 1\cr}$ 
map the first set of $(K,\tt,\ss)$ into the second and the third sets, 
respectively. 
The first set of $(K,\tt,\ss)$ is obtained from the Jain's construction
with a wave function $\chi_2^* (\chi_1)^2$,
the second set from the hierarchical construction, and the third set
is obtained by simply putting a $\nu=1$ QH state of electrons and a $\nu=-1/3$
QH state of holes together.

The above three sets of $(K,\tt,\ss)$s describe the same spin-polarized
single layer 2/3 state. However a double layer or a spin singlet 2/3 state
is described by
$$
K=\pmatrix{1&2\cr 2&1\cr}, \ \ \ \tt^T=(1,1),\ \ \ \ss^T=(1/2,1/2)
\eqno(4.3.15a)$$
which is inequevalent to the $(K,\tt,\ss)$ of the spin-polarized
single layer 2/3 state. The two kinds of the 2/3 states are separated
by a first order transition (or other FQH states) in clean systems.

\head{5. Classification of Abelian Hall States}
\taghead{5.}
In chapter 2 and 4, we have derived the effective theories and the form of 
$(K,\tt,\ss)$ for some FQH liquids from some specific construction schemes.
In this chapter we will derive the form of the $(K,\tt,\ss)$ from some
general principles that an electron system must satisfy. This allows us
to obtain a classification of most general abelian FQH states.

For a class of generalized hierarchical states, the
effective theory is given by \(4.2.6) with $(K,\tt,\ss)$ satisfying
$$\eqalign{K^h_{11}&={\rm odd\ integer},\cr
K^h_{II}\Bigm|_{I>1}&={\rm even\ integer},\cr
K^h_{IJ}&={\rm integer,\ \ for}\ I\neq J,\cr
 t^h_I&= \delta_{1I}\cr
 s^h_I&= \pm {K^h_{II}\o 2} \cr
}.\eqno(6.1)$$
We will call
this basis the hierarchical basis, as indicated by the superscript $h$. 
We also constructed the effective theory for the simple multi-layer
FQH states with $(K,\tt,\ss)$ having the form
$$\eqalign{
K^s_{II}&={\rm odd\ integer},\cr
K^s_{IJ}&={\rm integer,\ \ for}\ I\neq J,\cr
 t^s_I&= 1\cr
 s^s_I&=\pm  {K^s_{II}\o 2} \cr
}.\eqno(6.1a)$$
Such a basis will be called the symmetric basis, which is  indicated by the 
superscript $s$.

First we would like to mention that using the transformation \(4.3.8)
and the matrix $W_{IJ}=\delta_{IJ}- \delta_{I-1,J}$
in the group $SL({\rm dim}(K),Z)$, we can transform $(K^s,\tt^s)$ in the
symmetric basis satisfying \(6.1a)
into a one in the hierarchical basis satisfying \(6.1). $W^{-1}$ is also
an element of $SL({\rm dim}(K),Z)$ 
and will transform in the opposite direction.
Thus the matrices satisfying \(6.1) and \(6.1a) have a one-to-one
correspondence. However, in general the spin vectors in the symmetric
basis do not transform into the corresponding spin vector in the
hierarchical basis. To obtain the spin vectors in \(6.1) and \(6.1a) we have
assumed that the quasiparticles and the electrons always condense into 
Laughlin states in the ``first'' Landau level. If we allow them to form 
Laughlin states in higher Landau levels more general spin vectors
will be generated:
$$\eqalign{
 s^h_I&=\pm  {K^h_{II}\o 2} +{\rm integer} \cr
 s^s_I&=\pm  {K^s_{II}\o 2} +{\rm integer} \cr}
\eqno(6.1c)$$
In this case $(K,\tt,\ss)$ in the two basis will have a one-to-one
correspondence under the transformation $W$.

Even though the  $(K,\tt,\ss)$
given above are quite general, it is not clear whether these
$(K,\tt,\ss)$ cover all possible abelian FQH states or not.
In the following we will derive the effective theory of the most general 
abelian FQH states from some general principles. 
This discussion will help us understand why the $(K,\tt,\ss)$ have the
form described in \(6.1) and \(6.1a) and what the physical reasons are 
behind it. We will show that the generalized hierarchical
states and the multi-layer states represent the most general abelian FQH states
(if we generalize the spin vectors to the ones in \(6.1c)).

Our first working assumption is that the effective theory of abelian FQH
states is described by a Lagrangian of the form in
\(4.2.6). Certainly this assumption alone is not enough. For
arbitrary choices of 
$(K,\tt,\ss)$ the Lagrangian in \(4.2.6) may not describe an
electron system. So the problem we are facing is not how to derive the
effective theory of an electron system, but the reverse, how to determine
whether an effective theory is consistent with the underlying electron system or
not. This leads us to our second working assumption: In order for the
effective theory to describe an electron system, the effective theory must
contain $\ka$ independent electron operators, where $\ka$ denotes the rank of
$K$. (Here we ignore the possibility of electron pairing.)
We will discuss why we need $\ka$ electron operators later.

To implement our second assumption, 
it is convenient to work in the electron basis: 
$$\cL=-{1\over4\pi}\widetilde a_{I\mu} \widetilde K_{IJ} 
\veps^{\mu\nu\la} \partial_\nu \widetilde  a_{J\la} 
-{e\o 2\pi}\t t_I A_\mu  \veps^{\mu\nu\la}\part_\nu \t a_{I\la}
+s_I \om_i \t  \veps^{i\nu\la}\part_\nu \t a_{I\la}
+\t a_{I\mu}\t j_{eI}^\mu
.\eqno(6.2)$$
We have put the twiddle sign ($\sim$) to remind ourselves that we are in the
electron basis. The electron basis is defined by the requirement that
a source term that carries a unit $\t a_{I\mu}$ charge generates
an electron. Thus
$\t j_I$ in \(6.2) represents the current of the $I^{\rm
th}$ electron excitations above the ground state. 

In order for the electrons to have Fermi statistics, we must have
$$(\widetilde K^{-1})_{IJ}\equiv K_{IJ}=\cases{{\rm odd\ integer} &for $I=J$\cr
{\rm integer} &for $I\neq J$\cr}.\eqno(6.3)$$
Furthermore, in order for the electrons to carry unit charge, $\tilde \tt$ must
satisfy
$$\sum_J \left(\widetilde K^{-1}\right)_{IJ} \tilde t_J=1\quad{\rm for\,\,
all}\,\,I.\eqno(6.4)$$
This equation implies that
$$\tilde t_J = \sum_L \widetilde K_{JL}.\eqno(6.5)$$

To find the condition imposed on the spin vector $\t s_I$, let us put the
FQH state on a sphere. We know that the ground state on the sphere
form the singlet representation of the $su(2)$ rotation of the sphere.
Any neutral excitations above the ground state must carry integer
$\S$ spins.  As we add an electron to the system,
the electron excitation must form an $S^\S={N_\phi\o 2}+$integer representation
of the $su(2)$. From the relation $S^\S={N_\phi\o 2}+S_{ob}$ we find that
the orbital spin $S_{ob}$ of an electron excitation must be an integer.
Therefore (see \(4.2.16) )
$$S_{ob}^I= \sum_J (\t K^{-1})_{IJ} \t s_J + {1\o 2} (\t K^{-1})_{II}
={\rm integer}
\eqno(6.4a)$$

Given the effective theory in the electron basis, we now need to determine the
allowed quasiparticle excitations. A generic quasiparticle is
created by a source term
$$\cL= c_I a_{I\mu} j^\mu\eqno(6.6)$$
whose $a_I$-charge is equal to $c_I$.
Since the electron wave function must be single valued (even in the presence of
the quasiparticle), the phase induced by moving an electron around the
quasiparticle must be a multiple of $2\pi$. 
This requires $c_I$ to satisfy
$$\sum_{J=1}^\ka\left(\t K^{-1}\right)_{IJ} c_J
 =\ {\rm integer\ for\ all}\ \, I.\eqno(6.7)$$
(Note moving a quasiparticle described by $c_I$ around the one described by
$c'_I$ induces a phase $2\pi c_I (\t K^{-1})_{IJ} c'_J$.)
The set of all $c_J$s satisfying \(6.7) form a $\ka$-dimensional lattice, with
the basis vectors of the lattice ${\bf c}^{\ (L)}$, $L=1,2,...,\ka$.
The components of ${\bf c}^{\ (L)}$ are given by
$$c^{(L)}_J=\widetilde K_{JL}.\eqno(6.8)$$
The $a_{I\mu}$-charge ${\bf c}$ of
any allowed excitations, including the electron excitations, 
will be linear combinations of ${\bf c}^{(L)}$
with integer coefficients:
$${\bf c}=\sum_L l_L {\bf c}^{(L)}
\eqno(6.8a)$$
Such a quasiparticle (labeled by $\psi_\ll$) is created
by the following source term,
$$ \cL=j^\mu \t a_{I\mu} (\t K^{-1})_{IJ} l_J
\eqno(6.8b)$$

Now let us redefine the gauge fields and 
change to the so-called quasiparticle basis, in which
the fundamental quasiparticles described by ${\bf c}^{(L)}$ will carry
the unit charge of the new gauge fields. From \(6.8) we see that the
substitution $ a_I=\sum_J\widetilde K_{IJ}\tilde a_J$ in \(6.2) and \(6.8b)
will lead to
$$\cL=-{1\over4\pi} a_{I\mu} K_{IJ} \veps^{\mu\nu\la}
\partial_\nu a_{J\la} 
-{e\over2\pi} t_I A_\mu \veps^{\mu\nu\la} \part_\nu a_{I\la}
+s_I \om_i \veps^{i\nu\la} \part_\nu a_{I\la}
+ l_I a_{I\mu} j^\mu
\eqno(6.10)$$
The $(K,\tt,\ss)$ in \(6.10) are related to $(\t K,\t \tt,\t \ss)$ in \(6.2)
through
$$K=\t K^{-1}, \ \ \ \ \tt= K \t \tt, \ \ \ \ \ss= K \t \ss.
\eqno(6.10a)$$
\(6.10) is exactly the same as the effective
theory \(4.2.6) that we wrote down in chapter 4,
with $(K,\tt,\ss)$ satisfying \(6.1a) and \(6.1c)
as one can see from from \(6.4) and \(6.4a)

We have reached an important result: the most general
abelian FQH states of unpaired electrons are described by the effective theory
\(6.10) with $(K,\tt,\ss)$ satisfying \(6.1a). In other words, 
the topological orders in the
abelian FQH states are labeled by integer valued symmetric matrices with odd
diagonal elements and a spin vector $\ss$, 
up to an equivalency condition $K\sim WKW^T$ and $\ss \sim W\ss$
with $W$ an
element of $SL(\ka,Z)$ which leaves the vectors ${\bf t}=(1,1,\dots1)$
invariant. The determinant of $W$ must be equal to 1 in order to preserve the
charge quantization condition.

The above discussions also reveal that the condition $K^s_{II}=$odd
is a consequence of the Fermi statistics of the underlying electrons.
FQH liquids formed by bosons will have $K^s_{II}=$ even in the symmetric basis.

We would next like to say a few words about why  we need $\ka={\rm dim}(K)$ 
independent
electron operators. If we had less than $\ka$ electron operators, the condition
\(6.7) would become $\sum^\ka_{J=1}\left(\t K^{-1}\right)_{IJ}c_J=$
integer only for $I=1,\dots,\ka^\prime$, where $\ka^\prime<\ka$.
 This is not enough to
fix $c_J$ on a lattice. In this case, excitations with
arbitrarily small charge would be allowed. Such excitations may be continuously
connected to the ground state and would be, we believe, gapless. This argument
suggests that in order for the effective theory to have finite  energy gap, we
require the presence of $\ka$ different electron operators. (Of course, we also
require ${\rm det}\ K\neq0$ in order for the gauge fluctuations to have finite
gaps.)

We would like to remark that here we have only shown that a generic
abelian state must be described by $(K,\tt,\ss)$ that satisfies \(6.1)
or \(6.1a). However $(K,\tt,\ss)$ here merely appear as some parameters
in the effective theory. Only certain combinations are 
physically measurable. Thus different $(K,\tt,\ss)$ may describe the same 
topological orders. At least the $(K,\tt,\ss)$s related through the $SL(n,Z)$
transformation \(4.3.8) describe the same topological order. However,
it may be possible that $(K,\tt,\ss)$s with even different dimensions
might describe the same topological order. More details can be found
in \Ref{Hedge}. An attempt to describe topological orders using only measurable
quantities can be found in \Ref{Wtop}. Clearly more work needs to be
done to obtain a more complete and more satisfactory 
picture of abelian topological orders.

\head{6. Algebraic approach and non-abelian FQH liquids}
\taghead{6.}
So far in this article we have discussed only a small class of FQH liquids
-- abelian FQH liquids. It was first pointed out by Moore and Read\refto{MR}
that
another type of FQH liquids with quasiparticles of non-abelian statistics
is also possible. Non-abelian FQH liquids cannot be described by
abelian Chern-Simons effective theory; thus the approaches previously
used to study the edge states and quasiparticle quantum numbers can no longer
be used for non-abelian states. We need to develop new methods to
calculate the physical properties of non-abelian states.

At the moment, we know two ways to construct non-abelian FQH states.
The first method uses the parton construction and the second method
uses the conformal field theory (CFT). The parton construction allows
us to construct FQH states whose effective theory is a non-abelian
Chern-Simons theory \refto{NABW,NABBW,edgere}.
For example $\chi_1(\chi_2)^2$ is a $\nu=1/2$
non-abelian state described by a $U(1)\times SU(2)$ Chern-Simons theory.
The CFT approach allows us to construct more general non-abelian 
states \refto{MR,NABBW,WW,WWH}.
The effective theories of many non-abelian states obtained from CFT are
not known at this time.

In this chapter we will concentrate on the edge properties of non-abelian
FQH states. We will develop an algebraic method which allows us to
calculate the properties of the edge states for the FQH states constructed
using CFT. Our approach is motivated by an observation by Moore and 
Read\refto{MR} that
the Laughlin wave function can be written as a correlation function
between the charge operators (\ie the vertex operators)
in the $U(1)$ K-M algebra. The $U(1)$ K-M algebra
is a CFT which is also called the Gaussian model. 
(For readers not familiar with CFT, we suggest \Ref{CFT}.) Moore and Read's
observation links the Laughlin wave function to a CFT. We will call this CFT
the bulk CFT. From our discussion in chapter 3 we see that the
edge excitations of the Laughlin state are also described by a CFT --
the $U(1)$ K-M algebra. Such a CFT will be called the edge CFT of the
FQH liquid. It is very interesting to see that for a Laughlin state, the
bulk CFT is identical to the edge CFT.
It turns out that such a relation can be extended to more general abelian
and non-abelian states.
In this chapter we will explain how the bulk CFT and the edge CFT are connected
and how to use such a relation to calculate physical properties of non-abelian 
edge states.

However, we would like to stress that to relate the bulk CFT
that generates the bulk wave function to the edge
CFT that describes the edge
spectrum, we need to introduce a concept of
minimal CFT for the bulk CFT. This is because
once a FQH wave function
can be written as a correlation function in a certain CFT,
then the wave function can also be expressed as a
correlation in many other CFTs which contain
the original CFT. The minimal (bulk) CFT for a
FQH state not only reproduces the wave function,
it is also contained by any other CFT that reproduces the
wave function. 
As we will see later, it is this minimal bulk
CFT that is identical to the edge CFT that describes the edge 
excitations of the corresponding FQH state.
Thus it is very important to identify the minimal bulk CFT.

\subhead{6.1 Ideal Hamiltonians, zero energy states, and edge excitations}
In this section we will describe a microscopic theory for the 
edge excitations in the Laughlin states.
Let us consider an electron gas in first Landau level.
We choose 
the interaction between electrons to be $V(\vec r) \prop -\part^2\de(\vec r)$. 
Because A) $H_V=\sum_{ij} V(\vec r_i-\vec r_j)$ is
positive definite (\ie $\<\psi|H_V|\psi\> \geq 0$ for any $|\psi\>$)
and B) $H_V$ have zero expectation value 
for the $\nu=1/3$ Laughlin
wave function $$\Phi_3(\{ z_i\})=\prod_{i<j} (z_i-z_j)^3 \prod_k
e^{-{1\o 4}|z_k|^2}
\eqno(64.1)$$
Thus $\Phi_3$ is an exact ground state of our Hamiltonian with zero energy. 
Such Hamiltonian $H_V$ will be call the ideal Hamiltonian of the
1/3 Laughlin state.
However the Laughlin state
\(64.1) is not the only state with zero energy. One can easily check
that the following type of states all have zero energy:\refto{HE}
$$\Phi(\{ z_i\})=P(\{ z_i\})\Phi_3(\{ z_i\})
\eqno(64.2)$$
where $P(z_i)$ is a symmetric polynomial of $z_i$ so that
$\Phi(\{ z_i\})$ remains an anti-symmetric function.  In fact, the
reverse is also true: all the zero energy states are of the form in
\(64.2). This is because, in order for a fermion state to have zero
energy, $\Phi$ must vanish at least as fast as $(z_i-z_j)^3$ when any
two electrons $i$ and $j$ are brought together (the possibility of
$(z_i-z_j)^2$ is excluded by the fermion statistics). The
Laughlin wave function is zero only when $z_i=z_j$; therefore
$P=\Phi/\Phi_3$ is a finite function. Since $\Phi$ and $\Phi_3$ are
both anti-symmetric functions in the first Landau level, $P$ is
a finite symmetric holomorphic function that can only be a symmetric
polynomial.

Among all the states in \(64.2), the Laughlin state describes a circular
droplet with the smallest radius. All other states, having higher angular
momenta, are deformation and/or
inflation of the droplet of the Laughlin state. Thus the  states
generated by $P$, being low energy excitations above the
ground state $\Phi_3$, correspond to the edge excitations of the Laughlin
state.

Now let us study the structures of the zero energy space (\ie the space of
symmetric polynomials). We know that the space of symmetric polynomials is
generated by the following polynomials $s_n=\sum_i z_i^n$ (through
multiplication and addition).
Let $M_0=3{N(N-1)\o 2}$ be the total angular momentum
of the Laughlin state \(64.1). Then the state $\Phi$ will have an
angular momentum $M=\De M+M_0$ where $\De M$ is the order of the
symmetric polynomial $P$. Since we have only one order-zero and
order-one symmetric polynomial $s_0=1$ and $s_1=\sum_i z_i$, the
zero energy states for $\De M=0, 1$ are non-degenerate. However, when
$\De M=2$ we have two zero energy states generated by $P=s_2$ and
$P=s_1^2$. For general $\De M$ the degeneracy of the zero energy
states is given by $$\matrix{\De M: & 0 & 1 & 2 & 3 & 4 & 5 & 6\cr {\rm
degeneracy:}& 1 & 1 & 2 & 3 & 5 & 7 & 11 \cr }
\eqno(64.3)$$

Here we would like to point out that the
degeneracy in \(64.3) is exactly what we expected
from the macroscopic theory. We know for a circular droplet, the
angular momentum $\De M$ can be regarded as the momentum along the
edge $k=2\pi \De M /L$ where $L$ is the perimeter of the QH droplet.
According to the macroscopic theory the (neutral) edge excitations are
generated by the density operators $\r_k$. One can easily check that
the edge states generated by the density operators have same
degeneracies as those in \(64.3) for every $\De M$, \eg the two states
at $\De M=2$ are generated by $\r_{\ka_0}^2$ and $\r_{2\ka_0}$ where
$\ka_0=2\pi /L$. Therefore the space generated by the K-M algebra
\(2.7) and the space of the symmetric polynomials are identical.

There seems to be one problem in the above identification. 
The states generated by symmetric polynomials
$P$ all have zero energy, while the edge excitations in the macroscopic theory
have finite energies: $E\prop k \prop \De M$. This is because we did
not include the confining potential in our ideal Hamiltonian. A confining 
potential for the FQH droplet will lift the degeneracy and give edge excitation
an energy $E\prop \De M$.

Now let us ask a physical question. Do the symmetric polynomials
generate all the low energy states? If this is true, the space of the 
low energy edge excitations will be identical to the space of
symmetric polynomial (or the zero energy space) \refto{HE}.
Unfortunately, up to
now we do not have analytic proof of the above statement. This is
because, although states orthogonal to the states generated by the
symmetric polynomials have non-zero energies, it is not clear that
those energies remain finite in the thermodynamical limit.  It 
might be
possible that the energy gap approaches zero in the thermodynamical
limit. To resolve this problem, for now we have to rely on numerical
calculations. In Fig. 6.1 we present the energy spectrum of a system
of six electrons in the first 22 orbits for the Hamiltonian
introduced at the beginning of this chapter. The degeneracies of the zero
energy states at $M=45,...,51$ (or $\De M=0,...,6$) are found to be
$1,1,2,3,5,7,11$, which agrees with \(64.3).
More importantly, we can clearly see a finite
energy gap that separates all other states from the zero energy states. Thus,
the numerical results imply that all the low lying edge excitations of
the Laughlin state are generated by the symmetric polynomials
or the K-M algebra \(2.7).
Using the above assumption of the energy gap,
we see that the edge excitations
in an FQH liquid can be studied through the zero energy states of
the ideal Hamiltonian for that FQH liquid.
\vskip .2 in
\epsfysize=2.5truein
\centerline{ \epsffile{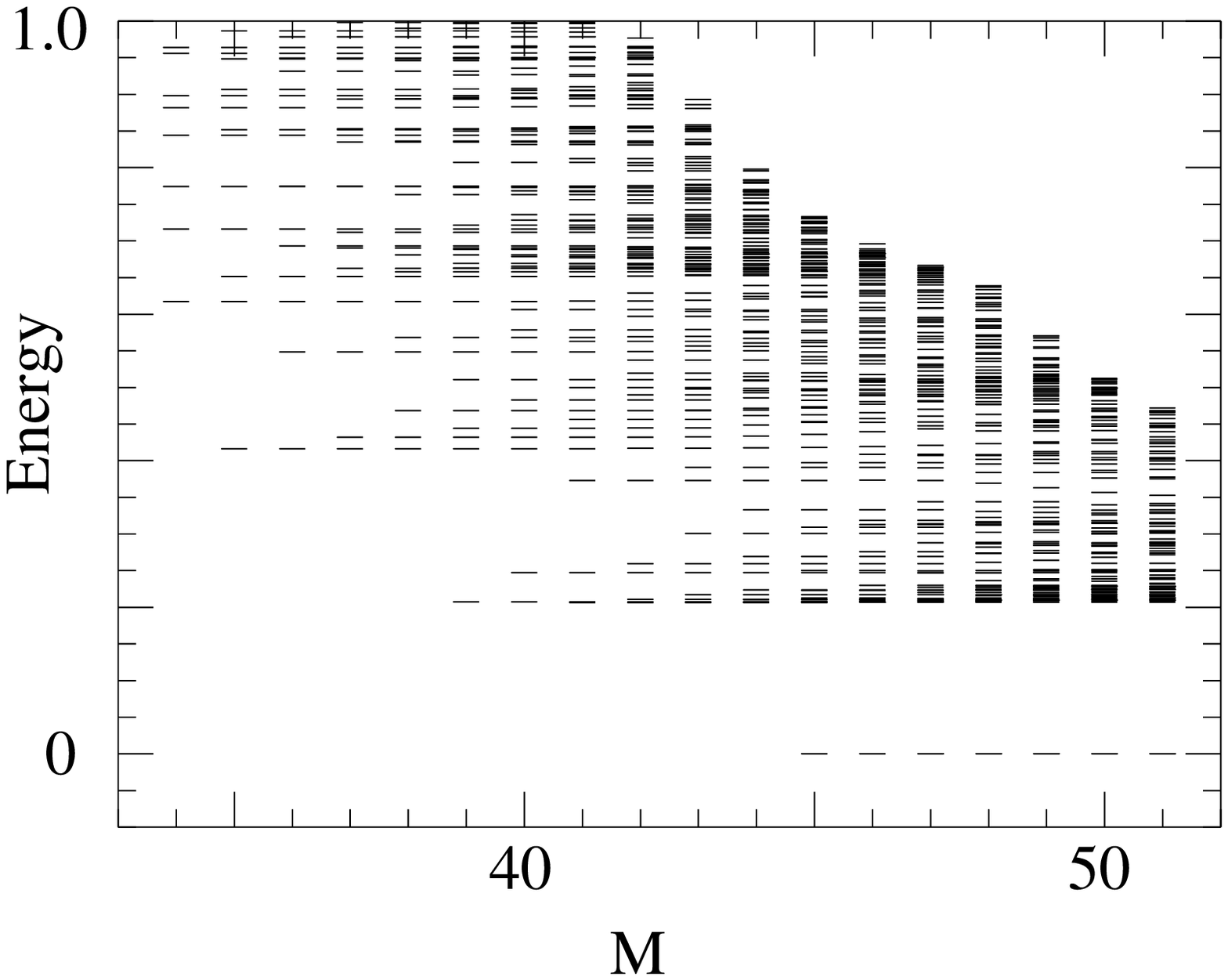} }
\caption{
Fig. 6.1: The energy spectrum of a system
of six electrons in the first 22 orbits for the Hamiltonian $H_V$.
The degeneracies of the zero energy states at $M=45,...,51$  are found to be
$1,1,2,3,5,7,11$.
}
\vskip .2 in

\subhead{6.2 Edge excitations and current algebra}
In this section we are going to study edge excitations
of the Laughlin state using a current algebra approach.
The algebraic approach developed here can be easily generalized
to study edge excitations of non-abelian states.

We note that the $\nu=1/m$ Laughlin state,
$$\prod_{i<j} (z_i-z_j)^m e^{-{1\o 4}\sum_i|z_i|^2}
\equiv \Phi_m(\{z_i\}) e^{-{1\o 4}\sum_i|z_i|^2}
\eqno(1)$$
is a zero-energy state of a Hamiltonian
with the following two-body interaction\refto{TK},
$$V(z_1-z_2)=-\sum_{n=1}^{(m-1)/2} C_n \part_{z_1^*}^{2n-1}
\de (z_1-z_2) \part_{z_1}^{2n-1}
\eqno(2)$$
where $C_n>0$. (The Haldane's pseudo-potentials\refto{sphere}
for the above
interaction are given by $V_{2l+1}>0$ for $2l+1<m$ and
 $V_{2l+1}=0$ for $2l+1\geq m$.)
Thus \(2) defines an ideal Hamiltonian for the $1/m$ Laughlin state.
Other zero-energy states of the ideal Hamiltonian form the edge excitation
of the 1/m state.
From \(2) we see that any anti-symmetric wave function of form
$$\t \Phi_m(\{z_i\}) e^{-{1\o 4}\sum_i|z_i|^2}
\eqno(3)$$
has zero energy, if and only if $\t \Phi_m$ does not contain
zeros of order less than $m$, \ie
$$\t\Phi_m(z_i)=O((z_1-z_2)^{m})
\eqno(4)$$
as $z_1\to z_2$.

Now the question is how to construct the Hilbert space of
these zero-energy states. One approach is to use the
symmetric polynomial discussed in the last section. Here
we will use a more complicated approach --
current algebra in CFT.  

Let us concentrate
on the holomorphic part of the wave function ($\Phi_m$ or
$\t\Phi_m$ in \(1) and \(3)). To construct the edge states we need
to construct holomorphic functions that satisfy \(4).
We notice that the Laughlin-Jastrow function
$\Phi_m$ can be written as a correlation of the vertex
operators (or primary fields), $\psi_e$, in the
Gaussian model as follows:\refto{MR,Fubini}
$$\Phi_m=
\lim_{z_\infty\to \infty} (z_\infty)^{2h_N}
\<\Psi_N(z_\infty) \prod_{i=1}^N \psi_e(z_i)\>~,
\eqno(5)$$
where
$$ \psi_e(z)=e^{i\sqrt{m} \phi(z)},\ \ \ \ \
 \Psi_N(z)=e^{-iN\sqrt{m} \phi(z)}~.
\eqno(6)$$
Here $N$ is the number of electrons, and
$h_N=mN^2/2$ is the conformal dimension of $\Psi_N$.
The scalar field, $\phi$, in the Gaussian model
is normalized so that $\< \phi(z)\phi(0) \>=\ln z$.
(Note the $\phi$ field here is just the $\phi$ field
we introduced in the $U(1)$ K-M algebra \(3.2) in section 3.1)
The factor $(z_\infty)^{2h_N}$ is included in \(5) so that the limit
$z_\infty\to \infty$ gives rise to a finite function \refto{com}.
Let $j(z)\equiv  \partial \phi$ be the $U(1)$ current in
the Gaussian model. Since $j$ is
a local operator, we find that the correlation
$$\t\Phi_m\propto
\< \oint dz \a(z) j(z)
\Psi_N(z_\infty)\prod_{i=1}^N \psi_e(z_i)\>~,
\eqno(7)$$
with an appropriately chosen holomorphic function $\a (z)$,
satisfies \(4) because $\psi_e(z)$ has the following operator
product expansion (OPE) as $z \to 0$:
$$\eqalign{
\psi_e(z)\psi_e(0)\propto & z^m e^{i2\sqrt{m} \phi(z)} +O(z^{m+1}) ~,\cr
j(z)\psi_e(0)\propto & {1\o z} \psi_e(0)+O(1) ~.\cr}
\eqno(8)$$
The OPE between $\psi_e$ guarantees the correlation to have m$^{th}$
orders zero as $z_i \to z_j$.
The integration contour of $z$ in \(7) encloses
all the $\psi_e(z_i)$ operators but not $\Psi_N$.
If $\a(z)$ has no poles except at $z_\infty$, then
\(7) is finite for finite $z_i$, since $z_\infty\to \infty$.
Therefore \(7) is a wave function with zero energy, hence
a wave function for
the edge excitations. Introducing
$$j(z)=\sum_{n=-\infty}^\infty {j_n\o (z-z_\infty)^{n+1} }
\eqno(9)$$
and choosing
$$\a(z)=(z-z_\infty)^{-n} \ \ \ \ \ \ (n\geq 0)~,
\eqno(10)$$
we find that, by shrinking the contour around $z_\infty$
and letting $z_\infty\to \infty$,
\(7) becomes
$$\Phi_m^{(n)}=\lim_{z_\infty\to \infty} (z_\infty)^{2h_N+2n}
\<\Psi_N^{(n)} (z_\infty) \prod_{i=1}^N \psi_e(z_i)\>~,
\eqno(11)$$
where
$$\Psi_N^{(n)} (z_\infty)= j_n \Psi_N(z_\infty)~.
\eqno(12)$$
Again the factor $z_\infty^{2h_N+2n}$ is included in \(12)
to ensure the existence of a finite limit \refto{com}.

The above discussion can be generalized to the case with
several insertions of the current
operator. The general statement is the following.
The ground state wave function can be written as a correlation
of $\psi_e(z_i)$ with a primary field $\Psi_N$ inserted
at infinity. To generate edge excitations above the ground
state, we simply replace the primary field $\Psi_N$
by its current descendants\refto{KZ},
which are generally of the form (with $n_1, n_2, \cdots, \geq 0$)
$$\Psi_N^{(n_1,n_2,...)} = (j_{-n_1}j_{-n_2}... )\Psi_N ~.
\eqno(13)$$
Let us call $l=\sum_i n_i$ the level of the descendant field
$\Psi_N^{(n_1,n_2,...)}$. The wave function
associated with it is given by
$$\Phi_m^{(n_1,n_2,..)}=\lim_{z_\infty\to \infty}
(z_\infty)^{2h_N+2l} \<\Psi_N^{(n_1,n_2,..)} (z_\infty)
\prod_{i=1}^N \psi_e(z_i)\>~.
\eqno(11a)$$
Such wave function has zero energy and can be identified
as an edge excitation.

We would like to show that all the
edge states generated by the level-$l$ descendant fields have
a total angular momentum $L=M_0+l$, where $M_0=mN(N-1)/2$ is the
total angular momentum of the ground state. We first note
that the angular momentum of the ground state
can be expressed in terms of conformal
dimensions, $h_e$ and $h_N$,
of $\psi_e$ and $\Psi_N$. Under a conformal transformation
$z\to w=f(z)$, the correlation of primary fields satisfies
$$\<\Psi_N(z_\infty) \prod_{i=1}^N \psi_e(z_i)\> =
(f'(z_\infty))^{h_N}\prod_i (f'(z_i))^{h_e}
\<\Psi_N(w_\infty) \prod_{i=1}^N \psi_e(w_i)\>~.
\eqno(13a)$$
Choosing $f(z)=\la z$, we have
$$\<\Psi_N(\la z_\infty) \prod_{i=1}^N \psi_e(\la z_i)\>
=\la^{-h_N-Nh_e} \<\Psi_N(z_\infty) \prod_{i=1}^N \psi_e(z_i)\>~,
\eqno(13b)$$
which implies that
$$\Phi_m(\la z_i)=\la^{h_N-Nh_e} \Phi_m(z_i)~.
\eqno(13c)$$
For $\la = e^{i\theta}$, this transformation is nothing
but a rotation by angle $\theta$ in the complex plane.
Thus the angular momentum of the ground state is
$$M_0=h_N-Nh_e~.
\eqno(13d)$$
With $h_N=mN^2/2$ and $h_e=m/2$, we see that $M_0=mN(N-1)/2$, as
expected for the Laughlin $1/m$-state. More generally,
since the dimension of a current descendant
$\Psi_N^{(n_1,n_2,...)}(z)$ is the sum $h_N+l$,\refto{KZ}
where $l$ is its level, we find that under $z\to w=\la z$
$$\Psi_N^{(n_1,n_2,...)}(z)=\la^{h_N+l}
\Psi_N^{(n_1,n_2,...)}(w)~;
\eqno(13e)$$
Thus, from \(11a) and \(13e) we can see
that the edge-excited state described by $\Phi_m^{(n_1,n_2,..)}$
carries an angular momentum of $L=h_N+l-Nh_e=M_0+l$.
As a general rule, valid for any FQH states that can be
generated by CFT, the angular momentum of an edge excitation
is equal to that of the ground state plus the level of the
descendant level of the associated insertion at infinity.

Note that the descendant fields $\Phi_m^{(n_1,n_2,..)}$
with different $(n_1,n_2,..)$ may not be linearly independent.
Using the standard approach in CFT, 
one can show that, from the OPE $j(z) j(0) =1/z^2$,
$j_n$ satisfies the $U(1)$ K-M algebra
$$[j_n,j_m]=2\pi n\de_{m+n}~.
\eqno(12a)$$
{}From the $U(1)$ K-M algebra \(12a), it is easy to
show that the number, $D_l$, of linearly independent
descendant fields, \(13), at level $l$ is given by the
partition number of $l$. The values of $D_l$ for small $l$
are given in \(64.3). 
The values of $D_l$ for generic $l$ can be
expressed in terms of the character 
of the $U(1)$ K-M algebra for the primary field $\Psi_N$ \refto{ch}:
$$\ch_N(\xi)\equiv \xi^{h_N} \sum_l D_l \xi^l
= \xi^{h_N} {1\over \prod_{n>0} (1-\xi^n)}\,.
\eqno(add1)
$$
If there is a one-to-one correspondence between the
edge states and the descendant fields of $\Psi_N$,
then we can use the above formula to obtain the number,
$D_L$, of edge excitations at any given angular
momentum $L$:
\def\Ch{{\rm Ch}}
$$\eqalign{
\Ch_N(\xi) &\equiv \sum_L D_L \xi^L
= \ch_N (\xi) \xi^{-Nh_e} \cr
&= {1\over \prod_{n>0} (1-\xi^n)}\, \xi^{M_0}\,.}
\eqno(add2)
$$

The one-to-one correspondence between the edge states
and the descendant fields of $\Psi_N$ means that
linearly independent descendant fields always generate linearly
independent wave functions of $z_i$ through \(11a).
However, we do not expect
this correspondence to be true for finite $N$ and
arbitrarily large level $l=\sum_k n_k$. On the other hand,
it is conceivable that this one-to-one correspondence
holds for any finite level $l$ when $N$ is very large or
in the limit $N\to \infty$. Though we do not know,
at present, how to prove
this statement within the CFT approach for a
generic abelian FQH state, its validity for the
$1/m$-states can be seen in the following way:
It is known\refto{HE,S} that the edge
states can be generated by multiplying
the ground state wave function by symmetric polynomials
of electron coordinates $z_i$. Mathematically,
the number of linearly independent symmetric
polynomials of degree $l$ is precisely given
by the partition number of the integer $l$,
which is just the number of linearly independent descendant fields.

In the above
we have generated the edge excitations of the 
Laughlin state by inserting the current operator. One may try to
generate more edge states by inserting  some other
operators, such as the energy-momentum tensor $T$,
because the insertions of the energy-momentum tensor also maintain
the structure of zeros \(4) of the wave functions. 
But one can show that the edge states 
generated by the energy-momentum tensor is contained
in those generated by the current since $T\propto j^2$.
For the Laughlin state, $j$ is the ``fundamental operator"
that generates all the edge excitations.

From the discussions in this section, 
we see a close relationship between the
(minimum) bulk CFT that generates the bulk wave function and the edge CFT
that generates the spectrum of the edge excitations, at least for
the $1/m$ Laughlin states.
In next section, we will show that this relationship
can be generalized to non-abelian FQH systems.

\subhead{6.3 Non-abelian FQH liquids and their edge excitations}
In this section we are going to study the physical properties of some
simple non-abelian FQH liquids,\refto{MR,hnab,WW,WWH} 
using the algebraic approach.  In general, the holomorphic part of 
the many-body wave function for a non-abelian FQH state
consists of two factors. One of them 
is of usual Laughlin-Jastrow form,
$\prod_{i<j} (z_i-z_j)^r$ ($r$ is a fraction or 
integer), which we call the $U(1)$ part. The connection of 
this part to the $U(1)$ K-M algebra (or the Gaussian model)
is well understood. 
Here we concentrate on the other part, which is 
not of the Laughlin-Jastrow form. We call it
the non-abelian part, because it is this part that
is presumably responsible to the
appearance of non-abelian statistics for quasiparticles.

Let us first consider a simple non-abelian FQH state,
namely the p-wave paired FQH state for spinless 
electrons discussed in \ref{MR}. The total wave 
function is given by 
the product of a Pfaffian wave function 
$\Phi_{Pf}$ and a Laughlin wave function $\Phi_m$
$$\eqalign{
\Phi_p=&\Phi_{Pf}\Phi_m \cr
\Phi_{Pf}=&\cA({1\o (z_1-z_2)}{1\o (z_3-z_4)}...) \cr
\Phi_m=& \left( \prod_{i<j} (z_i-z_j)^m \right) 
 e^{-{1\o 4} \sum_i |z_i|^2} \cr}
\eqno(7a)$$
where $\cA$ is the anti-symmetrization operator,
and $m$ is an even integer. The filling fraction of the  p-wave paired
state is $\nu=1/m$.

Let us first analyze the structure of zeros of this wave
function assuming, for simplicity, $m=2$. Let
$z_1=z_3+\de_1$ and $z_2=z_3+\de_2$, we find $\Phi_p$ has the
following expansion (we expand $\de_2$ first)
$$\eqalign{
\Phi_p
=& \sum_{k=odd}(\de_2)^k A_{k}(z_1,z_3,...) \cr
=& \sum_{k=odd}(\de_2)^k\sum_l (\de_1)^l A_{kl}(z_3,z_4,...) \cr}
\eqno(7b)$$
One can directly check that the coefficients 
$$A_{12}=0
\eqno(7c)$$
Therefore $\Phi$ is the exact ground state of Hamiltonian $H$
with the following three-body potential:
$$\eqalign{
V=&-V_1 \part_{z_2^*}    \de(z_2-z_3) \part_{z_2} 
        \part_{z_1^*}^2  \de(z_1-z_3) \part_{z_1}^2\cr}
\eqno(7d)$$
This is because the Hamiltonian $H$ is positive definite 
if $V_1>0$, and $\Phi$ is a zero-energy state of $H$.
It was checked 
numerically that $\Phi$ is the non-degenerate
ground state of $H$ (on a sphere) with a finite energy gap. 
This implies that the constraint \(7c)
uniquely fixes the ground state wave function \refto{hnab,RR}.

On a plane, the ground state of $H$ is identified as the minimum angular
momentum state with  zero energy. Numerical results indicate that
the ground state on the plane is non-degenerate \refto{hnab,RR}.
Other zero-energy states all
have higher angular momenta and are identified as 
the edge excitations above the ground state.

It was shown in \Ref{MR} that the so-called
non-abelian part, $\Phi_{Pf}$, can be written as a
correlation of the dimension-$1/2$ primary fields $\psi$ in the Ising
model. $\psi$ is just the fermion field in a free Majorana fermion theory. 
Therefore the total ground state wave 
function (the holomorphic part) can be written as
$$\Phi_{p}=\lim_{z_\infty\to \infty} (z_\infty)^{2h_N}
\< \Psi_N(z_\infty)
\prod_{i=1}^{N} \psi_{e}(z_i)\>~,
\eqno(14)$$
where
$$\psi_{e}(z)=\psi(z) e^{i\sqrt{m} \phi(z)}~,
\eqno(h5)$$
$$\Psi_N(z_\infty)=e^{-iN\sqrt{m} \phi(z_\infty)}~.
\eqno(15)$$
Here $\phi(z)$ is the chiral scalar field in the Gaussian model
whose correlation produces the $U(1)$ part $\Phi_m$. In \(14) 
$h_N=mN^2/2$ is the dimension of $\Psi_N$ and $N$ the total number
of the electrons.

Let $z_1=z_3+\de_1$ and $z_2=z_3+\de_2$ and assume $m=2$, 
we find the OPE of the operators $\psi_e$ has the following form
$$\eqalign{
   & \psi_e(z_1) \psi_e(z_2) \psi_e(z_3) \cr
=  & \psi_e(z_1)  \de_2 e^{2i\sqrt{m} \phi(z_3)} +...\cr
=  & \de_2 (\de_1)^4 \psi(z_3) e^{3i\sqrt{m} \phi(z_3)} +...\cr}
\eqno(b7)$$
We find that the lower-order zeros (such as the term
containing $\de_2^1 \de_1^2$) are absent as described by \(7c).
Thus the correlations between $\psi_e$ automatically generate wave functions
that satisfy \(7c) and become the zero energy states of the 
three-body Hamiltonian $H$.

From the above we see that the wave function of a p-wave paired FQH state
can be written as
a correlation in the $U(1)\times$Ising model. Since the bulk CFT of
the p-wave paired state is not the Gaussian model, this motivated Moore and
Read to suggest that the p-wave state is a non-abelian state.
Indeed it was shown that the quasiparticles in the p-wave state have
very different structures from those in abelian states \refto{MR,RR}.

To have some basic understanding of the new structures of the
quasiparticles in the non-abelian state, let us start with the
charge $e/2$ quasihole created by inserting a unit flux quantum:
$$\prod (\xi-z_i) \Phi_p(\{z_i\})
\eqno(b7a)$$
Due to the presence of the pairing wave function 
$\Phi_{Pf}=\cA \prod_{i=odd} {1\o z_i-z_{i+1}}$ we can split the
quasihole in \(b7a) into two excitations each carring a charge of
$e/4$:
$$\Phi(\{z_i\}; \xi_1,\xi_2)
=\cA \prod_{i=odd} {(z_i-\xi_1)(z_{i+1}-\xi_2)\o z_i-z_{i+1}}
\Phi_m(\{z_i\})
\eqno(b7b)$$
(Note $\Phi(\{z_i\}; \xi,\xi)$ is just the wave function in \(b7a).)
An interesting phenomenon happens when we create four $e/4$ quasiholes.
One can write down three different wave functions for
the same four quasiholes at the same positions:
$$\eqalign{
\Phi_1(\{z_i\}; \xi_1,\xi_2,\xi_3,\xi_4) =&\cA \prod_{i=odd} 
{(z_i-\xi_1)(z_{i+1}-\xi_2)(z_i-\xi_3)(z_{i+1}-\xi_4)\o z_i-z_{i+1}}
\Phi_m(\{z_i\})\cr
\Phi_2(\{z_i\}; \xi_1,\xi_2,\xi_3,\xi_4) =&\cA \prod_{i=odd} 
{(z_i-\xi_1)(z_{i+1}-\xi_3)(z_i-\xi_2)(z_{i+1}-\xi_4)\o z_i-z_{i+1}}
\Phi_m(\{z_i\})\cr
\Phi_3(\{z_i\}; \xi_1,\xi_2,\xi_3,\xi_4) =&\cA \prod_{i=odd} 
{(z_i-\xi_1)(z_{i+1}-\xi_2)(z_i-\xi_4)(z_{i+1}-\xi_3)\o z_i-z_{i+1}}
\Phi_m(\{z_i\})\cr }
\eqno(b7c)$$
However only two of them are linearly independent \refto{MR}.
It is a unique property of non-abelian states that
the four-quasihole states have a two-fold degeneracy
even when all the positions of the quasiholes are fixed. As we
exchange different pairs of quasiholes, non-abelian Berry's phases may be
generated. 

However, although we have found the existence
of the degeneracy, at the moment, 
we are still unable to calculate the non-abelian Berry's phase
due to the lack of the plasma analog for the pairing wave function $\Phi_p$.
But the numerical results in \ref{RR} do provide evidences for the 
existence of the non-abelian statistics in the p-wave state. 

In the following we will study the edge excitations of the p-wave state.
We find that the edge structure of the p-wave state is also
very different from that of abelian states. This provides additional
evidences for the existence of 
the new topological orders in the p-wave state.

From \(14) we see that the holomorphic part of the
ground state wave function can be written as a
correlation of $\psi_e$ with the operator $\Phi_N$ inserted at infinity.
Following the discussion in the last section, we can create edge excitations
by inserting other operators.
Here we can use both $\psi$- and $j$-insertions 
Repeating the derivation in the 
last section, we find that edge excitations [\ie the zero energy states of
the three body-Hamiltonian \(7d)] are generated 
by the descendant fields of $\Psi_N$:
$$\Phi_{p}^{edge}=\lim_{z_\infty\to \infty} 
(z_\infty)^{2h_N+2l} \< \Psi_N^{(n_1,..;m_1,..)}(z_\infty)
\prod_{i=1}^{N} \psi_{e}(z_i)\>~,
\eqno(16)$$
where
$$l=\sum_i n_i~+~\sum_k m_k
\eqno(16a)$$ 
is the level of the descendant field 
$$\Psi_N^{(n_1,..;m_1,..)}(z_\infty)
=(\psi_{-n_1}...j_{-m_1}...)\Psi_N(z_\infty)
\eqno(17)$$
and the $\psi_n$ are operators in the expansion of $\psi(z)$
$$
\psi(z)=\sum_n {\psi_n\o z^{n+{1\o 2}}}, \ \ \ \ n={\rm integer}+{1\o 2}
\eqno(17a)$$
Note that in order for the correlation to be non-zero,
$\Phi_N^{(n_1,..;m_1,..)}$ must contain even or odd 
numbers of $\psi_n$ operators if there are even or odd
number of electrons (\ie $N$=even or odd).

Not all descendant fields $\Phi_N^{(n_1,..;m_1,..)}$ with different indices are 
linearly independent. The linearly independent descendant fields
generated by $j_n$
can be calculated from the K-M algebra \(12a) as we did in the last section. 
The linearly independent
descendant fields generated 
by $\psi_n$ can be calculated from the fermion algebra satisfied by
$\psi_n$ fields
$$\{ \psi_n, \psi_m\}= 2 \pi \de_{n+m}
\eqno(17b)$$
The algebra \(17b) can be derived form the OPE ,
$$\psi(z_1)\psi(z_2)=-\psi(z_2)\psi(z_1)={1\o z_1-z_2}
\eqno(17ba)$$
Thus, the space of the descendant fields is isomorphic to
the direct product of the space of the K-M 
algebra generated by $j$ and the space of the Majorana fermion theory
generated by $\psi$. 
This is because $\psi$ and $j$ commute with each other. 
Similar to the pure $U(1)$ case in the last section,
the edge excitations generated by level-$l$ descendant fields 
can be shown to have angular momentum $L=M_0+l$, where
$M_0=h_N-Nh_e={m\o 2}N(N-1)-[{N\o 2}]$ is the angular
momentum of the ground state $\Phi_{p}$. Here we 
have used \(13d), and $h_e={m+1\o 2}$, 
$h_N={m\o 2}N^2$ (see \ref{WW}).

The number of the linear independent descendant fields at each level
can be easily calculated by counting the number of states in a free boson
theory (generated by $j$) and a free fermion theory (generated by $\psi$).
The results can expressed through characters.
The sectors with even and odd
$\psi_n$s should be considered separately.
Let us first concentrate on the space generated by $\psi_n$.
Notice that the character for the two states generated by $\psi_n$ is
$1+\xi^n$. The total character for both the even and the odd sectors is given by
$$\prod_{n =1/2}^{\infty} (1+\xi^n)=\ch_{even}(\xi) +\ch_{odd}(\xi)
\eqno(24a)$$
To obtain $\ch_{even,odd}$ separately we need to keep track of the number
of the fermion operators. This can be achieved by introducing a generalized
character $1+\xi^n \eta$ which leads to 
$$\ch(\xi,\eta)=\prod_{n =1/2}^{\infty} (1+\xi^n\eta)
\eqno(24ab)$$
It is easy to see that the power of $\eta$ counts the numbers of fermions.
Thus the characters for the even and the odd sectors are
$$\eqalign{
\ch_{even}=& {1\o 2} (\ch(\xi,1) +\ch(\xi,-1))
={1\o 2} \left( \prod_{n =1/2}^{\infty} (1+\xi^n) +
\prod_{n =1/2}^{\infty} (1-\xi^n) \right) \cr
\ch_{odd}=&{1\o 2} (\ch(\xi,1) -\ch(\xi,-1))
= {1\o 2} \left( \prod_{n =1/2}^{\infty} (1+\xi^n) -
\prod_{n =1/2}^{\infty} (1-\xi^n) \right) \cr}
\eqno(25)$$
The characters for edge excitations of the
p-wave paired state, after including the
Gaussian (or abelian) part generated by $j$,  can be written as 
products of the $U(1)$ character \(add1) and the non-abelian character
\(25), and are given by
$$\eqalign{
\Ch_{even}=& {1\o 2\prod_{n>0} (1-\xi^n)^2} 
\left( \prod_{n =1/2}^{\infty} (1+\xi^n) +
\prod_{n =1/2}^{\infty} (1-\xi^n) \right) 
\xi^{M_0^{(e)}} \cr
\Ch_{odd}=& {1\o 2\xi^{1/2} \prod_{n>0} (1-\xi^n)^2}
\left( \prod_{n =1/2}^{\infty} (1+\xi^n) -
\prod_{n =1/2}^{\infty} (1-\xi^n) \right)
\xi^{M_0^{(o)}} \cr}
\eqno(26)$$
where $M_0^{e,o}$ is the angular momentum for the ground state
with an even or odd number of electrons.

Having edge states generated in this way, two 
questions immediately come to our mind. First,
are the edge states generated with different 
descendants linearly independent to each other? 
Second, do the insertions with all descendant 
fields generated by $j$ and $\psi$ exhaust all 
possible edge states? These are very hard and important questions.
\(26) is actually a counting of linearly independent descendant 
fields. To apply \(26) to count the physical excitations at the edge
we need to address the above two questions.
Let us examine them in turns. 

For the first question, obviously the descendants of
primary fields with different dimensions (which are 
related to the angular momentum quantum numbers) 
generate linearly independent edge 
states.
The hard part of the question is whether different 
descendants at the same level 
will generate linearly independent states or not. Generally
in CFT, linearly independent descendants,
as operators, should have different (or linearly 
independent) sets of correlations which contain {\it arbitrary} 
numbers of electron operators $\psi_\pm$. 
But when the number $N$ of electrons is finite and 
fixed, one can {\it not} claim that different
descendant fields generate
linearly independent correlations,
in particular for descendant fields at arbitrarily 
large levels. So we suggest that the insertions
with different operators generate 
different edge excitations in the thermodynamic limit or
in the large-$N$ limit; though at the moment we do 
not know how to prove it within CFT. 

Now we turn to the question of whether the descendant fields 
discussed above can generate all possible edge excitations
in the system. We would like to first point out that the 
edge wave functions constructed above not only preserve 
the structure of zeros in the ground state as two or three electrons 
approach each other, they may also preserve the structure 
of higher-order zeros for four or more electrons 
approaching each other. If the wave functions generated 
by the descendant fields do not exhaust the zero-energy 
sector of a three-body Hamiltonian, in principle it is possible 
to construct a more restrictive Hamiltonian that contains
additional two-body, 
three-body, four-body, \etc interactions, so that 
the new Hamiltonian makes the wave functions 
constructed above be and exhaust its zero-energy 
states. Thus, the question of whether 
the space of edge states contains more states depends
on the dynamics of electron interactions, and cannot be 
addressed by merely studying the wave functions. 
In the following, we will assume that the Hamiltonian 
satisfies the above conditions. 

Under the above two assumptions we see that there is a one-to-one
correspondence between the descendant 
fields and the edge states (in the large $N$ limit). Thus
one can use the known results \(26) about the descendant 
fields derived from the CFT
to obtain the 
spectrum of low-lying edge states. 
Numerical diagonalizations for small 
systems have been done to test the predictions \refto{hnab}. 
It was found that indeed the numerical results shows
the violation of the suggested correspondence 
for finite $N$ at large level $l$. On the other hand,
they verify the validity of the correspondence
at the levels less than a certain number of order 
$N$. 

From \(26) one finds
the numbers of states (NOS) of the low lying edge excitations at total
angular momenta $M_0+\De M$ (where $M_0$ is the angular momentum of the
ground state) \refto{hnab}
$$\matrix{       \De M :&0&1&2&3& 4& 5& 6& 7&  8& \hbox{sector}\cr
             \hbox{NOS}:&1&1&3&5&10&16&28&43& 70& N=\hbox{even}\cr
             \hbox{NOS}:&1&2&4&7&13&21&35&55& 86& N=\hbox{odd} \cr
}
\eqno(p6)$$
The spectrum \(p6) is very different from the edge spectrums of the
abelian states.
The spectrum of one branch of edge excitations in abelian states is
given by
$$\matrix{    \De M: &0&1&2&3&4&5& 6& 7& 8\cr
\hbox{NOS}: &1&1&2&3&5&7&11&15&22\cr}
\eqno(p5)$$
while the  spectrum of two branches of abelian edge excitations is
$$\matrix{    \De M: &0&1&2& 3& 4&5 & 6&  7&  8\cr
\hbox{NOS}: &1&2&5&10&20&36&65&110&180\cr}
\eqno(p7)$$
It is clear that in addition to its special dependence on $N$, 
the number of the edge excitations in the p-wave state is more
then one branch but less than two branches of edge excitations in
abelian states.
The specific heat (per unit length) of the p-wave edge state turns out
to be
$3/2$ times of ${\pi \o 6}{T\o v}$ -- the specific heat of one branch of edge 
excitations of the abelian states.
In this sense, we may say that
the p-wave state contains one and a half branches of edge excitations. 

In the above we merely discussed the spectrum of the edge excitations.
We can go one step further. Both the Gaussian model and the 
Majorana fermion theory are equipped with a natural definition of
inner product in their Hilbert space. We may assume the natural
inner product in the CFT happens to be the physical inner product
between the edge wave functions generated by the corresponding operators
(in the large $N$ limit).
Under this assumption we may use the known correlation functions
of the CFT to calculate the correlation functions of physical operators
near the edge. For example, the electron operator is given by
$\psi_e$ in \(h5). From the the CFT results $\<\psi_e(z)\psi_e(0)\>=1/z^{m+1}$
we find the electron propagator on the edge to have the form
$$G_e(x,t)\sim {1\o (x-vt)^{g_e}}, \ \ \ \ \ g_e=m+1
\eqno(p8)$$
This result can be confirmed through a
calculation of the electron occupation number $n_l$ in the single particle
state $|l\>$.
Here $l$ is the angular momentum. 
For the p-wave state of 10 electrons with $m=2$,
it was found \refto{hnab} that $n_{l_0}:n_{l_0-1}:...=1:3.06:5.86:8.63:...$.
The theoretical predictions of $n_l$ calculated from the Green function
\(p8) for $g_e=m+1=3$ are given by\refto{edgere}
$1:3:6:10:...$. Here $l_0$  is the angular momentum
of the last occupied orbit ($l_0=17$ for 10 electrons).
The agreement suggests the validity of our assumption about the
correspondence between the inner products \refto{hnab}.

Next let us consider the quasiparticle operator. The CFT that
describes  a p-wave edge state -- the $U(1)\times Ising$ model --
contains the following local operators: $e^{i\a\phi}$, $\psi$ and $\si$.
Here $\si$ is the disorder operator in the $Ising$ model
which changes the boundary condition of the fermion $\psi$. 
Not all the above operators are physical,
\ie create an allowed excitation in the electron system. A physical 
quasiparticle operator
must have a single-valued correlation function with the electron operator.
This condition is closely related to the single-value property
 of the electron wave
function in the presence of the quasiparticle. The condition can be expressed
through the operator product expansion between the electron and quasiparticle
operators:
$$\psi_e(w_1)\psi_q(w_2)\prop (w_1-w_2)^\ga \hat O(w_2) 
\eqno(17aaa)$$
where we require $\ga$ to be an integer. From the OPE
in the $U(1)\times Ising$ model:
$$\eqalign{
e^{i\a\phi(w)}e^{i\be\phi(0)}\prop& w^{{1\o 4}[(\a+\be)^2-\a^2-\be^2]}
e^{i(\a+\be)\phi(0)}\cr
\psi(w)\si(0)\prop &w^{1\o 2}\mu(0)\cr}
\eqno(17ax)$$
we find that the following operators can be identified as quasiparticle
operators:
$\psi_q(x)= e^{i{1\o 2 \sqrt{m}}\phi(x)}\si(x)$.
and $\psi'_q(x)=e^{i{1\o \sqrt{m}}\phi(x)} $.
$\psi_q$ carries charge ${e\o 2m}$ and corresponds to the 
non-abelian quasiparticle
discussed above \refto{MR}.
$\psi'_q$ carries charge ${1\o m}$ and has an
abelian statistics $\th={\pi\o m}$. $\psi'_q$ is created by inserting a unit
flux. 
From the CFT result $\<\si(w)\si(0)\>\prop w^{1/8}$, we find
$\<\psi_q(x,t)\psi_q(0)\>\prop (x-vt)^{-g_q}$ with $g_q={m+2\o 8m}$.
The exponents $g_e$ and $g_q$ can be measured in tunneling experiments between
the edges and provide an experimental test of the non-abelian states.

Another important non-abelian FQH state is the Haldane-Rezayi (HR)
state for spin-1/2 electrons \refto{HR}. The HR state is important because
it is a candidate for the $\nu=5/2$ FQH state observed in
experiments\refto{5/2}.
The HR state is a d-wave-paired 
spin-singlet FQH state:
$$\eqalign{
\Phi_{HR}(z_i,w_i)=&\Phi_m(z_i,w_i)\Phi_{ds}(z_i,w_i) \cr
\Phi_{ds}(z_i,w_i)=& \cA_{z,w}\left( {1\o (z_1-w_1)^2} 
{1\o (z_2-w_2)^2}...\right) \cr
\Phi_m(z_i,w_i)=& \left( \prod_{i<j} (z_i-z_j)^m 
\prod_{i<j} (w_i-w_j)^m \prod_{i,j} (z_i-w_j)^m \right) 
e^{-{1\o 4} \sum_i (|z_i|^2+|w_i|^2)} \cr}
\eqno(11h)$$
which has a filling fraction $1/m$ with $m$ an even 
integer. 
Here $z_i$ ($w_i$) are the coordinates 
of the spin-up (-down) electrons, and the operator
$\cA_{z,w}$ performs separate anti-symmetrizations 
among $z_i$s and among $w_i$s. 

The non-abelian part of the 
bulk wave function of the HR state was shown to be described
by a CFT with central charge $c=-2$ \refto{WW}.
Using this $c=-2$ CFT and the $U(1)$ Gaussian model for the abelian part,
one can calculate the spectrum of edge excitations.
One finds that the edge excitations of an
$N$-electron system with total spin $s$ and a fixed $S_z$ component
$\si$ are described by the following character
$$\Ch_{N,s}(\xi) 
={1-\xi^{2s+1} \o \prod_n (1-\xi^n)^2}~
\xi^{M_0^{(s)}}~,
\eqno(24h)$$
where 
$$M_0^{(s)}=h_{N,s}-Nh_e={m\o 2}N(N-1)+{1\o 8} [(4s+1)^2-1]-N
\eqno(24ha)$$ 
is the minimum angular momentum 
in the spin-$s$ sector. Here $h_{N,s}={m\o 2}N^2+{1\o 8} [(4s+1)^2-1]$,
and $h_e=1+{m\o 2}$.
According to \(24h), 
the number of edge excitations in the spin-$s$ sector 
at angular momentum $L=M_0^s+l$,
is given by
$$\matrix{l:& 0&1&2&3& 4& 5 & \hbox{spin}  \cr
            & 1&1&3&5&10&16 & s=0\         \cr
            & 1&2&4&8&15&26 & s=1/2        \cr
            & 1&2&5&9&18&31 & s=1\         \cr
            & 1&2&5&10&19&34 & s=3/2        \cr
         }
\eqno(4.1hh)$$
which again has very different structures from the abelian edge states.

In summary non-abelian FQH states are a fascinating class of topological fluids,
which very likely exist in real experimental samples (such as the $\nu=5/2$
state). Our understanding for the non-abelian is quite limited.
In particular we do not understand the general mathematical structures
behind the non-abelian topological orders, and this limits our ability
to calculate the physics properties of non-abelian states. 

\head{7. Remarks and acknowledgments}
\taghead{7.}
Many important aspects of topological fluids were not covered. 
I recommand two review artcles \Ref{AZ} (which is fun to read)
and \Ref{FS} that address
some similar issues discussed in this paper
with different emphasizes. They also cover some subjects which is not
discussed here. Many early papers on topological properties of QH states
can be found in the book in \Ref{Sbook}.
In this
paper I include only what I believe to be the most fundamental properties
of topological orders in FQH liquids, in particular those properties
that might have experimental consequences. 
Other important and/or actively studied properties of FQH fluids
include:
\item{1)}
Phase transitions and phase structures of FQH systems \refto{phase};
\item{2)}
Hierarchical structures in multi-layer FQH systems \refto{mlh};
\item{3)}
Superfluid mode and symmetry breaking in some double layer FQH 
states \refto{mmm};
\item{4)}
Structure of ground states of topological fluids on closed 
space \refto{torus,Wtop,WN};
For an abelian FQH state characterized by $K$-matrix $K$, one can show
it has a $({\rm det}K)^g$-fold degeneracy on the genus $g$ Riemann surface;
\item{5)}
$W_{\infty}$ algebra in FQH states \refto{Winf};
\item{6)}
Spin dynamics and symmetries in FQH states \refto{spin};
\item{7)}
... ...

FQH liquids demonstrate extremely rich internal structures. The more we probe,
the more we are fascinated by the endless richness that the nature reveals to 
us.

Most of the work reviewed here are results of pleasant 
collaborations with A. Zee, D.H. Lee,  B. Blok, Q. Niu, Y.S. Wu
and Y. Hatsugai. I would like to thank them for their ideas, 
their imaginations, and their encouragements.
I also greatly benefited from discussions with F.D.M Haldane, F. Wilczek,
P.A. Lee, R.B. Laughlin, 
M. Stone, D. Tsui, E. Fradkin, N. Read, and E.H. Rezayi. 
I would like to thank them for sharing their insights with me.
This work is supported by NSF grant No. DMR-94-11574.
I would also like to  acknowledge the support from
A.P. Sloan Foundation.

\references

\refis{TSG}
D.C. Tsui, H.L. Stormer, and A.C. Gossard, \prl 48, 1559, 1982.

\refis{Lfqh} 
R.B. Laughlin, \prl 50, 1395, 1983.

\refis{LG}
R. Laughlin, \pr B23, 5632, 1981.

\refis{HH}
F.D.M. Haldane, \prl 51, 605, 1983; \\
B. I. Halperin, \prl 52, 1583, 1984; \\
S. Girvin, \prb 29, 6012, 1984;\\ 
A.H. MacDonald and D.B. Murray, \prb 32, 2707, 1985;\\ 
M.P.A. Fisher and D.H. Lee, \prl 63, 903, 1989;\\ 
J.K. Jain, \prl 63, 199, 1989; \pr B41, 7653, 1991.

\refis{H}
B.I. Halperin, \pr B25, 2185, 1982.

\refis{WG} X.G. Wen, \pr B43, 11025, 1991.

\refis{M} A.H. MacDonald, \prl 64, 220, 1990.

\refis{Wtop}
X.G. Wen,
\journal Int. J. Mod. Phys. B, 2, 239, 1990; \pr B40, 7387, 1989.

\refis{WN} X.G. Wen and Q. Niu,  \pr B41, 9377, 1990.

\refis{BW}
B. Blok and X.G. Wen, \pr B42, 8133, 1990; \pr B42, 8145, 1990.

\refis{FZ} J. Fr\"ohlich and A. Zee, \np  B364, 517, 1991.

\refis{MR}
G.~Moore and N.~Read, \np B360, 362, 1991.

\refis{NABW}
X.G. Wen,  \prl 66, 802, 1991.

\refis{TOP}
S. Elitzur,
G. Moore, A. Schwimmer and N. Seiberg, \np B326, 108, 1989.

\refis{NAB}
E. Witten, \journal Comm. Math. Phys., 121, 351, 1989;\\
J. Fr\"ohlich and C. King, \journal Comm. Math. Phys., 126, 167, 1989.

\refis{NABBW}
B. Blok and X.G. Wen, \np B374, 615, 1992.

\refis{WH}
X.G. Wen, \journal Mod. Phys. Lett. B, 5, 39, 1991.

\refis{WT}
X.G. Wen, \pr B44, 5708, 1991;\\
C.~L.~Kane and M.~P.~A.~Fisher
\prl  68, 1220, 1992;
\prb 46, 15233, 1992.  

\refis{LW}
D.H. Lee and X.G. Wen, \prl 66, 1765, 1991.

\refis{WCL}
X.G. Wen, \pr B41, 12838, 1990.

\refis{class} X.G. Wen and A. Zee, \pr B46, 2290, 1992.

\refis{HE}
F.D.M. Haldane,
\journal Bulletin of APS, 35, 254, 1990.

\refis{S}
M. Stone,
\prb 42, 8399, 1990.
\journal Annals of Physics, 207, 38, 1991;
\journal Int. J. Mod. Phys. B, 5, 509, 1991.

\refis{KM}
See, for example, \\ 
P. Goddard and D. Olive, ``Workshop on Unified String Theories, 1985",
eds.~M. Green and D. Gross (World Scientific, Singapore), p.~214; \journal
Inter. J. Mod. Phys., 1, 303, 1986; \\  
V.G. Kac, ``Infinite dimensional Lie
algebra", (Birkhauser, Boston, 1983).

\refis{T}
S. Tomonaga, \journal Prog. Theor. Phys. (Kyoto), 5, 544, 1950.

\refis{GL0} 
S. M. Girvin and A. H. MacDonald, \prl 58, 1252, 1987.

\refis{GL} 
S. C. Zhang, T. H. Hansson and S. Kivelson, \prl 62, 82, 1989; \\ 
N. Read, \prl 62, 86, 1989; \\  
Z.F. Ezawa and A. Iwazaki, \pr B43, 2637, 1991.

\refis{CB}
See for example, D. Gross \etal, \np B256, 253, 1985; \\ 
R. Floreanini and R. Jackiw, \prl 59, 1873, 1988.

\refis{JM}
M.D. Johnson and A.H. MacDonald, \prl 67, 2060, 1991.

\refis{spin}
T. Chakraborty, and F.C. Zhang, \prb 29, 7032, 1984;
\prb 30, 7320, 1984;\\
R.C. Clark, \etal, \prl 60, 1747, 1988; \prl 62, 1536, 1989; \\ 
S.~L. Sondhi and S.~A. Kivelson, \pr B46, 13319, 1992; \\ 
A. Balatsky and M. Stone, \pr B43, 8038, 1991.

\refis{FK}
J. Fr\"ohlich and T. Kerler, \np B354, 369, 1991.

\refis{WZR}
X.G. Wen and A. Zee, \np B15, 135, 1990.

\refis{cmhi}
D.H. Lee and X.G. Wen, \prb 49, 11066, 1994.

\refis{WWH} X.G. Wen, Y.S. Wu, and Y. Hatsugai, \np B422, 476, 1994.

\refis{R}
N. Read, \prl 65, 1502, 1990.

\refis{MR}  
G. Moore and N. Read, \np B360, 362, 1991.

\refis{RR}  
N. Read and E.H. Rezayi, Yale and CSU preprint, May 1993.

\refis{Web} F.P. Milliken, C.P. Umbach and R.A. Webb, unpublished.

\refis{CW} C. Chamon and X.G. Wen, \pr B49, 8227, 1994.

\refis{B}
C.W.J. Beenakker, \prl 64, 216, 1990.

\refis{edgere}
X.G. Wen, \journal Int. J. Mod. Phys., B6, 1711, 1992.

\refis{torus} 
F.D.M. Haldane and E.H. Rezayi, \pr B31, 2529, 1985; \\
T. Einarsson, \journal Mod. Phys. Lett. B, 5, 675, 1991; \\
D. Li,  \journal Mod. Phys. Lett. B, 7, 1103, 1993; \\
E. Keski-Vakkuri, and X.G. Wen, \journal Int. J. Mod. Phys., B7, 4227, 1993; \\
G. Christofano, G. Maiella, R. Musto and F. Nicodemi, 
\pl B262, 88, 1991; 
\journal Mod. Phy. Lett., A6, 1779, 1991; 
{\it ibid} {\bf A6}, 2985 (1991); {\it ibid} {\bf A7}, 2583 (1992).

\refis{Winf}
A. Cappelli, C. A. Trugenberger and G. R. Zemba,
               \np B396, 465, 1993;
               \pl B306, 100, 1993; \\
A. Cappelli, V.G. Dunne, C.A. Trugenberger and G. Zemba,
\np 398B, 531, 1993; \\
S. Iso, D. Karabali and B. Sakita, \pl B296, 143, 1992; \\
D. Karabali, \np B419, 437, 1994; \np B428, 531, 1994.

\refis{Fubini}
S. Fubini, \journal Int. J. Mod. Phys., A5, 3553, 1990; \\
S. Fubini and C.A. L\"utken, \journal Mod. Phys. Lett., A6, 487, 1991.

\refis{CFT}
For readers not familiar with CFT, we recommend the reprint
book, ``{\it Conformal Invariance and Applications to Statistical 
Mechanics}'', ed. C. Itzykson, H. Saleur and J.B. Zuber,
(World Scientific, 1988); and the review article 
by P. Ginsparg, in Lectures at Les Houches 
Summer School (1988), Vol. XLIX, ed. E. Brez\'{i}n and J.
Zinn-Just\'{i}n (North Holland, 1989).

\refis{KZ}
V.G. Knizhnik and A.B. Zamolochikov,
\np B247, 83, 1984. 

\refis{HR}
F.D.M. Haldane and E.H. Rezayi, \prl 60, 956, 1988;
{\bf 60}, E1886 (1988).

\refis{ch}
A. Rocha-Caridi, {\it ``Vacuum Vector Representations
of the Virasoro Algebra''}, in {\it Vertex Operators in 
Mathematics and Physics''}, MSRI Publications \# 3
(Springer, Heidelberg, 1984), p. 451.

\refis{com} Consider an operator $\Psi$ of dimension
$h$. The correlation function $\<\Psi(z) ...\>$ (here ``..."
representing other operators) as $z\to \infty$ is 
proportional to $\<\Psi(z)\Psi^\dag(0)\>(1+o(z^{-1}))$,
where $\Psi^\dag$ is the conjugate of $\Psi$.
Thus $\<\Psi(z) ...\> \prop z^{-2h}$ as $z\to \infty$.

\refis{TK} S.A. Trugman and S. Kivelson, \pr B26, 3682, 1985; \\
V.L. Pokrovsky and A.L. Talapov, \journal J. Phys. C, 18, L691, 1985.

\refis{sphere} F.D.M. Haldane, \prl 51, 605, 1983.

\refis{anyon} 
 A. Fetter, C. Hanna and R. Laughlin, \pr B39, 9679, 1989; \\
 W. Chen, F. Wilczek, E. Witten and B. Halperin,
\journal J. Mod. Phys. , B3, 1001, 1989; \\
X.G. Wen and A. Zee, \pr B44, 274, 1991.

\refis{cspin}
V. Kalmeyer and R.B. Laughlin, \prl 59, 2095, 1987; \\
X.G. Wen, F. Wilczek and A. Zee, \pr B39, 11413, 1989.

\refis{srvb} 
S.A. Kivelson, D.S. Rokhsar and J.P. Sethna, \prb 35, 8865, 1987; \\
D.S. Rokhsar and S.A. Kivelson, \prl 61, 2376, 1988; \\
N. Read and B. Chakraborty, \pr B40, 7133, 1989; \\
X.G. Wen, \pr, B44, 2664, 1991.

\refis{topcs} 
X.G. Wen, \journal Int. J. Mod. Pys., B5, 1641, 1991.

\refis{spinv}
 X.G. Wen and A. Zee, \prl 69, 953, 1992; (E) \prl 69, 3000, 1992. 

\refis{Hmult} 
 B. I. Halperin, \journal  Helv. Phys. Acta,  56, 75, 1983.

\refis{dual} 
M.P.A. Fisher and D.H. Lee, in \ref{HH}; \\
X.G. Wen and A. Zee, \journal Int. J. Mod. Phys., B4, 437, 1990.

\refis{ASW} 
 D. Arovas, J. R. Schrieffer, and F. Wilczek, \prl 53, 722, 1984.

\refis{mmm} 
H. Fertig, \pr B40, 1087, 1989; \\
A.H. MacDonald, P.M. Platzman, and G.S.
Boebinger, \prl 65, 775, 1990;  \\
L. Brey, \prl 65, 903, 1990;   \\
S. Sondhi, A. Karlhede, S.A. Kivelson and E.H. Rezayi, \pr B47, 16419, 1993.

\refis{WZmmm} 
S.Q. Murphy, J.P. Eisenstein, G.S. Boebinger, L.N. Pfeiffer and K.W. West,
\prl 72, 728, 1994; \\
X.G. Wen and A. Zee, \prl 69, 1811, 1992; \pr B47, 2265,  1993; \\
K. Yang, K. Moon, L. Zheng, A.H. MacDonald, S.M. Girvin, D. Yoshioka and S.-C.
Zhang, \prl 72, 732, 1994; \\
N. Read, preprint (cond-mat/9501010).

\refis{niu} 
 Q. Niu and D.J. Thouless, \pr B35, 2188, 1987.

\refis{EdgeT}
S.A. Trugman, \pr B27, 7539, 1983; \\
A.H. MacDonald, and P. Streda, \pr B29, 1616, 1984; \\
 P. Streda, J. Kucera and A.H. MacDonald, \prl 59, 1973, 1987; \\
M. Buttiker, \pr B38, 9375, 1988; \\
J.K. Jain and S.A. Kivelson, \pr B37, 4276, 1988; \prl 60, 1542, 1988.

\refis{EdgeE}
P.L. McEuen, \etal, \prl 64, 2062, 1990; \\
A.M. Chang and J.E. Cunningham,
\journal Solid State Comm., 72, 651, 1989; \\
L.P. Kouwenhoven \etal, \prl 64, 69, 1990; \\
J.K. Wang and V.J. Goldman, \prl 67, 749, 1991.

\refis{KFP} 
C. L. Kane, M.P.A. Fisher and J. Polchinski, \prl 72, 4129, 1994.
C. L. Kane, and M.P.A. Fisher, preprint (cond-mat/9409028)

\refis{mp} 
 V.K. Talyanskii \etal, \journal Surface Science, 229, 40, 1990; \\
M. Wassermeier \etal, \pr B41, 10287, 1990.

\refis{mp2/3} 
R.C. Ashoori, H. Stormer, L. Pfeiffer, K. Baldwin and K. West,
\pr B45, 3894, 1992.

\refis{elesta} 
D. B. Chklovskii, B. I. Shklovskii, L. I. Glazman,
\pr B46, 4026, 1992; \\
J. Dempsey, B.Y. Gelfand, and B.I. Halperin, \prl 70, 3639, 1993.

\refis{EG} 
T. Einarsson, S.~L. Sondhi, S.~M. Girvin, and D.~P. Arovas,
preprint (cond-mat/9411078).

\refis{ELi}
T. Einarsson, in \ref{torus}; \\
D. Li, in \ref{torus}.

\refis{hnab} 
X.G. Wen, \prl 70, 355, 1993.

\refis{Hedge}
F.D.M. Haldane, Princeton preprint (cond-mat/9501007)

\refis{WW} 
X.G. Wen and Y.S. Wu, \np B419, 455, 1994.

\refis{WWH} 
X.G. Wen, Y.S. Wu, and Y. Hatsugai, \np B422, 476, 1994.

\refis{5/2} 
R. Willett, J.P. Eisenstein, H.L. Str\"ormer, D.C. Tsui,
A.C. Gossard, and J.H. English, \prl 59, 1776, 1987.

\refis{phase}
 H.P. Wei, D.C. Tsui, M. Paalanen, and A.M.M. Pruisken,
\prl 61, 1294, 1988; \\
S. Koch, R. Haug, K. v. Klitzing, and K. Ploog,
\prl 67, 883, 1991; \\
  H.P Wei, S.W. Hwang, D.C. Tsui, and A.M.M. Pruisken,
 \journal Surf. Sci., 229, 34, 1990; \\
 H.W. Jiang, C.E. Johnson, K.L. Wang, and S.T. Hannahs, \prl 71, 1439, 1993; \\
  A.M.M. Pruisken, \prl 61, 1297, 1988; \\
S.A. Kivelson, D-H Lee and S-C Zhang, \pr B46, 2223, 1992; \\
  D-H Lee, Z. Wang and S.A. Kivelson, \prl 70, 4130, 1993; \\
Ziqiang Wang, Dung-Hai Lee, and Xiao-Gang Wen, \prl 72, 2454, 1994.

\refis{mlh}
Y.W. Suen {\it et\ al.,\/} \prl  68, 1379, 1992; \\
 J.P. Eisenstein {\it et\ al.,\/} \prl  68, 1383, 1992.

\refis{FS} J. Fr\"ohlich and U.M. Studer,
\journal Rev. of Mod. Phys., 65, 733, 1993;\\
J. Fr\"ohlich, U.M. Studer and E. Thiran, 
lectures presented by J.F. at the 1994 Les Houches Summer
School ``Fluctuating Geometries in Statistical Mechanics and Field Theory'',
cond-mat/9508062.

\refis{AZ}
A. Zee, ``Quantum Hall Fluids'', cond-mat/9501022.

\refis{Sbook} 
M. Stone, Ed. {\it Quantum Hall effects} (World Scientific, Singapore, 1991).

\endreferences

\vfil\eject

\head{Figure captions}

\item{Fig 1.1} A 1D crystal passing an impurity will generate a narrow
band noise in the voltage drop.
\item{Fig 1.2} A FQH fluid passing through a constriction will generate
narrow band noises due to the back scattering of the quasiparticles.
\item{Fig 6.1}
The energy spectrum of a system
of six electrons in the first 22 orbits for the three-body Hamiltonian.
The degeneracies of the zero energy states at $M=45,...,51$  are found to be
$1,1,2,3,5,7,11$.

\end